\documentclass[%
nofootinbib,
 amsmath,amssymb,
prc,
]{revtex4-2}

\usepackage{graphicx}
\usepackage{dcolumn}
\usepackage{bm}
\usepackage{hyperref}

\usepackage{color,soul}

\DeclareMathOperator{\Div}{div}

\renewcommand{\vec}{\mathbf}

\newcommand{\pFa}{{p}_{F_{\alpha}}}

\newcommand{\thp}{\mbox{$\theta_{\vec p}^{(\alpha)}$}}
\newcommand{\thpQ}{\mbox{$\theta_{{\vec p}+{\vec Q}_{\alpha}}^{(\alpha)}$}}

\newcommand{\thpsQ}{\mbox{$\theta_{{\vec p} ' +{\vec Q}_{\alpha'}}^{(\alpha')}$}}

\newcommand{\vp}{{\vec p}}         
\newcommand{\vps}{{\vp '}}         
\newcommand{\vQ}{{\vec Q}}         
\newcommand{\vQa}{\vQ_{\alpha}}   
\newcommand{\vQas}{\vQ_{\alpha'}}   
\newcommand{\vQb}{\vQ_{\beta}}   

\renewcommand{\vr}{{\vec r}}

\newcommand{\mN}{\mbox{$\mathcal N$}}
\newcommand{\mF}{\mbox{$\mathcal F$}}

\newcommand{\uup}{\mbox{$u_{\vec p}^{(\alpha)}$}}
\newcommand{\vvp}{\mbox{$v_{\vec p}^{(\alpha)}$}}
\newcommand{\uupsq}{\mbox{$u_{\vec p}^{(\alpha)2}$}}
\newcommand{\vvpsq}{\mbox{$v_{\vec p}^{(\alpha)2}$}}

\newcommand{\fp}{\mbox{$\mathfrak{f}_{\vec p}^{(\alpha)}$}}
\newcommand{\np}{\mbox{$n_{\vec p}^{(\alpha)}$}}

\begin{document}
\title{Diffusion in superfluid Fermi mixtures: General formalism}
\date{\today}
\author{ O. A. Goglichidze}
\email{goglichidze@gmail.com}
\author{M. E. Gusakov}
\affiliation{Ioffe Institute, Politekhnicheskaya 26, 194021 St. Petersburg, Russia}
 


\begin{abstract}
	With neutron star applications in mind, 
	we developed a theory 
	of diffusion in 
	mixtures of superfluid, strongly interacting Fermi liquids.
	By employing the Landau theory of Fermi liquids, we determined matrices that relate the 
	currents of different particle species, their momentum densities, and the partial entropy 
	currents to each other.
	Using these results,
	and applying the quasiclassical 
	kinetic equation
	for the Bogoliubov excitations, 
	we derived general expressions for the diffusion coefficients, 
	which properly incorporate all the Fermi liquid effects and 
	depend on the momentum transfer rates between different particle species.
	The developed framework can be used as a starting point for systematic 
	calculations of the diffusion coefficients (as well as other kinetic coefficients)
	in superfluid Fermi mixtures,
	particularly,
	in superfluid neutron stars.
\end{abstract}

\maketitle

\section{Introduction}

{The present work aims at developing a 
	framework for studying the transport properties (in particular, diffusion)
	in superfluid strongly interacting Fermi mixtures 
	and at elucidating the role of the Fermi-liquid effects in shaping these properties.
	Although this problem is quite general, its solution 
	is of particular 
	relevance to
	the physics of neutron stars (NSs),
	and we will always keep NSs in mind when discussing it below.
	
	Neutron stars are compact objects with a mass 
	of about $M\approx 1.4 M_\odot$ 
	and a radius $R \approx 12$~km ($M_\odot$ is the solar mass). 
	The density in their inner layers exceeds the nuclear saturation density, $2.8 \times 10^{14}$ 
	g cm$^{-3}$, making these objects unique astrophysical laboratories for  studying  superdense 
	matter and testing such fundamental physical theories as the theory of  strong interactions, 
	general relativity theory, and many-body quantum theory \cite{hpy07}. 
	Despite the fact that NSs were discovered more than 50 years ago, the equation of state,  and 
	even 
	the composition of matter in their deepest layers are still 
	not well understood.
	Various  theoretical models predict a composition ranging from a purely nucleon one (neutrons, 
	protons, and electrons with admixture of muons) to nucleon-hyperon and quark matter. 
	Laboratory studies of such dense and strongly degenerate matter are not feasible and the only 
	way 
	of testing the theories is to compare NS observations with  predictions from theoretical models.
	
	An important and 
	often
	crucial feature of not too hot NSs is the presence of baryon 
	superfluidity/superconductivity in their interiors (see, e.g., Ref. 
	\cite{LombardoSchulze2001,sc19}).
	Superfluidity of baryons (e.g., neutrons and protons in the case of the simplest NS matter 
	composition) has a dramatic impact on the NS dynamics by substantially modifying the  
	\mbox{(magneto-)hydrodynamic} equations  
	\cite{Khalatnikov_book,Khalatnikov1957,MendellLindblom1991,gas11, GusakovDommes2016, rw20, 
	DommesGusakov2021} and strongly affecting 
	the 
	dissipative properties of NS matter (see, e.g., the review \cite{SchmittShternin2018}).
	
	One of the potentially interesting dissipative mechanisms in NSs is associated with particle 
	diffusion.
	Indeed, since the  NS matter consists of a number of different strongly interacting particle 
	species, the departure from 
	the diffusion
	equilibrium can lead to effective dissipation of the mechanical (and magnetic) energy through 
	the 
	diffusion currents (see, e.g., \cite{LL6}). 
	The diffusion coefficients in a normal (nonsuperfluid) mixture of Fermi-liquids
	were calculated by Anderson et al.\ \cite{AndersonPethickQuader1987}.
	Later, Yakovlev and Shalybkov \cite{YakovlevShalybkov1990,YakovlevShalybkov1991}, 
	focusing mainly on the related mechanism of the electric conductivity, 
	outlined the derivation of the expression for the diffusion tensor in the presence of the 
	magnetic 
	field. 
	When calculating the momentum transfer rates, these authors adopted the free-particle model 
	\cite{YakovlevShalybkov1991b}.
	Subsequently, their calculations were improved by taking into account nuclear in-medium effects 
	\cite{Shternin2008,ShterninVidana2013,ShterninBaldo2020}.
	The diffusion and electrical conductivity were also analyzed for quark matter in Refs.\
	\cite{HeiselbergPethick1993,AlfordNishimuraSedrakian2014}.
	An importance of the diffusion effects was revealed for the evolution of the magnetic field
	in NSs in Refs.\
	\cite{GoldreichReisenegger1992,ShalybkovUrpin1995,UrpinShalybkov1999,HoyosReiseneggerValdivia2008,HoyosReiseneggerValdivia2010,GlampedakisJonesSamuelsson2011,BeloborodovLi2016,CastilloReiseneggerValdivia2017,PassamontiEtAl2017,GusakovKantorOfengeim2017,KantorGusakov2018,GusakovKantorOfengeim2020}.
	Moreover, recently it was argued that, under certain circumstances, the particle diffusion may
	become a leading dissipative agent for damping of  neutron star oscillations  
	\cite{KraavGusakovKantor2021}.

	Most of the works mentioned above were devoted to studying  nonsuperfluid Fermi mixtures. 
	Since NS matter is generally superfluid and superconducting, 
	the results obtained in these works are of limited scope.
	Cooper pairing of particles in a mixture leads to several important effects that can 
	potentially 
	affect diffusion.
	First of all, superfluidity leads to the appearance of the energy gap in the dispersion 
	relation 
	for the elementary Bogoliubov thermal excitations.	
	An impact of this effect on the diffusion coefficients was analyzed in Ref.\ 
	\cite{OstgaardYakovlev1992}.
	Second, the number of Bogoliubov excitations is not conserved in the collisions (see, e.g., 
	\cite{BhattacharyyaPethickSmith1977}), 
	which significantly complicates
	all calculations related to particle scatterings.
	Third, the superconductivity (i.e., the superfluidity of charged particles) 
	noticeably
	modifies 
	the screening properties 
	of a 
	mixture
	and affects
	the electromagnetic interactions between different charged particle  
	species 
	\cite{HeiselbergPethick1993, ShterninYakovlev2007,ShterninYakovlev2008,Shternin2018}.
	All these effects substantially complicate the collision integral.	
	Note, however, that in a strongly interacting  Fermi mixture, 
	besides the dissipative interaction described by the collision integral, 
	one should also account for the nondissipative  interaction between particles, the so called 
	Fermi-liquid effects.
	To our best knowledge, the influence of these effects on the particle diffusion in superfluid 
	mixtures has not been studied in the literature.
	Still, the  theory of transport processes in such systems cannot be  developed in a consistent 
	way 
	without taking Fermi-liquid effects into account. 
	The aim of the present paper is to fill this gap and to introduce a formalism allowing one to 
	calculate the main transport coefficients, in particular, diffusion in  strongly interacting 
	superfluid Fermi mixtures within the framework of 
	the Landau Fermi-liquid theory.

	The paper is organized as follows. 
	Section \ref{sec:sf_currents} provides basic definitions and notations.
	It also briefly discusses, following Ref.\ \cite{GusakovKantorHaensel2009b}, 
	the method of deriving the superfluid entrainment matrix,
	as this method shares many common features with the approach adopted in the present work.
	In Sec. \ref{sec:normal_currents}, the expressions for the normal currents 
	are obtained for a simplified problem 
	ignoring the dissipative interaction between 
	different particle species.
	In Sec. \ref{sec:diffusion}, the same problem is considered in the framework of kinetic theory.
	The relation between the normal currents and the chemical potential gradients is found, the  
	expressions for the diffusion coefficients are derived, and the entropy generation equation is 
	presented.
	Section \ref{sec:electric_field} generalizes the results obtained above to charged mixtures.
	Section \ref{sec:conclusions} presents summary of our results.
	The paper also contains a number of Appendices. 
	In Appendix \ref{ap:coeffs}, the various ``entrainment matrices'' 
	introduced in the paper, are given in different limiting cases.
	Appendices \ref{ap:hydrodynamics} and \ref{ap:non-rel-hydr} describe, respectively, the 
	general 
	equations of the relativistic superfluid hydrodynamics 
	and the same equations in the limit of small fluid velocities. 
	In Appendix \ref{sec:col_mat} we discuss the effective interaction Hamiltonian for Bogoliubov 
	excitations.
	Finally, Appendix \ref{sec:Integral} presents the collision integrals for 
	Bogoliubov excitations, as well as the formal derivation of the expressions for
	the momentum transfer rates. 
	
	Throughout this paper we will use the system of units in which the Planck constant $\hbar$, 
	the  Boltzmann constant $k_B$, and the normalization volume $V$ equal unity ($\hbar = k_B  
	=V=1$).  
	However, we will 
	not set 
	the speed of light $c$ equal to 1,
	since we 
	will be dealing mostly with the nonrelativistic hydrodynamic velocities. 
	
}

\section{Superfluid currents in a Fermi-liquid mixture}
\label{sec:sf_currents}

Let us  consider a mixture of two interacting superfluid Fermi liquids which we  label by 
the indices
``$n$'' and ``$p$''.  
In spite of the obvious association with neutrons and protons, up to Sec.
\ref{sec:electric_field}, we will assume that both constituents are uncharged fluids. 
In what follows, the indices $\alpha$, $\alpha'$, and $\alpha''$ 
run over particle species. 
The index $\beta$ labels the particle 
species different from the species $\alpha$ ($\beta \neq \alpha$).
We deal only with spin-unpolarized matter.
This allows us to disregard the spin dependence of various quantities and 
treat them as 
spin-averaged functions whenever 
possible.

One of the key features of the hydrodynamics and kinetics of superfluid mixtures is the so-called 
entrainment effect, which manifests itself in the fact that superfluid currents are, generally, not 
parallel to superfluid velocities \cite{AndreevBashkin1976}. 
In particular, in the case of a two-component nonrelativistic mixture, the 
currents can be represented as 
\begin{eqnarray}
		{\vec J}_{ n} &=& (\rho_{ n} - \rho_{ nn}-
		\rho_{ np}) \, {\vec V}_{ q} +
		\rho_{ nn} \,  {\vec V}_{s{ n} } +
		\rho_{ np} \, {\vec V}_{s{ p} },
		\label{eq:Jn_nonrel} \\
		{\vec J}_{ p} &=&
		(\rho_{ p} - \rho_{ pp}-
		\rho_{ pn}) \, {\vec V}_{ q} +
		\rho_{ pp} \, {\vec {\vec V}}_{s{ p}} +
		\rho_{ pn} \, {\vec {\vec V}}_{s{ n} },
		\label{eqJp_nonrel}
\end{eqnarray}
where ${\vec J}_{\alpha}$ is the mass-current density ($\alpha=n, p$), $\vec{V}_{ s\alpha}$ is 
the superfluid 
velocity, $\vec{V}_{ q}$ is the velocity of 
thermal excitations (normal liquid component), and $\rho_{\alpha\alpha'}$ is the so-called 
Andreev-Bashkin 
matrix {(also known as entrainment or mass-density matrix)}.
The elements of this matrix
were calculated
for both nonrelativistic and relativistic mixtures 
in a series of papers
{(see, e.g., Refs.\  
\cite{BorumandJoyntKluzniak1996,GusakovHaensel2005,ChamelHaensel2006,GusakovKantorHaensel2009,
GusakovKantorHaensel2009b, son16, Leinson2018,Leinson2018,ChamelAllard2019,AllardChamel2021}).}
In this section we briefly outline the calculation of the entrainment matrix based on the 
relativistic Landau Fermi-liquid theory \cite{BaymChin1976}, 
closely following the work of Gusakov et al.\ \cite{GusakovKantorHaensel2009b}.

In the case of relativistic fluids, 
it is more convenient to work with the particle current densities $\vec{j}_\alpha$ instead of
the mass current densities ${\vec J}_{\alpha}$. The former can be represented as 
\cite{GusakovKantorHaensel2009}
\begin{equation}
	\label{eq:j_def}
	\vec{j}_\alpha = \left(n_\alpha - \sum_{\alpha'} \mu_{\alpha'} Y_{\alpha\alpha'}\right)\vec{u} + c^2 \sum_{\alpha'} Y_{\alpha\alpha'} \vec{Q}_{\alpha'},
\end{equation}
where $\vec{u}$ is the spatial component of the four-velocity $u^\mu$, normalized by condition 
$u^\mu u_\mu = -c^2$, 
and describing the motion of normal liquid component; 
$n_\alpha$ and $\mu_\alpha$ are, respectively, the number density and relativistic 
chemical potential 
of particle species $\alpha$ measured in the frame, in which 
$u^\mu =  (c, 0, 0, 0)$.
Finally, $\vec{Q}_{\alpha}$ is the half Cooper-pair momentum.
In the nonrelativistic limit,
\begin{equation}
		\vec{u} = \vec{V}_{ q}, \ \ \ \ \  \rho_ {\alpha\alpha'} = m_\alpha m_{\alpha'} c^2 Y_{\alpha\alpha'},
	\end{equation}
where $m_\alpha$ is the 
bare mass
of particle species $\alpha$.

\subsection{Basic definitions}

Throughout the paper we assume that $Q_\alpha/p_{F\alpha} \ll 1$, as well as that 
$Q_\alpha/m_\alpha \ll c$, 
where $p_{F\alpha}$ is the Fermi momentum for particle species 
$\alpha$. 	
In this case,
the energy {density $E$} of the system can be represented as 
\cite{GusakovKantorHaensel2009b,Gusakov2010}
\begin{align}
		&E- \sum_\alpha \breve\mu_\alpha n_\alpha =
		\sum_{\vp \sigma \alpha}
		\left(\varepsilon_{0}^{(\alpha)}\left(\vp + \vQa \right) - \breve\mu_\alpha\right) \,
		\left( \mN_{\vp + \vQa }^{(\alpha)} - \thpQ \right)
		\nonumber \\
		&+ \frac{1}{2}  \sum_{\vp \vps  \sigma \sigma ' \alpha \alpha '}
		f^{\alpha \alpha'}\left(\vp + \vQa, \vps + \vQas \right) 
		\left( \mN_{\vp + \vQa }^{(\alpha)} - \thpQ \right)
		\left( \mN_{\vps + \vQas }^{(\alpha')} - \thpsQ \right)
		\nonumber \\
		&- \sum_{\vp \alpha}  \Delta_{\vp}^{(\alpha)}
		\uup \vvp 
 		\left(  1- \mF_{\vp + \vQa}^{(\alpha)}
		- \mF_{-\vp + \vQa }^{(\alpha)}\right).
		\label{eq:energy}
\end{align}
Here,
$\mN_{\vp + \vQa }^{(\alpha)}$ and $\mF_{\vp + \vQa}^{(\alpha)}$ are, respectively, the 
distribution functions for Landau quasiparticles and Bogoliubov excitations of particle 
species $\alpha$, 
$\thp = \theta \left( \pFa - |\vp| \right)$, $\theta(x)$ is the step function,
$\varepsilon_{0}^{(\alpha)} \left(\vp \right)$ is the first variation of the system energy 
{density} $E$ 
with respect to the distribution function $\mN_{\vp + \vQa }^{(\alpha)}$ 
calculated for the normal Fermi-mixture, 
$f^{\alpha \alpha'} \left(\vp, \vps \right)$ is the second variation, also called the 
(spin-averaged) Landau quasiparticle interaction function \cite{PinesNozieres},
$\breve{\mu}_\alpha$ is the nonequilibrium analog of  the chemical potential ${\mu}_\alpha$
{(to be specified below)},
$ \Delta_{\vp}^{(\alpha)}$ is the Fourier component of the superfluid order parameter, and
$\uup$ and $\vvp$ are the Bogoliubov coherence factors related by the normalization condition,
\begin{equation}
	\label{eq:upvp1}
	\uupsq+\vvpsq = 1.
\end{equation}
The expression \eqref{eq:energy} formally contains summations over the spin indices $\sigma$ and 
$\sigma'$. However, since the matter is assumed to be spin-unpolarized, we omit these indices 
in the distribution functions.
The 
quasiparticle distribution function $\mN_{\vp + \vQa }^{(\alpha)}$ 
is related to the corresponding distribution function for Bogoliubov excitations $\mF_{\vp + 
\vQa}^{(\alpha)}$ by the equation   
\begin{equation}
    		\label{eq:Np_def}
    		\mN_{\vp + \vQa}^{(\alpha)}  
    		= \vvpsq +
    		\uupsq \, \mF_{\vp + \vQa }^{(\alpha)}
    		- \vvpsq \, \mF_{-\vp + \vQa }^{(\alpha)}.
\end{equation}

The Bogoliubov coherence factors can be found by minimizing the quantity $E- \sum_\alpha 
\breve\mu_\alpha n_\alpha$ with respect to $\uup$, taking into account the condition 
\eqref{eq:upvp1} and treating the distribution functions  $\mF_{\vp + \vQa }^{(\alpha)}$ as fixed 
parameters \cite{Gusakov2010}.%
\footnote{ 
{We emphasize that the
vectors $\vQa$ should also be
treated 
fixed
when varying 
all the thermodynamic potentials 
considered in the present paper.}
}
The result is
\begin{equation}
     		\uupsq = \frac{1}{2} \, \left( 1 + { H_{\vp + \vQa }^{(\alpha)} +H_{-\vp + \vQa }^{(\alpha)} \over 2 \mathfrak{E}_{\vp + \vQa }^{(\alpha)} + H_{-\vp + \vQa }^{(\alpha)} - H_{\vp + \vQa }^{(\alpha)}} \right),
    		\label{upQ}    
\end{equation}
where 
\begin{equation}
		\mathfrak{E}_{\vp + \vQa }^{(\alpha)}
		= \frac{\delta \left( E  - \sum_\alpha \breve\mu_\alpha n_\alpha\right) }{\delta \mF_{\vp + \vQa }^{(\alpha)} }
		= \frac{1}{2} \, \left( H_{\vp + \vQa }^{(\alpha)} -H_{-\vp + \vQa }^{(\alpha)} \right)  
    		+ \sqrt{ \frac{1}{4}  \left( H_{\vp + \vQa }^{(\alpha)} +H_{-\vp + \vQa }^{(\alpha)}\right)^2 + \Delta_{\vec p}^{(\alpha) 2} }
    		\label{distribution1} 
\end{equation}
is the energy of a Bogoliubov excitation, 
while
\begin{equation}
   		 H_{\vp+\vQa}^{(\alpha)} = 
   		 \varepsilon_0^{(\alpha)}\left(\vp + \vQa \right)  - \breve\mu_\alpha
    		+ \sum_{\vps\sigma' \alpha'}
    		f^{\alpha \alpha'}(\vp+\vQa, \vps + \vQas)
     		\left(  \mN_{\vps + \vQas}^{(\alpha')} - \thpsQ \right)
		\label{eq:localenergy}
\end{equation}
is the quantity that formally coincides with 
the energy of a Landau quasiparticle in a nonsuperfluid matter 
\cite{PinesNozieres,GusakovHaensel2005}. 
It is easy to see that the Bogoliubov coherence factors are even in $\vec{p}$: $\uup = u_{-\vec 
p}^{(\alpha)}$, $\vvp = v_{-\vec p}^{(\alpha)}$. 
The nonequilibrium chemical potential $\breve{\mu}_\alpha$ 
is 
determined 
from
the requirement that, for a given (generally, nonequilibrium) 
distribution  function of Bogoliubov excitations
$\mF_{\vp + \vQa}^{(\alpha)}$
the summation of the quasiparticle distribution function $\mN_{\vp + \vQa}^{(\alpha)}$ over the 
quantum states 
gives
the particle number density,
\begin{equation}
		n_{\alpha} = \sum_{\vp \sigma} \mN_{\vp + \vQa}^{(\alpha)}.
\end{equation}
In the vicinity of the Fermi surface the absolute values of the arguments of the function  
$f^{\alpha \alpha'}(\vp,\vps)$ can be approximately put equal
to $p \approx \pFa$ while the function itself can be expanded into Legendre polynomials $P_l (\cos 
\Theta)$:
\begin{equation} 
		f^{\alpha \alpha'} \left( {\vec p}, {\vec p}' \right) = \sum_l f_l^{\alpha \alpha'} P_l (\cos \Theta),
\end{equation}
where $\Theta$ is the angle between the vectors $\vec{p}$ and $\vec{p}'$,
$f_l^{\alpha \alpha'}$ are the symmetric Landau parameters ($f_l^{\alpha \alpha'} = 
f_l^{\alpha'\alpha}$).
{The Landau parameters and the particle effective mass, defined as
\begin{equation}
	\label{eq:m_eff_def}
	m_\alpha^*=\frac{p_{F \alpha}}{v_{F \alpha}},
\end{equation}
are related by the equation 
\cite{GusakovKantorHaensel2009}
\begin{equation}
		\label{eq:m_eff_eq}
		{\mu_\alpha \over  m^*_\alpha c^2} = 1 - \sum_{\alpha'} {G_{\alpha\alpha'} \mu_{\alpha'} \over  n_\alpha c^2}.
\end{equation}
In these formulas $v_{F\alpha}$ is the Fermi velocity and 
the symmetric matrix $G_{\alpha\alpha'}$ 
is given by
}
\begin{equation}
	\label{eq:G_matrix}
		G_{\alpha\alpha'} = \frac{1}{9\pi^4} p_{F\alpha}^2 p_{F\alpha'}^2 f_1^{\alpha\alpha'}.
\end{equation}
The Fermi momentum $p_{F\alpha}$ is expressed through the particle number density by the standard 
{formula:
$p_{F\alpha} = (3 \pi^2 n_\alpha)^{1/3}$.}
The equilibrium distribution function for Bogoliubov excitations can be found from minimization of  
the thermodynamic potential 
\begin{equation}
		\label{eq:F_potential}
		F = E  - \sum_\alpha \mu_\alpha n_\alpha - TS
	\end{equation}
with respect to $\mF_{\vp + \vQa }^{(\alpha)}$. 
{Here $T$ is the temperature, 
$S$ is the system entropy {density}, and 
$E  - \sum_\alpha \mu_\alpha n_\alpha$ is given by the expression \eqref{eq:energy}, 
where
the chemical potential $\breve\mu_\alpha$, taken in equilibrium, 
is denoted as $\mu_\alpha$.
}
The entropy {density} $S$ is given by the standard combinatorial expression
\begin{equation}
		\label{eq:entropy}
		S =   \sum_{{\vec p} \sigma \alpha} \sigma_{\vp + \vQa}^{(\alpha)} ,
\end{equation}	
where  
\begin{gather}
		\label{eq:entropy_dens}
		\sigma_{\vp + \vQa}^{(\alpha)}  = 
  		  -\left( 1- \mF_{\vp + \vQa}^{(\alpha)} \right) \ln \left( 1- \mF_{\vp + \vQa}^{(\alpha)}\right)
  		  -  \mF_{{\vec p}+ {\vec Q}_{\alpha}}^{(\alpha)}  \ln  \mF_{\vp + \vQa}^{(\alpha)}.
\end{gather}
Taking the variation of $F$
and equating the result to zero, one obtains the standard Fermi-Dirac distribution function 
\cite{GusakovKantorHaensel2009}
\begin{equation}
     		\mF_{\vp + \vQa }^{(\alpha)} = 
    		\frac{1}{ 1 + { e}^{\mathfrak{E}_{\vp + \vQa }^{(\alpha)}/T}}.
    		\label{eq:Fp} 
\end{equation}
This is the equilibrium distribution function for Bogoliubov thermal excitations
in the reference frame, in which 
the normal velocity $\vec{u}$ vanishes, $\vec{u}=0$.
The equilibrium distribution function in an arbitrary frame
can be calculated by minimizing
the thermodynamic potential \eqref{eq:F_potential}
{with an additional
constraint
that fixes the momentum density associated with Bogoliubov excitations; 
see Sec.\ \ref{sec:normal_currents} for details.}

Throughout this paper, we make the assumption that all the pairing gaps
	$\Delta_{\vp}^{(\alpha)}$ 
	are isotropic, meaning they do not depend on the direction of the vector 
	$\vp$. 
	This assumption requires further clarification.
	As mentioned in the Introduction, neutron stars serve as one of the primary applications for 
	the results obtained in this study.
	It is widely accepted that neutrons in the cores of neutron stars form Cooper pairs 
	in the triplet $^3P_2$ state (see, however, Ref.\ \cite{KrotscheckEtAl2023} for an alternative 
	viewpoint). In such a state, the energy gap is anisotropic (see, e.g., Ref.\
	\cite{YakovlevLevenfishShibanov1999}), and it depends on the direction of the quantization axis.
	The presence of this additional preferred direction significantly complicates 
	the analysis,
	leading to the diffusion and other coefficients obtained in the paper 
	becoming tensors rather than scalars.
To avoid this difficulty,
it is usually assumed (e.g., Refs. \cite{BaikoHaenselYakovlev2001, 
	HaenselLevenfishYakovlev2002c,ManuelSarkarTolos2014, GusakovHaensel2005, 
	GusakovKantorHaensel2009b})
that the matter of the neutron star cores consists of a collection 
of microscopic domains with randomly oriented quantization axes (see, however, Refs.\
\cite{Leinson2017, Leinson2018}).
Then, after averaging over the volume containing a large number of domains, the preferred direction 
will disappear.
In this case,  
it is reasonable to expect that
accounting for 
the microscopic anisotropy of the energy gap does not result in 
conceptually new effects 
and can be treated within the formalism developed 
for the isotropic gaps. 
The only thing that needs to be done is to establish a relation between a given anisotropic gap and 
the effective isotropic gap
(see Sec.\ \ref{entrmatr} below).
%
This is the strategy we choose to adopt in the present work, which aligns with the method used 
in calculating the entrainment matrix components in Refs.\ \cite{GusakovHaensel2005, 
GusakovKantorHaensel2009b}.
The authors of the recent work \cite{AllardChamel2021} also restrict themselves to considering the 
isotropic neutron gaps.
As pointed out by Leinson \cite{Leinson2017}, who studied the anisotropic gaps, 
such an approach can be helpful 
in the case of small hydrodynamic velocities, which is relevant to our study. 
A similar 
approach
is commonly employed in calculating transport coefficients \cite{BaikoHaenselYakovlev2001, 
HaenselLevenfishYakovlev2002c,ManuelSarkarTolos2014}, 
while
the method of averaging the 
angle-dependent neutron gap (not related to domain averaging) is often used to simplify 
first-principle calculations of the gap itself 
(see, e.g., 
Refs.\ \cite{LombardoSchulze2001,DeanHjorthJensen2003, ding_etal16}).

\subsection{Calculation of the relativistic entrainment matrix}
\label{entrmatr}

Throughout this paper
we work 
under the assumption 
that 
it is always possible to choose a reference frame
in which all the hydrodynamic velocities in the system are small, 
in particular,
$Q_\alpha/p_{F\alpha} \ll 1$.%
%
\footnote{For the sake of brevity, 
the velocity $\vec{u}$ as well as the momenta $\vec{Q}_\alpha$ are further 
referred to as the hydrodynamic velocities.}
%
Physically, this means that all the relative velocities are assumed to be small.
Restricting ourselves to the linear approximation in $Q_\alpha/p_{ F\alpha}$,
we can 
rewrite the 
``energy''
\eqref{eq:localenergy} as
\begin{equation}
		H_{\vp + \vQa}^{(\alpha)} = \varepsilon_{\vp}^{(\alpha)}  + \Delta H_{\vp}^{(\alpha)},
 	   \label{eq:expand1}
\end{equation}
where 
$\varepsilon_{\vp}^{(\alpha)}$ is, by definition, 
the energy
$H_{\vp + \vQa}^{(\alpha)}$ in the system {\it without 
superfluid 
currents} ($\vec{Q}_{\alpha'}=0$) and 
the linear dependence on $\vec{Q}_{\alpha'}$ is 
encoded
in the second term, $\Delta H_{\vp}^{(\alpha)}$.
In the absence of superfluid currents, the first term in the expression \eqref{eq:localenergy} 
is of the order $\sim T + \Delta_\vp^{(\alpha)}$,
while the second term 
{$\sim f^{\alpha \alpha'} \left( {\vec p}_{ F \alpha}, {\vec p}'_{ F 
\alpha'} \right) n_{\alpha'} (T^2 + \Delta_\vp^{(\alpha)2}) / \mu_\alpha^2$} 
\cite{GusakovHaensel2005}. 
Therefore, the contribution of the second term in  the expression \eqref{eq:localenergy} 
to the energy $\varepsilon_{\vp}^{(\alpha)}$ is negligibly small 
{and}
one can 
write 
{$\varepsilon_{\vp}^{(\alpha)} = \varepsilon_0^{(\alpha)} ({\vp}) - \breve\mu_\alpha$ }.
The second term in the expansion \eqref{eq:expand1},
generally, can be 
presented
as%
\footnote{Note that this representation is general as long as the system does not contain any 
vectors other than $\vQa$. In particular, Eq. \eqref{eq:dH} is not correct if at least one of 
the paring gaps is treated as anisotropic (see Ref.\ \cite{GusakovHaensel2005} for more details). 
Replacing the anisotropic gap with an effective isotropic one allows us to avoid this complication. 
 }
\begin{equation}
		\label{eq:dH}
		\Delta H_{\vec p}^{(\alpha)} = \sum_{\alpha '} \gamma_{\alpha \alpha'} 
	  	\vp  {\vec Q}_{\alpha'}  ,
\end{equation}
where $\gamma_{\alpha\alpha'}$ is a matrix to be determined below.

The energy of Bogoliubov excitations and the distribution functions can be expanded as well:
\begin{align}
    		\mathfrak{E}_{\vp + \vQa}^{(\alpha)}
    		&= E_{\vec p}^{(\alpha)} + \Delta H_{\vec p}^{(\alpha)},
    		\label{eq:E_exp} \\
    		\mF_{\vp + \vQa}^{(\alpha)} 
    		&= \fp + {\partial \fp \over \partial E_{\vec p}^{(\alpha)}} \,\, \Delta H_{\vec p}^{(\alpha)},  
    		\label{eq:F_exp} \\
		\mN_{{\vec p}+{\vec Q}_{\alpha}}^{(\alpha)} &=
		\np + {\partial \fp \over \partial E_{\vec p}^{(\alpha)}}  \Delta H_{\vec p}^{(\alpha)},
		\label{eq:n_exp}
\end{align}
where $E_{\vec p}^{(\alpha)}$ is  the Bogoliubov excitation energy, 
while $\fp$ and $\np$ are the distribution functions for Bogoliubov excitations 
and the Landau quasiparticles in the absence of superfluid currents.
These quantities are given by the following well-known expressions 
{(see, e.g., 
Ref.\ \cite{GusakovHaensel2005}):}
\begin{align}
		&E_{\vec p}^{(\alpha)}  = \sqrt{ \varepsilon_{\vp}^{(\alpha)2} + \Delta_\vp^{(\alpha)2} }, \\
		\label{eq:fp_eq_cl}
		& \fp = \frac{1}{1+e^{E_{\vec p}^{(\alpha)}/T}}, \\
		&\np = \vvpsq + \left(\uupsq - \vvpsq \right)\fp.
\end{align}

Substituting \eqref{eq:expand1} and \eqref{eq:n_exp} into \eqref{eq:localenergy}, one obtains, with 
the accuracy to the terms linear in $\vec{Q}_\alpha$,
\begin{align}
    		&\Delta H_{\vec p}^{(\alpha)} = { {\vec p  {\vec Q}_{\alpha}} \over m_\alpha^\ast} 
		+\sum_{{\vec p}' \sigma' \alpha'}  f^{\alpha \alpha'} \left( {\vec p}, {\vec p}' \right)  \left( {\partial \mathfrak{f}_{{\vec p}'}^{(\alpha')} \over \partial E_{{\vec p}'}^{(\alpha')} } \Delta H_{{\vec p}'}^{(\alpha')} - {\partial \theta_{{\vec p}'}^{(\alpha')} \over \partial {\vec p}'}  {\vec Q}_{\alpha'} \right) .
    		\label{eq:dH_eq}
\end{align}
We have already neglected the terms 
$\sim (T^2 + \Delta_\vp^{(\alpha)2}) / \mu_\alpha^2$ in this expression.
The functions in the parentheses have a sharp maximum near the Fermi surface of particle species 
$\alpha'$ 
(at $p \sim p_{F\alpha'})$,
so that the sums in Eq. \eqref{eq:dH_eq} can be 
approximately calculated as
\begin{align}
    		\sum_{{\vec p}' \sigma'}  f^{\alpha \alpha'} \left( {\vec p}, {\vec p}' \right)  {\partial \mathfrak{f}_{{\vec p}'}^{(\alpha')} \over \partial E_{{\vec p}'}^{(\alpha')} }
   		 \Delta H_{{\vec p}'}^{(\alpha')} 
    		&=  -  {G_{\alpha\alpha'} \over n_\alpha} m^*_{\alpha'} \Phi_{\alpha'}  \Delta H_{{\vec p}}^{(\alpha')}  ,
    		\label{eq:intF} 
    		\\
   		 \sum_{{\vec p}' \sigma'}  f^{\alpha \alpha'} \left( {\vec p}, {\vec p}' \right)  {\partial \theta_{{\vec p}'}^{(\alpha')} \over \partial {\vec p}'}  {\vec Q}_{\alpha'}
    		&= -  {G_{\alpha\alpha'} \over n_\alpha} \vec{p} \, {\vec Q}_{\alpha'}.
    		\label{eq:intQ}
\end{align}
For the same reason we have
\begin{equation}
		\label{eq:int_pp}
		\sum_{{\vec p} \sigma} p^i p^k {\partial \mathfrak{f}_{{\vec p}}^{(\alpha)} \over \partial 
		E_{{\vec p}}^{(\alpha)} } = - m_\alpha^* n_\alpha \Phi_\alpha \delta^{ij}.
\end{equation}
This equality is extensively used in the rest of the paper.
The function $\Phi_\alpha$ in {Eqs.  \eqref{eq:intF} and }\eqref{eq:int_pp} is defined as
\begin{equation}
		\label{eq:Phi}
		\Phi_\alpha =- \frac{\pi^2}{m^*_\alpha p_{F\alpha}} \sum_{\vec{p}\sigma} \frac{\partial \mathfrak{f}_{{\vec p}}^{(\alpha)}}{\partial E_{\vec p}^{(\alpha)}}.
\end{equation}
It changes from $\Phi_\alpha = 0$ at $T = 0$ to $\Phi_\alpha = 1$ at $T \geq T_{c\alpha}$, where 
$T_{c\alpha}$ is the critical temperature for transition of
particle species $\alpha={ n}$, ${ p}$ into the superfluid state.

\begin{figure*}
		\includegraphics[width=0.45\textwidth,trim= 1.0cm 0.3cm 3.0cm 0.0cm, clip = true]{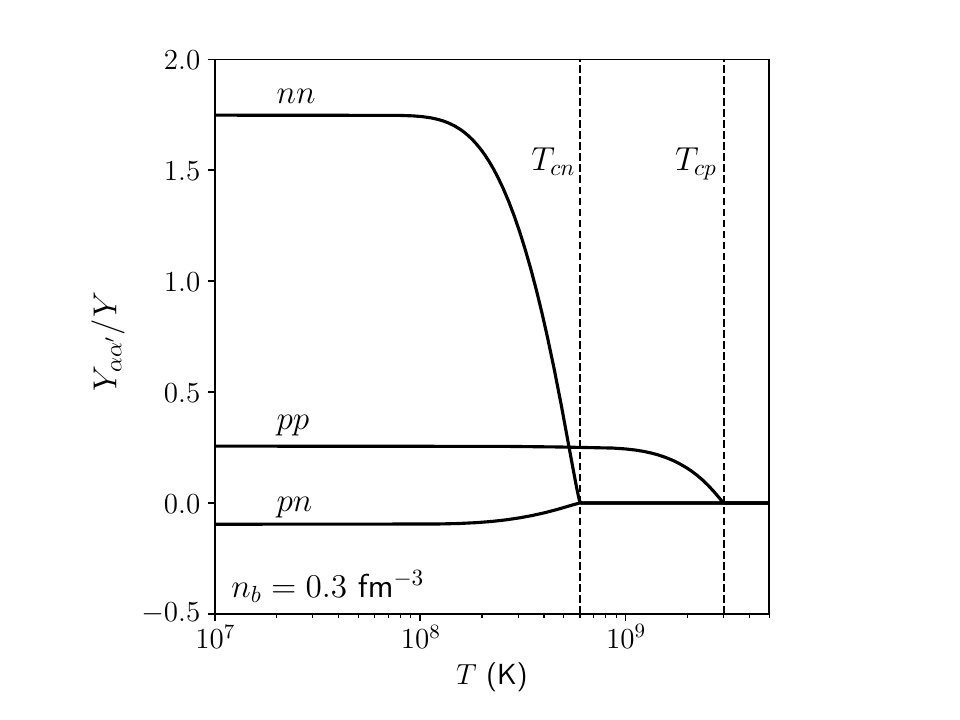}
		\includegraphics[width=0.45\textwidth,trim= 1.0cm 0.3cm 3.0cm 0.0cm, clip = true]{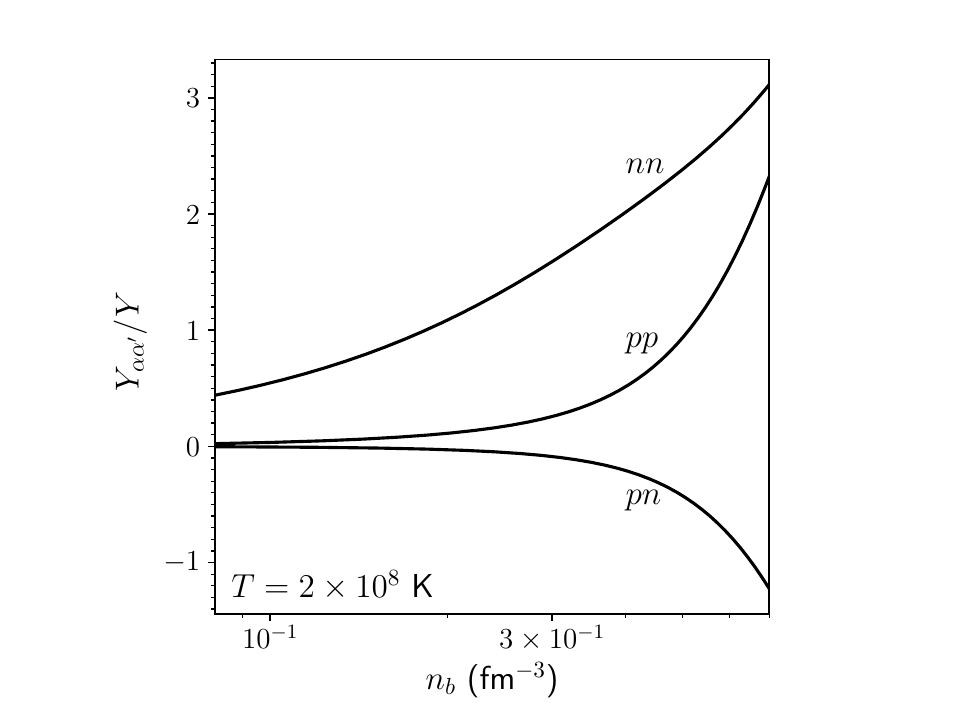}
		\caption{\label{fig:YaaT} 
		The coefficients $Y_{\alpha\alpha'}$ normalized by $Y = 10^{41}$ cm$^{-3}$erg$^{-1}$ 
		as  functions of the temperature $T$ (left panel) and the baryon number density $n_b$ 
		(right 
		panel). 
		The critical temperatures are chosen to be 
		$T_{ cn}=6\times10^8$ K and $T_{ cp}=3\times10^9$ K (shown by the vertical dashed lines 
		in the left panel). The left panel is plotted for 
		$n_{b} = 0.3$ 
		fm$^{-3}$, the right panel is plotted for 
		$T = 2\times10^8$~K.
		}
	\end{figure*}

Plugging Eqs.\ \eqref{eq:dH}, \eqref{eq:intF}, and \eqref{eq:intQ} into the expression 
\eqref{eq:dH_eq}, one can obtain a system of equations on $\gamma_{\alpha\alpha'}$:
\begin{equation}
		\label{eq:gamma_eq}
		\gamma_{\alpha\alpha'} = {\delta_{\alpha\alpha'} \over m^*_\alpha} + {G_{\alpha\alpha'} \over n_\alpha} - \sum_{\alpha''} {G_{\alpha\alpha''} \over n_\alpha} m^*_{\alpha''} \Phi_{\alpha''} 		\gamma_{\alpha'' \alpha'}.
\end{equation}
The solution to these equations is 
\begin{align}
		\label{eq:gamma_aa}
		&\gamma_{\alpha\alpha} = \frac{(n_\alpha + G_{\alpha\alpha} m^*_\alpha)(n_\beta + G_{\beta\beta} m^*_\beta \Phi_\beta) - G_{\alpha\beta}^2 m^*_\alpha m^*_\beta \Phi_\beta}{m^*_\alpha \mathcal S}, \\
	\label{eq:gamma_ab}
		&\gamma_{\alpha\beta} = \frac{G_{\alpha\beta}n_\beta(1-\Phi_\beta)}{\mathcal S},
\end{align}
where $\beta \neq \alpha$ and 
\begin{equation}
		\label{eq:S_det}
		\mathcal S = (n_{ n} + G_{ nn}  m^*_{ n}  \Phi_{ n} )(n_{ p}  + G_{ pp} m^*_{ p} \Phi_{ p} ) - G_{ np} ^2 m^*_{ n}   m^*_{ p}  \Phi_{ n} \Phi_{ p} .
\end{equation}
To calculate the particle current density, one can use the standard ``nonsuperfluid'' formula 
\cite{Leggett1965,Gusakov2010,AllardChamel2021}
\begin{equation}
		\label{eq:J}
		{\vec j}_{\alpha} = \sum_{{\vp \sigma}} 
		{\partial H_{\vp + \vQa}^{(\alpha)} \over \partial {\vec p}}
		\,\, \mN_{\vp + \vQa}^{(\alpha)}.
\end{equation}
Plugging Eqs.\ \eqref{eq:expand1}, \eqref{eq:dH},  and \eqref{eq:n_exp} into this 
formula and using the expression \eqref{eq:int_pp}, 
one obtains Eq.\ \eqref{eq:j_def} with
\begin{equation}
		\label{eq:Y}
		Y_{\alpha\alpha'} = {\gamma_{\alpha\alpha'} n_\alpha \over c^2} (1-\Phi_\alpha).
\end{equation}
It is easy to see that the matrix $Y_{\alpha\alpha'}$ is symmetric. 
{Figure \ref{fig:YaaT} shows the behavior of the coefficients $Y_{\alpha\alpha'}$ as functions 
of temperature $T$ and baryon  number density, $n_b=n_n+n_p$.
It is plotted for 
neutron star matter composed of protons, neutrons, and electrons
assuming the BSk24 equation of state 
\cite{GorielyChamelPearson2013}. The Landau parameters {and the functions $\Phi_\alpha$ [see 
	Eq. \eqref{eq:Phi}]} were calculated as described in 
Ref.\ \cite{KantorGusakov2020}; see this reference for more details.
Following Refs.\ \cite{BaikoHaenselYakovlev2001, GusakovHaensel2005}, 
the effective neutron gap was taken to be equal to the minimum value of the angle-dependent $^3P_2$ 
gap with the projection of the total angular momentum of a pair $m_J=0$ (see Appendix A in Ref.\ 
\cite{GusakovHaensel2005} for details).
}

The particle current density \eqref{eq:J} can alternatively  be represented as \cite{Gusakov2010}
\begin{equation}
		\label{eq:j_2010}
		{\vec j}_{\alpha} = c^2 \sum_{\alpha'} Y_{\alpha\alpha'}^0 \left[ \vec{Q}_{\alpha'} + {1 \over n_{\alpha'}} \sum_{\vec{p}\sigma} \vec{p} \mF_{{\vec p}+{\vec Q}_{\alpha'}}^{(\alpha')} \right],
\end{equation}
where the coefficients
\begin{equation}
		Y_{\alpha\alpha'}^0 = \delta_{\alpha\alpha'} { n_\alpha \over  m_\alpha^* c^2} + 
		{G_{\alpha\alpha'} \over c^2}	
		\label{Yik0}
\end{equation}
are equal to the matrix elements $Y_{\alpha\alpha'}$ taken at $T=0$.
On the other hand, the momentum density 
of particle species $\alpha$ equals \cite{Gusakov2010} 
\begin{equation}
		\label{eq:p_2010}
		{\vec P}_{\alpha} =  \sum_{\vec{p} \sigma } (\vec{p} + \vec{Q}_{\alpha}) \mN_{{\vec p}+{\vec Q}_{\alpha}}^{(\alpha)} = n_\alpha \vec{Q}_{\alpha} +  \sum_{\vec{p}\sigma} \vec{p} \mF_{{\vec p}+{\vec Q}_{\alpha}}^{(\alpha)}.
\end{equation}
Using Eqs.\ \eqref{eq:j_2010} and \eqref{eq:p_2010} together with the relation 
\eqref{eq:m_eff_eq}, one can easily verify that
\begin{equation}
		\label{eq:PJ}
		\sum_{\alpha } {\vec P}_{\alpha} = \sum_\alpha {\mu_\alpha \over c^2} \vec{j}_\alpha,
\end{equation}	
{that is, the total momentum density coincides with the total mass-current density in the 
system, the expected result.}

{
Note
that the expressions 
(\ref{eq:J}) and (\ref{eq:j_2010}) for particle current 
densities, as well as the expression (\ref{eq:p_2010}) for the momentum densities, are 
quite general and can be used even if the system is {\it not} in complete thermodynamic equilibrium 
\cite{Gusakov2010}
(i.e., $\mF_{{\vec p}+{\vec Q}_{\alpha}}^{(\alpha)}$ in these equations is not necessarily the 
Fermi-Dirac 
distribution function).
We will make use of this fact 
in the subsequent presentation.
}

\section{Normal currents in a superfluid 
	mixture of Fermi liquids}
\label{sec:normal_currents}

All the calculations in the previous section were done in the frame of reference comoving with the 
normal liquid component.
In other words, it was implicitly assumed that thermal excitations of both 
species, ``$n$'' and ``$p$'', move with identical 
{\it normal} (nonsuperfluid) velocity. 
How should the results of the previous section be modified if we assume that
these velocities differ for 
{the ``$n$'' and ``$p$''}
species?
In the present section we will try to 
address
this question for an idealized system, in which
normal velocities of two species differ, but the dissipative interaction (due to collisions)
between them is absent. At the same time, we will allow both species to interact
with a heat bath so that the system can be described with a single temperature.
A similar but more general problem (allowing for dissipative interaction between the two species)
will be considered in Sec. \ref{sec:diffusion}
within the
kinetic theory.
It will be demonstrated that this problem, formulated in the language of kinetic theory, is 
directly related to the diffusion in superfluid mixtures.	
{

As it is discussed in the previous section, we assume that there always exists 
a reference frame in which 
the velocities of all the particle species
are small. 	
In the system with two different normal currents it is convenient to work
in this frame and minimize the thermodynamic potential
\begin{equation}
		\label{eq:F_potential_np_tmp}
	\tilde{F} = E-\sum_\alpha \mu_{\alpha 1} n_\alpha  -TS - \sum_{\alpha}
		\vec{P}_{\alpha}{\vec V}_{{ q}\alpha},
\end{equation}
where $\vec{P}_{\alpha}$ is the momentum density of particle species $\alpha$
given by Eq.\ \eqref{eq:p_2010}; the vectors ${\vec V}_{{ q}\alpha}$ 
are abstract Lagrange multipliers whose physical meaning will be clarified 
{below}
and $\mu_{\alpha 1}$ are some chemical potentials 
in the chosen 
reference frame
(also Lagrange multipliers,  
required  to 
keep the total number of particles $\alpha$ fixed).
The last 
term in the right-hand side of the
expression \eqref{eq:F_potential_np_tmp} indicates that our thermodynamic
potential is minimized at fixed 
{momentum densities of the ``$n$'' and ``$p$'' particle species.}
Using the definition \eqref{eq:p_2010}, 
the 
thermodynamic potential \eqref{eq:F_potential_np_tmp}
can be rewritten as
\begin{equation}
\label{eq:F_potential_np}
E-\sum_\alpha \mu_\alpha n_\alpha  -TS - \sum_{\alpha}
\vec{\mathcal P}_{\alpha}{\vec V}_{{ q}\alpha},
\end{equation}
where $\vec{\mathcal P}_{\alpha} \equiv
\sum_{{\vec p}\sigma} \vp \mathcal{F}_{{\vec p}+{\vec Q}_\alpha}^{(\alpha)}$
and we introduced the new chemical potential, 
$\mu_\alpha  \equiv \mu_{\alpha 1}+\vec Q_\alpha {\vec V}_{{ q}\alpha}$.

There is one point to be made here before moving on.
All the thermodynamic variables appearing in the hydrodynamic equations 
are usually {defined (measured)} in the reference frame comoving with the fluid.
When a mixture of few superfluids 
is considered,
one usually chooses the frame comoving 
with the normal  (nonsuperfluid) liquid component 
(see, e.g., Ref.\ \cite{AndreevBashkin1976}).
In particular, the chemical potentials entering  the formula \eqref{eq:j_def} are assumed to be the 
same as those arising in the second law of thermodynamics \eqref{eq:ap:dE} 
written down in the comoving reference frame.
However, 
allowing for the two independent normal velocity fields, 
the standard definition of the comoving frame loses its meaning
and we are forced to work in a more general reference frame. 
This means, in particular, that
the chemical potentials $\mu_\alpha$ 
entering the expression 
\eqref{eq:F_potential_np}, are not the 
same potentials as those 
appearing 
in the formulas \eqref{eq:j_def} and \eqref{eq:ap:dE}.
However, 
since these potentials are scalars with respect to spatial 
{transformations,}
the difference between chemical potentials measured in different reference frames 
can depend only on bilinear combinations of the hydrodynamic velocities.
Thus, this difference is quadratically  small by assumption
(see the beginning of Sec. \ref{entrmatr}) and can be ignored in the subsequent analysis.

Minimizing the thermodynamic potential \eqref{eq:F_potential_np} 
by calculating the  variation 
of the energy density $E$, the number densities $n_{\alpha}$, and the quantities $\vec{\mathcal  
P}_{\alpha}$ with respect to variation of the distribution  function $\mF_{\vp + \vQa}^{(\alpha)}$,
one obtains%
\begin{equation}
		\mF_{\vp + \vQa}^{(\alpha)} =
		\frac{1}{ 1 + {e}^{(\mathfrak{E}_{\vp + \vQa}^{(\alpha)}-{\vec p}{\vec V}_{{ q}\alpha})/T}}.
		\label{eq:Fp1} 
\end{equation}
As we just emphasized, 	
we work in a reference frame where all the hydrodynamic velocities are small.
Therefore, assuming that
$Q_\alpha/p_{ F \alpha}$ and
$ { V}_{{ q}\alpha}/ v_{ F \alpha} \ll 1$,%
\footnote{{Further it will be argued  that 
$\vec{V}_{q\alpha} \approx \vec{u}$.}}
all the quantities  in this expression can  be expanded in powers of $\vQa$ and ${\vec V}_{{ 
q}\alpha}$.
In this way, the expressions 
\eqref{eq:F_exp} and \eqref{eq:n_exp} 
should be replaced with
\begin{eqnarray}
		    \mF_{\vp + \vQa}^{(\alpha)} &=&
 		   \fp + {\partial \fp \over \partial E_{\vp}^{(\alpha)}} 
		    \left(\Delta H_{\vp}^{(\alpha)}-{\vp}{\vec V}_{{ q}\alpha}\right), 
 		   \label{eq:F_exp_nc}\\
 		   \mN_{\vp + \vQa}^{(\alpha)} &=&
 		   \np + {\partial \fp \over \partial E_{\vp}^{(\alpha)}} \left(\Delta H_{\vp}^{(\alpha)} 
 		   -{\vp}{\vec V}_{{ q}\alpha}\right),
		    \label{eq:N_exp_nc}
\end{eqnarray}		
where we made use of the expansion \eqref{eq:E_exp} for
the energy $\mathfrak{E}_{ \vp + \vQa}^{(\alpha)}$. 
The quantity $\Delta H_{\vec p}^{(\alpha)}$ in Eq.\ \eqref{eq:E_exp} now, generally,
depends on 
both the vectors ${\vec Q}_{\alpha}$ and ${\vec V}_{{ q}\alpha}$,
and 
can be written as
\begin{equation}
		\label{eq:dH_pn}
		\Delta H_{\vec p}^{(\alpha)} = \sum_{\alpha '}  \left(\gamma_{\alpha \alpha'}
		 \vp  {\vec Q}_{\alpha'}  +K_{\alpha\alpha'}  {\vec p}{\vec V}_{{ q}\alpha'}\right).
\end{equation}
Plugging Eqs.\  \eqref{eq:expand1},   \eqref{eq:N_exp_nc}, 
and \eqref{eq:dH_pn} into 
\eqref{eq:localenergy}, and using (again, as in Sec. \ref{entrmatr}) Eqs.\ \eqref{eq:intF} and 
\eqref{eq:intQ}, one obtains, besides Eqs.\ \eqref{eq:gamma_eq}, the equations for the 
coefficients $K_{\alpha\alpha'}$:
\begin{equation}
		\label{eq:K_eq}
		K_{\alpha\alpha'} = {G_{\alpha\alpha'} m^*_{\alpha'} { \Phi_{\alpha'} }\over n_\alpha} - \sum_{\alpha''} {G_{\alpha\alpha''} \over n_\alpha} m^*_{\alpha''} \Phi_{\alpha''} K_{\alpha''\alpha'}.
\end{equation}
The solution to these equations is 
\begin{align}
		\label{eq:K_aa}
		&K_{\alpha\alpha} = \frac{ G_{\alpha\alpha} m^*_\alpha \Phi_\alpha(n_\beta + G_{\beta\beta} m^*_\beta \Phi_\beta) - G_{\alpha\beta}^2 m^*_\alpha m^*_\beta \Phi_\alpha \Phi_\beta}{\mathcal S}, \\
		\label{eq:K_ab}
		&K_{\alpha\beta} = \frac{G_{\alpha\beta} m^*_\beta n_\beta\Phi_\beta}{\mathcal S},
\end{align}	
where { we recall that} $\beta \neq \alpha$ and $\mathcal S$ is given by the expression \eqref{eq:S_det}.

Now one can calculate the particle current density $\vec{j}_\alpha$. 
Substituting Eqs.\ \eqref{eq:expand1}, 
\eqref{eq:N_exp_nc}, and \eqref{eq:dH_pn} 
into \eqref{eq:J}, one obtains
\begin{equation}
		\label{eq:j_nc}
		\vec{j}_\alpha = R_{\alpha\alpha} \vec{V}_{ q\alpha} + R_{\alpha\beta} \vec{V}_{ q\beta} + c^2 Y_{\alpha\alpha}\vec{Q}_\alpha + c^2 Y_{\alpha\beta}\vec{Q}_\beta,
	\end{equation}
where the elements of the superfluid entrainment matrix $Y_{\alpha\alpha'}$ are given by Eqs.\ 
\eqref{eq:Y} and the coefficients $R_{\alpha\alpha'}$ are defined as
\begin{align}
		\label{eq:Naa}
		& R_{\alpha\alpha} = n_\alpha K_{\alpha\alpha} (1-\Phi_{\alpha}) + n_{\alpha} \Phi_{\alpha},
		\\
		\label{eq:Nab}
		&R_{\alpha\beta} =n_{\alpha} K_{\alpha\beta}(1-\Phi_{\alpha}). 
\end{align}
This set of coefficients 
is, to our best knowledge, 
introduced in the literature for the first time. 
In analogy to the matrix $Y_{\alpha\alpha'}$, we call the coefficients $R_{\alpha\alpha'}$
the {\it normal entrainment matrix}.
Alternatively, the matrix $R_{\alpha\alpha'}$ can be expressed through 
$\gamma_{\alpha\alpha'}$ as
\begin{equation}
		\label{eq:Naa_alt}
		R_{\alpha\alpha'} = n_{\alpha'}m^*_{\alpha'} \Phi_{\alpha'} \gamma_{\alpha' \alpha}.
\end{equation}
It can be verified that the obtained coefficients satisfy the following {\it sum rule}
\begin{equation}
    		\label{eq:part_cons_rel}
		\mu_\alpha Y_{\alpha\alpha} + \mu_\beta Y_{\alpha\beta} + R_{\alpha\alpha} + R_{\alpha\beta} = n_\alpha,
\end{equation}
which is the 
generalization to finite temperatures of the sum rule derived 
in Ref.\ \cite{GusakovKantorHaensel2009}.
This sum rule is 
related to
particle number conservation.
Indeed, if one introduces the superfluid velocities according to the definitions, 
$\vec{V}_{ s \alpha} \equiv \vQa c^2/\mu_\alpha$,%
%
\footnote{Note that,
for a 
relativistic equation of state, the superfluid velocity 
introduced this way does not obey the potentiality condition,
${\rm curl} \vec{V}_{\rm s\alpha} \neq 0$.
This is in contrast to  
the vectors $\vQa$, which are proportional (with constant coefficient) 
to the gradient of the superfluid order parameter phase [see expression \eqref{eq:ap:Q_def}], 
and thus satisfy the potentiality condition. }
%
and assumes that all the fluid components move with one and the same velocity, i.e.,  
$\vec{V}_{ qn} = \vec{V}_{ qp} = \vec{V}_{ sn} = \vec{V}_{ sp} = \vec{u}$, 
then [in view of Eq.\ \eqref{eq:part_cons_rel}] the expression \eqref{eq:j_nc} 
will reduce to 
$\vec{j}_\alpha = n_\alpha \vec{u}$, as expected.
One can obtain another useful 
{relation} 
from 
{Eq.\ \eqref{eq:part_cons_rel} by}
using the definitions \eqref{eq:Y}, \eqref{eq:Naa}, and \eqref{eq:Nab}:
\begin{equation}
		\label{eq:gamma_k_rel}
		{\mu_\alpha \over c^2} \gamma_{\alpha\alpha} + {\mu_\beta \over c^2} \gamma_{\alpha\beta} + K_{\alpha\alpha} + K_{\alpha\beta} = 1.
\end{equation}

Setting  $\vec{V}_{ q n} = \vec{V}_{ q p} = \vec{u}$ in Eq.\ \eqref{eq:j_nc} and taking into 
account Eq.\ \eqref{eq:part_cons_rel}, one derives 
[cf. Eq. \eqref{eq:j_def}]
\begin{equation}
		\label{eq:j_nc_eq}
		\vec{j}_{\alpha,0} = n_{ q\alpha} \vec{u} + c^2 \sum_{\alpha'} Y_{\alpha\alpha'} \vec{Q}_{\alpha'},
\end{equation}
where 
\begin{equation}
		\label{eq:N_qa}
		n_{ q\alpha} = n_\alpha - \sum_{\alpha'} \mu_{\alpha'} Y_{\alpha\alpha'} = \sum_{\alpha'} R_{\alpha \alpha'}.
\end{equation}
Therefore, 
when
the vectors $\vec{V}_{ q n}$ and  $\vec{V}_{ q p}$ 
coincide,
they
have 
the meaning of the normal velocity of thermal excitations.
In turn, the quantity $n_{ q\alpha}$ can be interpreted as the normal density of
particle species $\alpha$.
Note, however, that in a strongly interacting mixture $n_{ q\alpha}$ can {become} negative; 
see the discussion at the end of the present section.

If one of the constituents, say ``$n$'', is nonsuperfluid ($\Phi_{ n} =1$), 
then the  particle current densities reduce to
\begin{equation}
		\vec{j}_{ n} = n_{ n} \vec{V}_{ qn}, \ \ \ \  \vec{j}_{ p} = R_{ pp} \vec{V}_{ qp} + R_{ 
		pn} \vec{V}_{ qn} + c^2 Y_{ pp}\vec{Q}_{ p};
\end{equation}
the values of the matrix elements 
in this limit are given
in Appendix 
\ref{ap:coeffs}.
In the case of a completely nonsuperfluid mixture, they further simplify as
\begin{equation}
		\label{eq:j_nc_nsf}
		\vec{j}_{ n} = n_{ n} \vec{V}_{ qn}, \ \ \ \  \vec{j}_{ p} = n_{ p} \vec{V}_{ qp}.
\end{equation}
Thus, in the nonsuperfluid  mixture, 
the vectors $\vec{V}_{ q\alpha}$ 
have the meaning of the
hydrodynamic velocities of the corresponding particle species. 
In the superfluid mixture the 
interpretation of these vectors is more complicated,
see the end of Sec. \ref{sec:diffusion} for a discussion.

On the other hand, if one of the species, say ``$p$'', is in the regime of strong superfluidity
($\Phi_{ p} \rightarrow 0$, $T\ll T_{{ cp}}$),
then 
{one has for the currents}
\begin{align}
		\vec{j}_{ n} &=  R_{ nn} \vec{V}_{ qn} + c^2 Y_{ nn}\vec{Q}_{ n}  + c^2 Y_{ np}\vec{Q}_{ p},
		\\
		 \vec{j}_{ p} &= R_{ pn} \vec{V}_{ qn} + c^2 Y_{ pn}\vec{Q}_{ n}  + c^2 Y_{ pp}\vec{Q}_{ p}.
\end{align}	
Again, the relevant matrix elements are given in Appendix \ref{ap:coeffs}.
As one can see, the Fermi-liquid effects 
lead to the finite normal current density of particle species ``$p$''
even at $T\ll T_{ cp}$.

One can notice that the matrix $R_{\alpha\alpha'}$, in contrast to the matrix $Y_{\alpha\alpha'}$,  
is not symmetric. 
This fact, however, does not point to some defect of the theory  since the vectors $\vec{V}_{ 
q\alpha}$, in contrast to the vectors $\vQa$, are not momenta.
To symmetrize the expression \eqref{eq:j_nc},
we introduce the momentum densities $\vec{P}_{ q\alpha}$, associated with the normal fluid 
motions instead of the vectors $\vec{V}_{ q\alpha}$.
Using Eq.\ \eqref{eq:p_2010}, they are defined as
\begin{equation}
		\vec{P}_{ q\alpha} = \
		 \sum_{\vec{p}\sigma} \vec{p} \mF_{{\vec p}+{\vec Q}_{\alpha}}^{(\alpha)} |_{\vec{Q}_{ n}=\vec{Q}_{ p} =0}.
\end{equation} 
Now, with the help of Eq.\ \eqref{eq:j_2010}, one can express the particle current densities 
through the momentum variables $\vec{P}_{ q\alpha}/n_\alpha$:
\begin{equation}
		\label{eq:j_trh_p}
		\vec{j}_\alpha =  {c^2 Y_{\alpha\alpha}^0} \frac{\vec{P}_{{ q}\alpha}}{n_\alpha} + {c^2 Y_{\alpha\beta}^0 } \frac{\vec{P}_{{ q}\beta}}{n_\beta} + c^2 Y_{\alpha\alpha}\vec{Q}_\alpha + c^2 Y_{\alpha\beta}\vec{Q}_\beta.
	\end{equation}
In this expression, the matrix $Y_{\alpha\alpha'}^0$, which is a special case of the symmetric 
matrix $Y_{\alpha\alpha'}$ [see Eq.\ \eqref{Yik0}], is obviously symmetric.

\begin{figure*}
	\includegraphics[width=0.45\textwidth,trim= 1.0cm 0.3cm 3.0cm 0.0cm, clip = true]{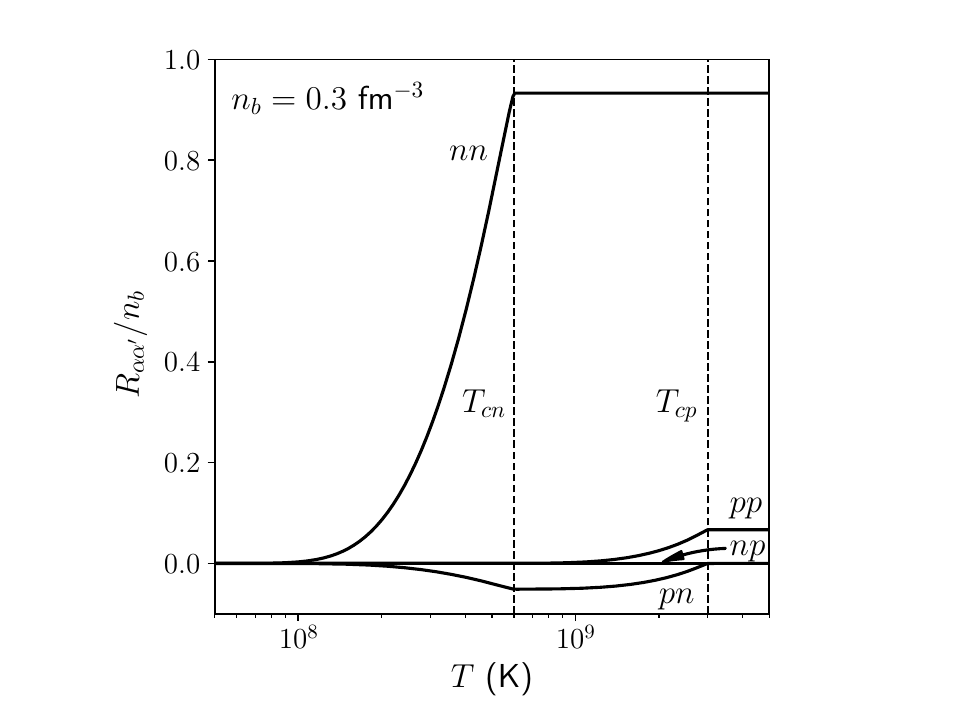}
	\includegraphics[width=0.45\textwidth,trim= 1.0cm 0.3cm 3.0cm 0.0cm, clip = true]{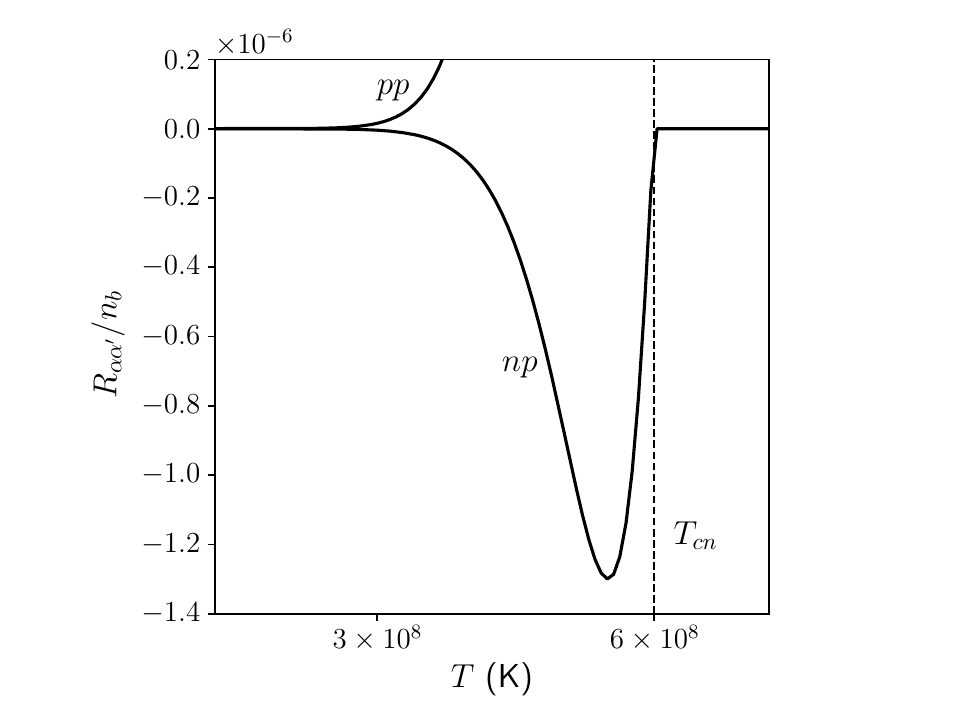}
	\caption{\label{fig:Raas} 	
		Left panel: The coefficients $R_{\alpha\alpha'}$ normalized to the baryon density $n_{ b}$
		as  functions of temperature 
		are shown for $n_{ b}$ = 0.3 fm$^{-3}$, $T_{ cp} = 2\times10^9$ K,  $T_{ cn} = 6\times10^8$ 
		K, and for the BSk24 equation of state.
		Right panel:  Zoomed-in version of the plot in the left panel near $T\approx T_{cn}$.
	   }
\end{figure*}
\begin{figure*}
	\includegraphics[width=0.45\textwidth,trim= 1.0cm 0.3cm 3.0cm 0.0cm, clip = true]{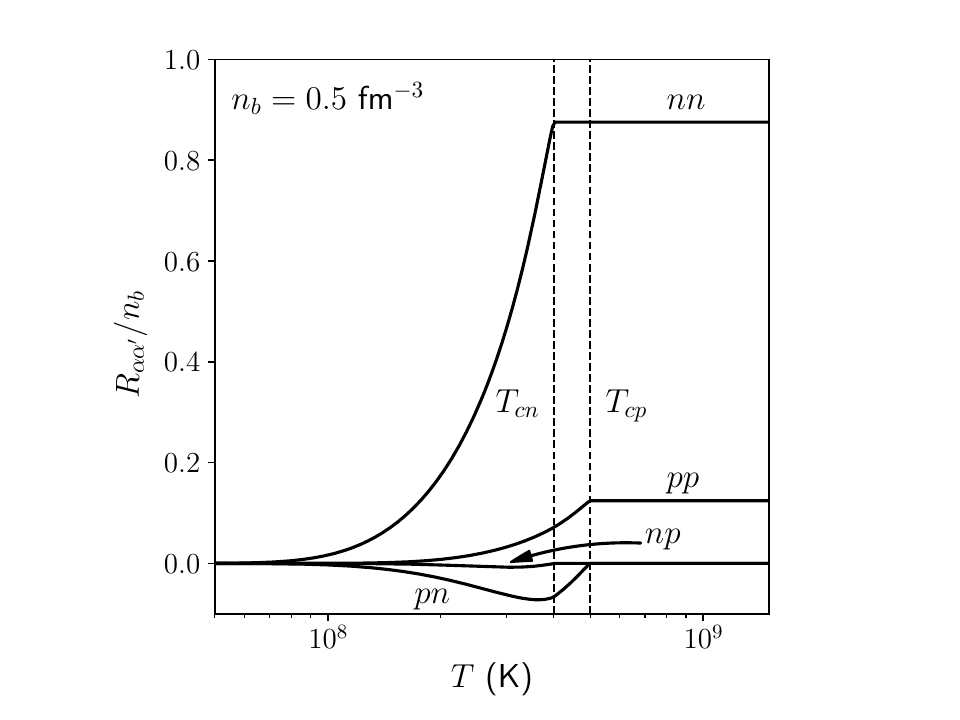}
	\includegraphics[width=0.45\textwidth,trim= 1.0cm 0.3cm 3.0cm 0.0cm, clip = true]{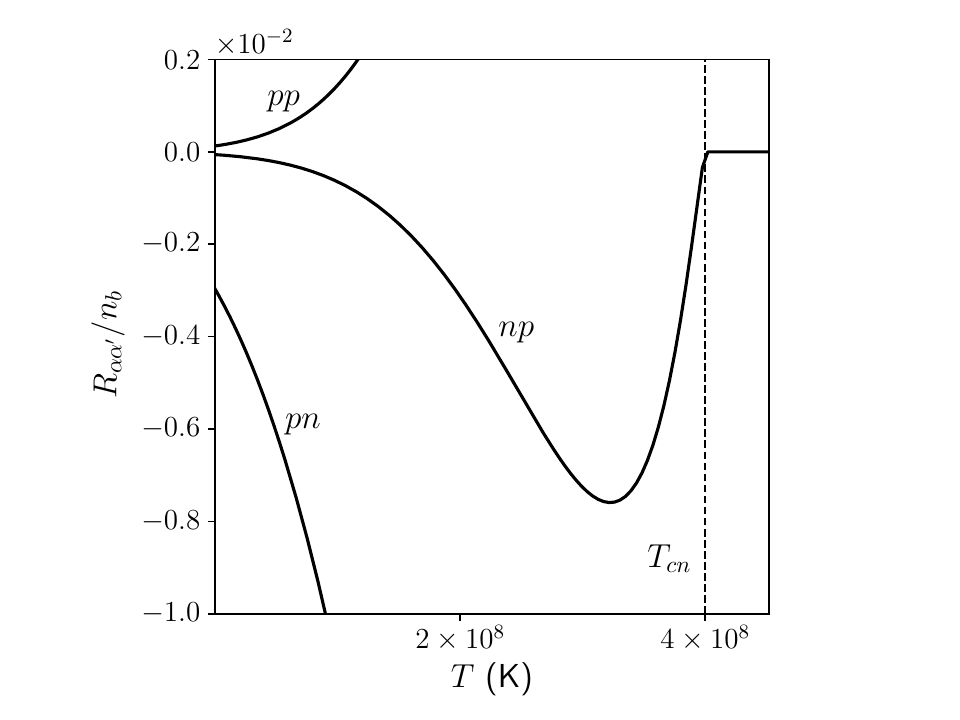}
	\caption{\label{fig:Raas2} 		
		The same as in Fig. \ref{fig:Raas} but for $n_{ b}$ = 0.5 fm$^{-3}$, $T_{ cp} = 4\times10^8$ K,  and $T_{ cn} = 5\times10^8$ K. 
	}
\end{figure*}
\begin{figure*}
	\includegraphics[width=0.45\textwidth,trim= 1.0cm 0.3cm 3.0cm 0.0cm, clip = true]{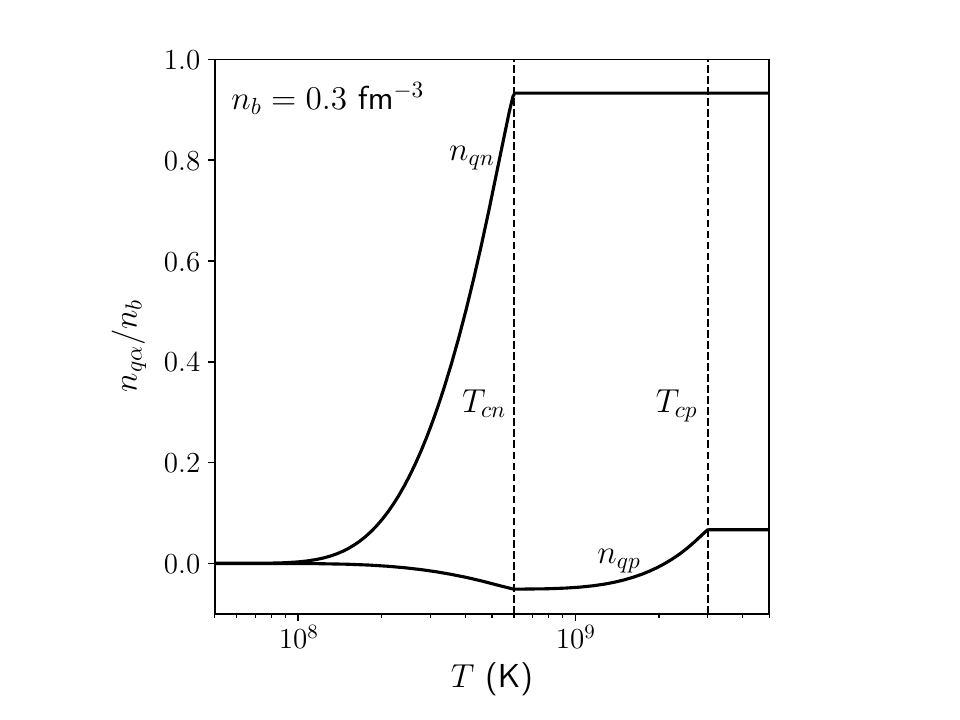}
	\includegraphics[width=0.45\textwidth,trim= 1.0cm 0.3cm 3.0cm 0.0cm, clip = true]{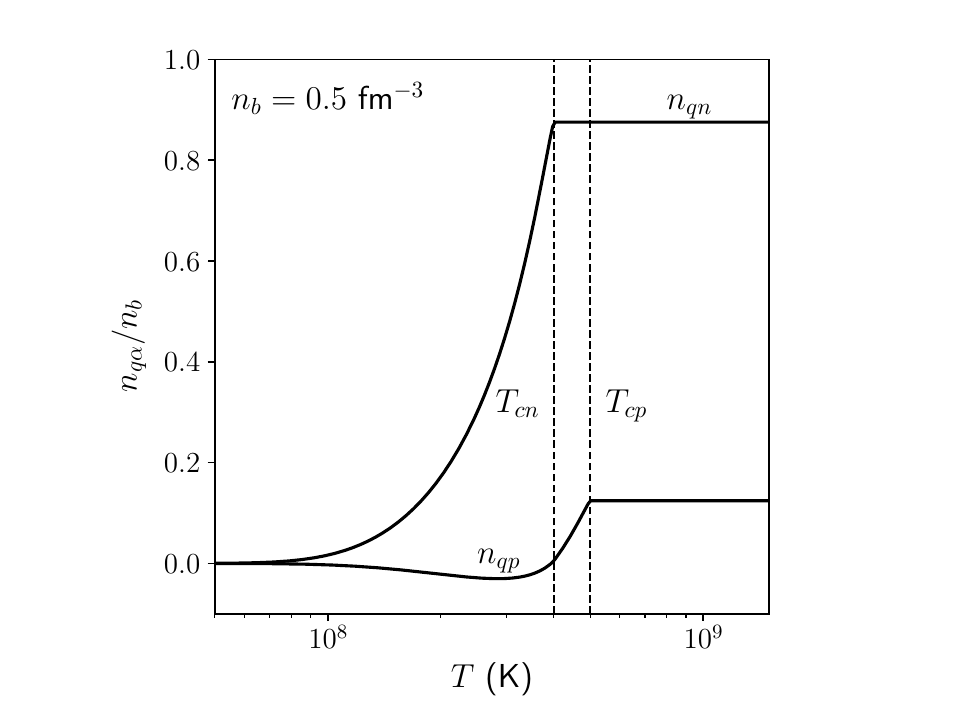}
	\caption{
		\label{fig:nqa}	
		The {normal number densities} $n_{ q\alpha}$ normalized {to} the baryon density 
		$n_{ b}$ as functions 
		of temperature plotted for $n_{ b}$ = 0.3 fm$^{-3}$, $T_{ cp} = 2\times10^9$ K,  $T_{ cn} = 
		6\times10^8$ K (left panel), and  $n_{ b}$ = 0.5 fm$^{-3}$, $T_{ cp} = 4\times10^8$ K,  
		$T_{ cn} = 5\times10^8$ K (right panel).
	}
\end{figure*}

The coefficients $R_{\alpha\alpha'}$, 
calculated
for
the same model of neutron-star matter as the matrix $Y_{\alpha\alpha'}$ in Fig.\ \ref{fig:YaaT},
are shown 
as functions of temperature $T$
in Fig.\ \ref{fig:Raas} for $n_b=0.3$~fm$^{-3}$
and in Fig.\ \ref{fig:Raas2} for $n_b=0.5$~fm$^{-3}$.
}
One can see that all the coefficients vanish when the temperature tends to zero.  
This is a reasonable result, since there are no temperature excitations in the zero-temperature 
limit.
Considering the vicinity of the critical temperature $T_{ cn}$, 
one can notice that the behavior of the matrix elements $R_{\alpha\alpha'}$ and $Y_{\alpha\alpha'}$ 
is 
different. 
If $T\geq T_{ cn}$, 
both the non-diagonal coefficients $Y_{ pn}=Y_{ np}$ 
vanish together with the coefficient $Y_{ nn}$. 
This is expected, since then the particle species ``$n$'' is nonsuperfluid (normal) 
so that
the
current density 
$\vec{j}_{ p}$ 
cannot depend on $\vec{Q}_{ n}$.
As for the matrix $R_{\alpha\alpha'}$ , only the element $R_{ np}$ vanishes 
at 
$T>T_{ cn}$.

{ The normal number densities $n_{ q\alpha}$
are shown in Fig.\ \ref{fig:nqa}	
as functions of temperature for the same equation of state BSk24.
The function $n_{ qn}$ behaves exactly as one would expect. 
It  equals $n_{ n}$ for $T\geq T_{ cn}$ and exponentially decreases to zero 
at $T \ll T_{ cn}$.
In contrast to $n_{ qn}$, the behavior of the normal density $n_{ qp}$ 
is more counterintuitive.
It starts from $n_{ p}$ at $T=T_{ cp}$ and then rapidly drops to negative values 
with decreasing temperature.
After reaching a minimum, it 
begins
to increase, approaching zero at $T\rightarrow 0$.  
The negativity of $n_{ qp}$ is related to negativity of the coefficient $G_{ pn}$ 
for the chosen equation of state 
[see Eq.\ \eqref{eq:G_matrix} and Eq.\ \eqref{nqp2} below]. 
This feature, however, does not lead to any unphysical consequences.
In particular,
the system energy in the presence of 
particle currents always increases, as shown in 
what follows.

Assume, 
for simplicity,
that the only 
currents generated in the system are the normal ones, i.e., 
$\vec{Q}_{ n} = \vec{Q}_{ p} = 0$.
Then, in the linear approximation,
one can express  $\vec{V}_{ q\alpha}$ through $\vec{P}_{ q\alpha}$ as
\begin{equation}
	\vec{V}_{ q\alpha} = \sum_{\alpha'} M_{\alpha \alpha'} \vec{P}_{ q\alpha'},
	\label{Vqa1}
\end{equation} 
where the elements of the matrix $M_{\alpha \alpha'}$ 
can be calculated by equating Eqs.\ 
\eqref{eq:j_nc} and \eqref{eq:j_trh_p}
{, and using the expression \eqref{eq:RR_expr} for the determinant of the matrix 
$R_{\alpha\alpha'}$}:
\begin{equation}
	M_{\alpha\alpha'} = \frac{n_\alpha \delta_{\alpha\alpha'} + G_{\alpha\alpha'} m_\alpha^* 
	\Phi_\alpha}{n_\alpha n_{\alpha'} m_\alpha^* \Phi_\alpha}.
\end{equation}
Now, using the fact that the thermodynamic potential \eqref{eq:F_potential_np} is stationary with 
respect to variations 
of $n_\alpha$, $S$, and $\vec{\mathcal P}_{\alpha}=\vec{P}_{q\alpha}$, 
one has $d E= \sum_\alpha \mu_\alpha dn_\alpha + T dS + \sum_\alpha \vec{V}_{ q\alpha} d \vec{P}_{ 
q\alpha}$,
and hence [see Eq.\ (\ref{Vqa1})]

\begin{equation}
	E \approx E_0(n_{\alpha}, S) + \frac{1}{2} 
	\sum_{\alpha\alpha'} M_{\alpha\alpha'} \vec{P}_{ q\alpha} \vec{P}_{ q\alpha'},
\end{equation}
where $E_0(n_{\alpha}, S)$ is the energy density of the system in the absence of currents.
To make the system energy with currents larger than that without currents,
the matrix $M_{\alpha\alpha'}$ should be positive-definite, 
which implies
\begin{align}
	\label{eq:stab_rel}
	&{\frac{n_\alpha}{m_\alpha^*}  + G_{\alpha\alpha}  \Phi_\alpha} > 0, \ \ \ \ \mathcal S > 0.
\end{align}
An equivalent set of constraints has been obtained, in particular, in Ref.\ 
\cite{ChamelHaensel2006} 
from the requirement 
of stability of superfluid Fermi mixture at $T=0$ with respect to spontaneous generation of 
superfluid 
currents in the system.
The conditions \eqref{eq:stab_rel} do not constrain the sign of the coefficient $G_{ np}$.
On the other hand, 
in the limit $T \ll T_{ cp}$, one obtains 
[cf.\  Eq.\ \eqref{eq:Rnp_Tcp0}]
\begin{equation}
	n_{ qp} \approx R_{ pn} \approx \frac{n_{ n} m_{ n}^* G_{ np} \Phi_{ n}}{n_{ n} + G_{ nn} m_{ 
	n}^*\Phi_{ n}},
	\label{nqp2}
\end{equation}
according to which the quantity $n_{ qp}$ has the same sign as the coefficient $G_{ np}$.
}

\section{Diffusion}
\label{sec:diffusion}

In the previous section, we obtained
the expression for particle current densities under the assumption that 
the dissipative interaction between different particle species can be neglected.
If one wants 
to allow for 
such interaction, 
one should work within the framework of transport  theory.
In this section, we demonstrate the close relation of the approach
developed in Sec. \ref{sec:normal_currents} to the diffusion theory of particles 
in a superfluid Fermi mixture.

\subsection{Basic equations}

Let us suppose that the characteristic wavenumber of our problem is $\mathfrak q$, 
while the characteristic frequency is $\omega$.
In the long wavelength/small frequency (``quasiclassical’') 
limit ($\mathfrak q v_{ F\alpha}, \ \omega \ll \Delta_\alpha$),  one can formulate the Boltzmann-like 
kinetic equation for the  Bogoliubov excitations \cite{AronovEtAl1981,Gusakov2010}:
\begin{equation}
		\label{eq:kin_eq}
		\frac{\partial \mF_{\vp + \vQa}^{(\alpha)}}{\partial t} + \frac{\partial \mathfrak{E}_{\vp + \vQa}^{(\alpha)}}{\partial \vp}\frac{\partial \mF_{\vp + \vQa}^{(\alpha)}}{\partial \vr} - \frac{\partial \mathfrak{E}_{\vp + \vQa}^{(\alpha)}}{\partial \vr} \frac{\partial \mF_{\vp + \vQa}^{(\alpha)}}{\partial \vp} = I_\alpha [  \mF ],
\end{equation}
where $ \mF_{\vp + \vQa}^{(\alpha)}$ is the distribution function of Bogoliubov excitations, 
$\mathfrak{E}_{\vp + \vQa}^{(\alpha)}$ is the energy of  a Bogoliubov excitation given by 
Eq.\ \eqref{distribution1},
$I^{(\alpha)}  [  \mF] $ is the collision integral (see Appendix \ref{sec:Integral}), and $\mF$ denotes 
the  set of  the distribution functions $\mF_{\vp + \vec{Q}_{ n}}^{( n)}$ and $\mF_{\vp + 
\vec{Q}_{ p}}^{( p)}$.
This equation should be supplemented with the continuity equations 
for particle species $\alpha={ n, \, p}$.
In what follows, we do not account for the chemical reactions between the species ``$n$'' and ``$p$'', 
assuming that the number of particles of each species is conserved separately. 
Thus, the continuity equations can be represented as
\begin{equation}
		\label{eq:cont_eq}
		{\partial n_\alpha \over \partial t} + \Div \vec{j}_\alpha = 0,
\end{equation}
where the particle current density $\vec{j}_\alpha$ can still be calculated with {the 
expressions
\eqref{eq:J} or \eqref{eq:j_2010}.}
It is worth noting that Eq.\ \eqref{eq:cont_eq} cannot be obtained from the kinetic equation 
\eqref{eq:kin_eq}. 
To derive it from the microphysical theory, one needs to consider the full system of kinetic 
equations 
for 
Landau quasiparticles \cite{bn69,AronovEtAl1981,Gusakov2010}.
In addition to Eqs.\ \eqref{eq:kin_eq} and \eqref{eq:cont_eq}, one also needs an equation 
describing the evolution of the superfluid component. 
For an uncharged mixture this ``superfluid'' equation takes the form \cite{AronovEtAl1981,Gusakov2010}
\begin{equation}
		\label{eq:sf_eq}
		\frac{\partial {\vec Q_{\alpha}}}{\partial t} = - \nabla  {\breve\mu_\alpha},
\end{equation}
where $\breve\mu_\alpha$ is the nonequilibrium analog 
of the corresponding chemical potential of particle species $\alpha$,
which has already been introduced in the expression \eqref{eq:energy}.

\subsection{Thermodynamic equilibrium} 

{First of all, 
we need 
to determine the equilibrium state to which dissipative corrections 
will be 
sought.	
Besides the usual thermodynamic variables (e.g., temperature and chemical potentials), the 
equilibrium 
state of a superfluid Fermi mixture  
is generally characterized by 
the velocity  $\vec{u}$ of normal part of the mixture (Bogoliubov thermal excitations)
and by the superfluid currents for each particle species \cite{Khalatnikov1957}.
These supercurrents can exist in the system without any dissipation 
until they reach some critical values (see, e.g., Refs.\ \cite{Landau_stat2,bardeen62,gk13}).
The equilibrium distribution function corresponding to this situation
has already been found in Sec.\ \ref{sec:normal_currents}.
Indeed, as discussed in Sec.\ \ref{sec:normal_currents}, 
when the vectors $\vec{V}_{ q n}$ and  $\vec{V}_{ q p}$ 
are equal to each other, they have the meaning of the normal velocity $\vec{u}$. 
In this case, corresponding to complete thermodynamic equilibrium, 
the equilibrium distribution function of Bogoliubov excitations 
takes the form [see Eq.\ \eqref{eq:Fp1} with $\vec{V}_{ q \alpha}=\vec{u}$] }
\begin{equation}
		\label{eq:mF0_comp}
		\mF_{\vp + \vQa ,0}^{(\alpha)} =
		\frac{1}{ 1 + { e}^{(\mathfrak{E}_{\vp + \vQa ,0}^{(\alpha)}-\vec{p}\, \vec{u})/T}},
\end{equation}
where
\begin{equation}
		\label{eq:E_p_eq}
		\mathfrak{E}_{\vp + \vQa , 0}^{(\alpha)} = E_{\vp}^{(\alpha)} + \vp \sum_{\alpha'} \gamma_{\alpha\alpha'} \pmb{w}_{\alpha'}+ \vp \vec{u}
\end{equation}
is the equilibrium energy of Bogoliubov excitations
and  $\pmb{w}_\alpha = [\vQa - (\mu_\alpha/c^2) \vec{u}]$
is the vector proportional to the difference between the superfluid and normal velocities.

To obtain Eq.\ \eqref{eq:E_p_eq} one needs to substitute Eq.\ \eqref{eq:dH_pn} into 
\eqref{eq:E_exp}, set $\vec{V}_{ q n}=\vec{V}_{ q p}= \vec{u}$, and apply the relation 
\eqref{eq:gamma_k_rel}.
Plugging \eqref{eq:E_p_eq} into the distribution function \eqref{eq:mF0_comp}, one finally gets
\begin{equation}
		\label{eq:n_{e}q_curr}
		\mF_{\vp + \vQa ,0}^{(\alpha)} =
		\frac{1}{ 1 + {e}^{\left\{{E}_{{\vec p}}^{(\alpha)} + \vec{p} \sum_{\alpha'} \gamma_{\alpha\alpha'} \pmb{w}_{\alpha'} \right\}/T}}.
\end{equation} 	 

The equilibrium particle  current density can be calculated by setting $\vec{V}_{ q 
n}=\vec{V}_{ q p}= \vec{u}$ in Eq.\ \eqref{eq:j_nc} 
and 
is given by the
expression \eqref{eq:j_nc_eq}.

\subsection{Perturbations of the thermodynamic equilibrium}

Let us now assume a small departure from the thermodynamic equilibrium. 
We restrict ourselves to considering the corrections
caused by the small gradients of macroscopic variables. 
In this case, the formal small parameter is the Knudsen number $\mathcal{K} = \ell \mathfrak q$, where 
$\ell$ is the mean free path of Bogoliubov excitations. 

To find an approximate solution to Eqs.\ \eqref{eq:kin_eq}--\eqref{eq:sf_eq}, we, following  
Chapman-Enskog method 
(see, e.g., Ref. \cite{ChapmanCowlingCercignani1970}), 
expand the distribution function for Bogoliubov excitations 
in the powers of the small parameter $\mathcal K$:
\begin{equation}
		\label{eq:mF_comp_tmp}
		\mF_{\vp + \vQa}^{(\alpha)} = \overline{\mF}_{\vp + \vQa,0}^{(\alpha)} + \bar{\mathfrak{f}}_1^{(\alpha)}.
\end{equation}		
Here the function $\overline{\mF}_{\vp + \vQa,0}^{(\alpha)}$ 
satisfies the equation 
\begin{equation}
		\label{eq:col_int_eq}
		I^{(\alpha)} [  \overline{\mF}_{\vp + \vQa,0}^{(\alpha)},  \overline{\mF}_{\vp + \vQb,0}^{(\beta)} ] = 0
\end{equation}
(see Appendix \ref{sec:Integral} for details) 
and  $ \bar{\mathfrak{f}}_1^{(\alpha)}$ is the small correction to be found below.
The function $\overline{\mF}_{\vp + \vQa,0}^{(\alpha)} $ 
is the Fermi-Dirac distribution 
function,
\begin{equation}
		\label{eq:ol_mF}
		\overline{\mF}_{\vp + \vQa,0}^{(\alpha)} =
		\frac{1}{ 1 + {\rm e}^{(\mathfrak{E}_{\vp + \vQa}^{(\alpha)}-\vec{p}\, \vec{u})/T}},
\end{equation}	
{where all the thermodynamic variables, 
as well as the hydrodynamic velocities can generally 
depend on time and space coordinates, and
\begin{equation}
		\label{eq:E_p}
		\mathfrak{E}_{\vp + \vQa}^{(\alpha)} = \mathfrak{E}_{\vp + \vQa,0}^{(\alpha)} + \Delta \mathfrak{E}_{\vp + \vQa}^{(\alpha)}
\end{equation}	
{
is the {\it local} Bogoliubov excitation energy,
which depends on the nonequilibrium distribution functions $\mF_{\vp + \vQa}^{(\alpha)}$ 
of all particle species.}
In the Landau theory of Fermi liquids
the energy $\mathfrak{E}_{\vp + \vQa}^{(\alpha)}$
generally 
differs
from the 
equilibrium energy $ \mathfrak{E}_{\vp + \vQa,0}^{(\alpha)}$, given by Eq.\ \eqref{eq:E_p_eq} 
\cite{PinesNozieres}.
The difference $\Delta 
\mathfrak{E}_{\vp + \vQa}^{(\alpha)}$ between these energies
should be found simultaneously (and self-consistently) 
with 
the function $\bar{\mathfrak{f}}_1^{(\alpha)}$; see Sec.\ \ref{currents} for details.	 
}

{

Let us transform the left-hand side of Eq. \eqref{eq:kin_eq} to a form 
more suitable for the subsequent analysis.
Plugging the expansion \eqref{eq:mF_comp_tmp} into  \eqref{eq:kin_eq}, we get
\begin{align}
		\nonumber
		&{ \partial \overline{\mF}_{\vp + \vQa, 0}^{(\alpha)} \over \partial  \mathfrak{E}_{\vp + \vQa}^{(\alpha)}}
	 	\left[ 
		{\partial E_{\vp}^{(\alpha)} \over \partial t}
		+ \vp \sum_{\alpha'} \gamma_{\alpha\alpha'}   {\partial \pmb w_{\alpha'} \over \partial t}
		+\vp  \sum_{\alpha'}  \pmb w_{\alpha'} {\partial \gamma_{\alpha\alpha'}   \over \partial t} 
		 -  { {E}_{{\vec p}}^{(\alpha)} + \vp \sum_{\alpha'} \gamma_{\alpha\alpha'}  \pmb w_{\alpha'} \over T}   {\partial T \over \partial t}   \right.
	  	\\
	  	&     
		- \left( \frac{\partial {E}_{{\vec p}}^{(\alpha)}}{\partial \vp} +  \sum_{\alpha'} \gamma_{\alpha\alpha'} \pmb w_{\alpha'} { + \vec{u} } \right)  { {E}_{{\vec p}}^{(\alpha)} + \vp \sum_{\alpha'} \gamma_{\alpha\alpha'} \pmb w_{\alpha'} \over  T}   \nabla T
		-   \left( \frac{\partial {E}_{{\vec p}}^{(\alpha)}}{\partial \vp} +  \sum_{\alpha'} \gamma_{\alpha\alpha'} \pmb w_{\alpha'}  \right) \nabla \left(\vp \vec{u} \right) 
		\nonumber 
		\\
		&
		\left. 
		 +  (\vec{u} \nabla) \left( {E}_{{\vec p}}^{(\alpha)} + \vp \sum_{\alpha'} \gamma_{\alpha\alpha'}  \pmb w_{\alpha'} \right)
		  \right]
		  + \mbox{derivatives of }  \Delta \mathfrak{E}_{\vp + \vQa}^{(\alpha)} 
		  + \mbox{derivatives of } \bar{\mathfrak{f}}_1^{(\alpha)} 
		= I^{(\alpha)} [  \mF ].
		\label{eq:kin_eq_tmp}
\end{align}

First of all, we note that the expression \eqref{eq:kin_eq_tmp} contains a number of terms with various derivatives of 
$\bar{\mathfrak{f}}_1^{(\alpha)}$ and $\Delta \mathfrak{E}_{\vp + \vQa }^{(\alpha)}$,
which we do not write down explicitly.
There terms 
are quadratically small in the parameter $\mathcal{K}$
and, therefore, they  should be omitted.
Further, choosing $n_\alpha$ and $S$ as the independent thermodynamic 
variables,%
%
\footnote{\label{fn:hydr_vels} 
Strictly speaking, besides 
$n_\alpha$ and $S$ 
any thermodynamic quantity in a superfluid matter will also generally depend on
the velocity difference squared 
$\pmb{w}_\alpha^2 = [\vQa - (\mu_\alpha/c^2) \vec{u}]^2$ 
(see, e.g., Ref. \cite{Khalatnikov_book} for a detailed discussion and Ref.\
\cite{Gusakov2016} for a relativistic generalization). 
However, since we  work in the linear approximation in hydrodynamic velocities,
this dependence can be ignored.} 
%
we substitute the relations
\begin{align}
		\label{eq:dEdt_expand}
		{\partial E_{\vp}^{(\alpha)} \over \partial t} &= \sum_{\alpha'} {\partial E_{\vp}^{(\alpha)} \over \partial n_{\alpha'}} {\partial n_{\alpha'} \over \partial t} + {\partial E_{\vp}^{(\alpha)} \over \partial S} { \partial S \over \partial t},  
		\\
		\label{eq:dTdt_expand}
	 	{\partial T \over \partial t} & = \sum_{\alpha'} {\partial T  \over \partial n_{\alpha'}} {\partial n_{\alpha'} \over \partial t} + {\partial T \over \partial S} { \partial S \over \partial t},
		\\
		\label{eq:ap:dgamma_aa}
		{\partial \gamma_{\alpha\alpha'} \over \partial t} & = \sum_{\alpha''} {\partial \gamma_{\alpha\alpha'}  \over \partial n_{\alpha''}} {\partial n_{\alpha''} \over \partial t} + {\partial 	\gamma_{\alpha\alpha'} \over \partial S} { \partial S \over \partial t}, \\
		\label{eq:ap:dmu}
		{\partial \mu_\alpha \over \partial t} & = \sum_{\alpha'} {\partial \mu_\alpha \over \partial n_{\alpha'}} {\partial n_{\alpha'} \over \partial t} + {\partial \mu_\alpha\over \partial S} { \partial S \over \partial t}
\end{align}
into Eq.\ \eqref{eq:kin_eq_tmp}.
To calculate the time derivatives of $n_{\alpha}$, we make use of the continuity equations
\eqref{eq:cont_eq}, where the particle current density is given by Eq.\ \eqref{eq:j_nc_eq}.
In order to calculate the time derivative of the entropy we apply the entropy equation 
\eqref{eq:dS_ideal}, which has a natural form and will be derived from the kinetic equation in 
Sec. \ref{sec:entropy}.
	
Besides the expansion in Knudsen number we should 
work in the linear approximation in hydrodynamic velocities.
Moreover, following the standard approach (see, e.g., Refs.\ \cite{LL6, Khalatnikov_book, 
Gusakov2007}), we will omit all the terms in Eq.\ (\ref{eq:kin_eq_tmp}) that explicitly depend on 
the 
velocity difference $\pmb{w}_\alpha$ (but not on its derivatives).%
\footnote{
Accounting for these terms would lead to a substantial increase in the number of transport 
coefficients, even for a one-constituent superfluid liquid (see, e.g., Ref.\ 
\cite{Putterman_book}). However, as argued in the literature \cite{LL6}, 
these terms are small in comparison to the retained ones
in the majority of applications.
Since obtaining the most general 
form of the (linearized) transport equation
is not 
among the goals of the present paper, we restrict ourselves to neglecting these terms in what 
follows.
	}
To simplify the calculations,	
we also consider a certain point in space where $\vec{u}=0$ at a given moment of time.
Clearly, this assumption does not lead to any loss of generality, 
since it can always be fulfilled by choosing an appropriate 
inertial reference 
frame.

To calculate the time derivative of the momentum  $\vec{Q}_\alpha$, we should use the 
superfluid equation \eqref{eq:sf_eq}.
Before using it, it is
necessary to relate the chemical potentials $\breve{\mu}_\alpha$ and $\mu_\alpha$.
Generally, they differ for two reasons. 
First, as discussed above, 
the chemical potential itself can be defined in 
various
ways 
(e.g., in different reference frames).
This difference
is of the second order smallness in hydrodynamic velocities 
and can be neglected (see Sec. \ref{sec:normal_currents}).
Second, the quantity $\breve{\mu}_\alpha$ may contain dissipative corrections caused by the 
departure of the distribution functions ${\mF}_{\vp + \vQa}^{(\alpha)} $ from those defined in 
local 
thermodynamic equilibrium.
However, these differences 
are small, 
$\sim  \mathcal{O}(\mathcal{K})$,
and should be neglected in the Chapman-Enskog method
after substituting Eq.\ \eqref{eq:sf_eq} into the left-hand side of the kinetic equation.
Thus, for our purposes, one could use Eq.\ \eqref{eq:sf_eq} 
with $\breve{\mu}_\alpha$ replaced by $\mu_\alpha$.

With these comments in mind, 
Eq.\ \eqref{eq:kin_eq_tmp} can be rewritten as
\begin{align}
		\nonumber
		- { \partial \fp \over \partial {E}_{{\vp}}^{(\alpha)}}
	 	& \left[ \sum_{\alpha'} \gamma_{\alpha\alpha'}   {\mu_{\alpha'} \over c^2}\vp{ \partial \vec{u}  \over \partial t}  + \sum_{\alpha'} \gamma_{\alpha\alpha'}  \vp  \nabla \mu_{\alpha'} 
		  {+} \sum_{\alpha'}  \left({ \partial E_{\vp}^{(\alpha)} \over\partial n_{\alpha'}} 
		 { -} { E_{\vp}^{(\alpha)} \over T} {\partial T \over  \partial n_{\alpha'}} \right)\Div \vec{ j}_{\alpha'} \right. 
		   \\
		  + & \left.  S \left({  \partial E_{\vp}^{(\alpha)} \over \partial S} - { E_{\vp}^{(\alpha)} \over T} {\partial T \over \partial S} \right)\Div  \vec{u} 
		  +   {\partial  E_{\vp}^{(\alpha)} \over \partial \vp}  \nabla \left(\vp \vec{u} \right) +  {\partial  E_{\vp}^{(\alpha)} \over \partial \vp} { E_{\vp}^{(\alpha)} \over T}  \nabla T 
  \right]  = I_\alpha  [  \mF ].
  	\label{eq:ap:kin_tmp2}
\end{align}	

It remains
just to exclude the time derivative of the velocity $\vec{u}$.
To this aim, let us multiply Eq.\ \eqref{eq:ap:kin_tmp2} by $\vp$ and sum the result over the 
quantum states.
Using Eqs.\ \eqref{eq:int_pp} and  \eqref{eq:Naa_alt},
we find
\begin{equation}
		\label{eq:mom_eq_a_lin}
		 \sum_{\alpha'} R_{\alpha'\alpha} {\mu_{\alpha'} \over c^2}  {\partial \vec{u} \over \partial t}     +   \sum_{\alpha'} R_{\alpha'\alpha} \nabla \mu_{\alpha'} + S_\alpha  \nabla T = \sum_{\vec{p} \sigma} \vp  I_\alpha [  \mF].
\end{equation}	
Note that some terms in Eq.\ \eqref{eq:ap:kin_tmp2} disappeared after the summation 
because they are antisymmetric 
in $\vec{p}$.	 
In Eq.\ \eqref{eq:mom_eq_a_lin},  $S_\alpha$ 
is the ``partial entropy density'' 
for particle species $\alpha$,
\begin{equation}
		\label{eq:part_entropy}
		S_\alpha =   \sum_{{\vec p} \sigma } \sigma_{\vp,0}^{(\alpha)},
\end{equation}	
where 
$ \sigma_{\vp,0}^{(\alpha)}$ is given by the expression 
\eqref{eq:entropy_dens} with the current-free equilibrium distribution function \eqref{eq:fp_eq_cl} 
instead of $\mF_{{\vec p}+ {\vec Q}_{\alpha}}^{(\alpha)}$. 
It can be shown that 
\begin{equation}
		\label{eq:sigma_0_rel}
		\sum_{\vp \sigma }   \sigma_{\vp,0}^{(\alpha)} = -  {1\over 3} \sum_{\vp\sigma } p  {  {E}_{\vp}^{(\alpha)} \over T} \frac{\partial {E}_{\vp}^{(\alpha)}}{\partial p} \frac{\partial \fp}{ \partial {E}_{{\vp}}^{(\alpha)}}.
\end{equation}	
From this equation it becomes apparent that
the last term in the left-hand side of Eq.\ \eqref{eq:mom_eq_a_lin} came 
from the last term in the left-hand side of Eq.\ \eqref{eq:ap:kin_tmp2}.	
The sum of the right-hand sides of Eq.\ \eqref{eq:mom_eq_a_lin} for all particle species
must vanish due to the momentum conservation (see Appendix \ref{sec:Integral}). 
As for the left-hand sides, summing them up and using the definition \eqref{eq:N_qa} we get%
%
\footnote{ One can arrive at the same equation 
from the phenomenological hydrodynamics. To do this, 
one needs to substitute Eq.\ \eqref{eq:ap:superfluid_eq_lin} into \eqref{eq:ap:Euler_lin}, 
accounting for the relation \eqref{eq:part_cons_rel} and setting the electrical
charges of all particle species to zero. }
%
\begin{equation}
		\label{eq:mom_eq_norm}
	 	\rho_{ q} {\partial \vec{u} \over \partial t}  =   -   n_{ qn} \nabla \mu_{n} - n_{ q p} \nabla \mu_{p} -  S\nabla T,
\end{equation}
where 
\begin{equation}
		\label{eq:rho_q}
	 	\rho_{ q} = 
		\sum_\alpha {\mu_\alpha \over c^2}n_{ q \alpha}.
\end{equation}

Equation \eqref{eq:mom_eq_norm} has the form of the linearized Euler equation, in which the 
quantity $\rho_{ q}$ plays the role of the density of normal (nonsuperfluid) component of the 
liquid.
However, it was argued in Sec.\ \ref{sec:normal_currents} 
that at least one of the normal number densities $n_{ q \alpha}$ can be negative. 
In spite of this, the quantity $\rho_{ q}$ appears to be always non-negative.
Indeed, substituting the definitions \eqref{eq:N_qa}, \eqref{eq:Naa_alt}, \eqref{eq:gamma_aa}, and 
\eqref{eq:gamma_ab} into \eqref{eq:rho_q}, 
and making use of the relation \eqref{eq:m_eff_eq}, one can represent $\rho_q$ as
\begin{equation}
	\rho_{ 	q} = \frac{n_{ n} n_{ p}}{\mathcal S} \left[n_{ n} m_{ n}^* \Phi_{ n} (1 - \Phi_{ p}) + n_{ p} m_{ p}^* \Phi_{ p} (1 - \Phi_{ n}) \right]
	+ \frac{\mathcal{S}_{\rm nsf}}{\mathcal S} \frac{n_{ n}\mu_{ n} + n_{ p}\mu_{ p}}{c^2} \Phi_{ n} \Phi_{ p } ,
\end{equation}
where $\mathcal S$ is given by Eq.\ \eqref{eq:S_det}, and ${\mathcal S}_{\rm nsf}$ is the 
value of $\mathcal S$ in a nonsuperfluid mixture  [see Eq.\ \eqref{Snsf}].
According to the stability 
constraint
\eqref{eq:stab_rel}, $\mathcal S > 0$ and, consequently, 
${\mathcal S}_{\rm nsf} > 0$.
Thus, the normal density $\rho_{q} \geq 0 $ and vanishes only at $T=0$, as expected.

Plugging the time derivative from Eq.\ \eqref{eq:mom_eq_norm} into \eqref{eq:ap:kin_tmp2}, 
one arrives at
	}
\begin{align}
		\nonumber
		- &{ \partial \fp \over \partial {E}_{\vp}^{(\alpha)}}
		 \left\{\sum_{\alpha'} \left[ \gamma_{\alpha\alpha'} - \left(\gamma_{\alpha\alpha}{\mu_\alpha \over c^2} + \gamma_{\alpha\beta}{\mu_\beta \over c^2}\right){n_{ q\alpha'} \over \rho_{ q}} \right] \vp \, \nabla\mu_{\alpha'} - 
		 \right.
		  \nonumber \\
		& \left.
		 -\left[ \left(\gamma_{\alpha\alpha}{\mu_\alpha \over c^2} + \gamma_{\alpha\beta}{\mu_\beta \over c^2}\right) {S \over \rho_{ q}} \vp -  { E_{\vp}^{(\alpha)} \over T}  {\partial  E_{\vp}^{(\alpha)} \over \partial \vp} \right] \nabla T \right. + 
		\nonumber \\
		 &  + \sum_{\alpha'}  \left({ \partial E_{\vp}^{(\alpha)} \over\partial n_{\alpha'}} - { E_{\vp}^{(\alpha)} \over T} {\partial T \over  \partial n_{\alpha'}} \right)\Div \vec j_{\alpha'} 
		 \nonumber \\
		& \left.
		 +    S\left({  \partial E_{\vp}^{(\alpha)} \over \partial S} - { E_{\vp}^{(\alpha)} \over T} {\partial T \over \partial S} \right)\Div \vec{u}  		 
		 +    {\partial  E_{\vp}^{(\alpha)} \over \partial \vp}     \nabla \left(\vp \vec{u}\right)   \right\} = I_\alpha [  \mF ].
\end{align}
This equation can be rewritten in a more canonical form as
\begin{align}
		\nonumber
		- &{ \partial \fp \over \partial {E}_{{\vec p}}^{(\alpha)}}
		 \left\{\sum_{\alpha'} \left[ \gamma_{\alpha\alpha'} - \left(\gamma_{\alpha\alpha}{\mu_\alpha \over c^2} + \gamma_{\alpha\beta}{\mu_\beta \over c^2}\right){n_{ q\alpha'} \over \rho_{ q}} \right] \vp \, \nabla\mu_{\alpha'} \right.
		 \\
		  & -\left[ \left(\gamma_{\alpha\alpha}{\mu_\alpha \over c^2} + \gamma_{\alpha\beta}{\mu_\beta \over c^2}\right) {S \over \rho_{ q}}\vp -  { E_{\vp}^{(\alpha)} \over T}  {\partial  E_{\vp}^{(\alpha)} \over \partial \vp} \right] \nabla T  + 
		\nonumber \\
		& \left.  + \left[ \sum_{\alpha'}  n_{\alpha'}\left({ \partial E_{\vec p}^{(\alpha)} \over\partial n_{\alpha'}} - { E_{\vec p}^{(\alpha)} \over T} {\partial T \over  \partial n_{\alpha'}} \right) +  S\left({  \partial E_{\vec p}^{(\alpha)} \over \partial S} - { E_{\vec p}^{(\alpha)} \over T} {\partial T \over \partial S} \right) + { 1\over 3} {\partial  E_{\vec p}^{(\alpha)} \over \partial \vp}  \vec{p} \right]\Div \vec{u}  + \right. \nonumber \\ 
		 & + \sum_{\alpha'}  \left({ \partial E_{\vec p}^{(\alpha)} \over\partial n_{\alpha'}} - { E_{\vec p}^{(\alpha)} \over T} {\partial T \over  \partial n_{\alpha'}} \right)\Div \left( \vec{j}_{\alpha'} - n_{\alpha'} \vec{u} \right) +  \nonumber \\
		&\left. +   
		{1\over 2}\left( {\partial  E_{\vec p}^{(\alpha)} \over \partial  p^i} {p}^j - {1\over 3} \delta^{ij}{\partial  E_{\vec p}^{(\alpha)} \over \partial \vec p} \vec{p} \right) \left( {\partial u^i \over \partial r^j} + {\partial u^j \over \partial r^i }- {2 \over 3} \delta^{ij}\Div\vec{u} \right)       \right\} = I_\alpha  [  \mF].
		 \label{eq:kin_eq_tot}
\end{align}
This is the general equation allowing one to study different transport processes 
 in strongly interacting superfluid Fermi mixtures, such as the thermal 
conductivity, bulk and shear viscosities, and particle diffusion.
In nonsuperfluid matter one has $\gamma_{\alpha\alpha} \rightarrow 1/m_\alpha^{*}$ and 
$\gamma_{\alpha\beta} \rightarrow 0$,
and thus
the in-medium effects in the left-hand side of Eq.\ \eqref{eq:kin_eq_tot} manifest themselves only 
through the renormalization of the particle masses, when $m_\alpha$ is replaced with $m_\alpha^*$.
In contrast, in a superfluid mixture,
Eq.\ \eqref{eq:kin_eq_tot} 
explicitly depends
on {\it all} the Landau parameters $f_1^{\alpha\alpha'}$ through the 
elements of the matrix $\gamma_{\alpha\alpha'}$ see Eqs.\ \eqref{eq:gamma_aa} and 
\eqref{eq:gamma_ab}.

\subsection{Diffusion currents}
\label{currents}

In the present paper we focus on the diffusion of particles 
produced by the gradients of chemical potentials.
Assuming all other gradients in Eq.\ \eqref{eq:kin_eq_tot} vanish, we find
\begin{align}
		- &{ \partial \fp \over \partial {E}_{\vp}^{(\alpha)}}
		 \sum_{\alpha'} \left[ \gamma_{\alpha\alpha'} - \left(\gamma_{\alpha\alpha}{\mu_\alpha \over c^2} + \gamma_{\alpha\beta}{\mu_\beta \over c^2}\right){n_{ q\alpha'} \over \rho_{ q}} \right] \vp \, \nabla\mu_{\alpha'} 
		 = I_\alpha [  \mF ].
		 \label{eq:kin_eq_mu}
\end{align}
Let us multiply this equation by 
$\vp$ and sum the result over quantum states.
Using the expressions \eqref{eq:int_pp} and  \eqref{eq:Naa_alt} 
together with the definitions \eqref{eq:N_qa} and \eqref{eq:rho_q}, we obtain%
\footnote{ The 
same equations can be obtained if one 
substitute the time derivative from Eq. 
\eqref{eq:mom_eq_norm} into Eqs. \eqref{eq:mom_eq_a_lin}, where $\nabla T$ is set to zero.}
\begin{gather}
		\label{eq:froce_bal_n}
		\frac{{\mathbb R}}{ \rho_{ q}}{\mu_{ n} \mu_{ p}\over c^2} \left( {\nabla {\mu}_{ n} \over \mu_{ n}} - {\nabla {\mu}_{ p} \over \mu_{ p}}\right) = \sum_{{\vec p} \sigma} \vp   \, I_{ n} [  \mF], 
		\\
		- \frac{{\mathbb R}}{ \rho_{ q}}{\mu_{ n} \mu_{ p}\over c^2} \left( {\nabla {\mu}_{ n} \over \mu_{ n}} - {\nabla {\mu}_{ p} \over \mu_{ p}}\right) = \sum_{{\vec p} \sigma} \vp   \, I_{ p} [  \mF],
		\label{eq:froce_bal_p}
\end{gather}
where $\mathbb R = R_{ nn}R_{ pp} - R_{ np}R_{ pn}$.

{
Our immediate goal will be
to find
the corrections $\bar{\mathfrak{f}}_1^{(\alpha)}$ to the {equilibrium} distribution functions 
\eqref{eq:mF_comp_tmp},
and to 
calculate 
the 
particle current densities using the expression \eqref{eq:J}.
Instead of the corrections $\bar{\mathfrak{f}}_1^{(\alpha)}$
it is convenient to introduce the functions $\phi_\alpha$ through the relation 
\cite{PinesNozieres}
\begin{equation}
		\label{eq:f1}
		\bar{\mathfrak{f}}_1^{(\alpha)} = -\frac{\partial \overline{\mF}_{\vp + \vQa,0}^{(\alpha)}}{ \partial \mathfrak{E}_{\vp+\vQa,0}^{(\alpha)}} \phi_\alpha
		= \frac{\overline{\mF}_{\vp + \vQa,0}^{(\alpha)} \left( 1- \overline{\mF}_{\vp + \vQa,0}^{(\alpha)}\right)}{T}  \phi_\alpha.
\end{equation}
Following Refs.\ \cite{ziman1960,YakovlevShalybkov1991},
we look for $\phi_\alpha$ in the form%
%
\footnote{It should be stressed that if a set of functions 
$\vec{V}_{i\alpha}(\vp)$ satisfies the transport equation, then another set $\vec{V}_{i\alpha}(\vp) 
+ \vec{a}$, where $\vec{a}$ is an arbitrary vector, would satisfy the same equation 
(see {also} Appendix \ref{sec:Integral}).
From the physical point of view, this ambiguity is related to the fact that the definition of 
the velocity $\vec{u}$ in the dissipative hydrodynamics is not unique (see, e.g., Ref.\ 
\cite{LL6}). 
Hence, 
an 
additional constraint 
should be applied to the solution. 
The choice of the constraint is discussed 
after the expression \eqref{eq:i_def}.}
%
\begin{equation}
		\label{eq:phi_def}
		\phi_\alpha =\vp \vec{V}_{i\alpha}(\vp),
\end{equation}	
where the vector $\vec{ V}_{i\alpha}(\vp)$ is assumed to be a smooth function of $\vp$.	
In a highly degenerate matter,  collisions occur mostly between 
particles
in the vicinity of the corresponding Fermi surfaces,
where the derivative $-\partial 
\overline{\mF}_{\vp + \vQa,0}^{(\alpha)} / \partial \mathfrak{E}_{\vp+\vQa,0}^{(\alpha)}$ has a 
sharp maximum, while the function $\vec{V}_{i\alpha}(\vp)$ is expected to vary only slightly. 
Hence, 
to 
a first approximation one may treat $\vec{V}_{i\alpha}$ as a constant vector.  
In the present paper,  following Refs.\ \cite{ziman1960,YakovlevShalybkov1991}, 
we will work within this 
approximation. 
	
As was already mentioned above, besides the correction $\bar{\mathfrak{f}}_1^{(\alpha)}$, one also 
needs to determine the correction $\Delta \mathfrak{E}_{\vp+\vQa}^{(\alpha)}$ to the {equilibrium} 
Bogoliubov 
excitation 
energy 
[see the expression \eqref{eq:E_p}].
To do this, 
we make 
use of the self-consistency relation  \eqref{eq:localenergy}, exactly as was done 
in Sec. \ref{sec:normal_currents}.
Working in the linear approximation in vectors $\vec{V}_{i\alpha}$, 
the energy correction can be presented as 
\begin{equation}
		\label{eq:dE_alt}
		 \Delta \mathfrak{E}_{\vp}^{(\alpha)} = \vec{p} \sum_{\alpha'}   \tilde{K}_{\alpha\alpha'}\vec{V}_{ i\alpha'},
\end{equation}
where the coefficients $\tilde{K}_{\alpha\alpha'}$ are yet to be determined. 
Substituting this expression together with expressions \eqref{eq:f1} and \eqref{eq:phi_def} into 
the distribution function \eqref{eq:mF_comp_tmp} and linearizing the result with respect to 
$\vec{V}_{i\alpha}$, one gets
\begin{equation}
		\label{eq:dF_dif_1}
		\mF_{\vp+\vQa}^{(\alpha)} \approx \fp  + { \partial \fp \over \partial {E}_{\vp}^{(\alpha)}} \vp \sum_{\alpha'}  \gamma_{\alpha\alpha'} \pmb w_{\alpha'} +{ \partial \fp \over \partial {E}_{\vp}^{(\alpha)}} \vp \sum_{\alpha'}   \tilde{K}_{\alpha\alpha'}\vec{V}_{ i\alpha'} - { \partial \fp \over \partial {E}_{\vp}^{(\alpha)}} \vp \vec{V}_{ i\alpha}.
\end{equation}
Out of the thermodynamic equilibrium the Bogoliubov coherence factors are still given by Eqs.\ 
\eqref{eq:upvp1} and \eqref{upQ} \cite{AronovEtAl1981,Gusakov2010}.
Hence,  the distribution functions $\mN_{\vp + \vQa}^{(\alpha)} $ and 
$\mF_{\vp+\vQa}^{(\alpha)}$ are still related by the expression \eqref{eq:Np_def}.
Then, linearizing Eq.\ \eqref{upQ} and using the fact 
that 
Bogoliubov coherence factors are even, while the 
representation \eqref{eq:dE_alt} is odd with respect to transformation 
$\vp \rightarrow -\vp$, one can show that 
$\Delta H_{\vp}^{(\alpha)} = \Delta \mathfrak{E}_{\vp}^{(\alpha)}$,
where $\Delta H_{\vp}^{(\alpha)}$ is the nonequilibrium correction to the quantity   
$H_{\vp+\vQa}^{(\alpha)}$, given by Eq.\  \eqref{eq:localenergy}.
Plugging all the expansions into the expression \eqref{eq:localenergy}, 
one can finally obtain the equations for the coefficients $\tilde{K}_{\alpha\alpha'}$; 
the detailed 
calculation is similar to that presented
in Sec. \ref{sec:normal_currents}.
As a consequence,
the resulting equations coincides with Eq.\ \eqref{eq:K_eq}, hence
the matrix elements $\tilde{K}_{\alpha\alpha'}$ coincide with 
$K_{\alpha\alpha'}$, 
and are given by the expressions \eqref{eq:K_aa} and \eqref{eq:K_ab}.  
The fact that $\tilde{K}_{\alpha\alpha'}=K_{\alpha\alpha'}$ 
allows us to use the expansion \eqref{eq:E_exp} with $\Delta H_{\vec p}^{(\alpha)}$ given by 
Eq.\ 
\eqref{eq:dH_pn},
where by $\vec{V}_{ q\alpha}$ one should understand
the sum $\vec{u} + \vec{V}_{ i\alpha}$,
i.e., $\vec{V}_{ q\alpha}=\vec{u} + \vec{V}_{ i\alpha}$.

Now, substituting the expansion \eqref{eq:dF_dif_1} into Eq.\ \eqref{eq:J} and using the relation 
\eqref{eq:gamma_k_rel} together with the definitions \eqref{eq:Y}, \eqref{eq:Naa}, \eqref{eq:Nab}, 
and \eqref{eq:N_qa}, one obtains
\begin{align}
		\label{eq:j_nc_kin}
		\vec{j}_\alpha & = R_{\alpha\alpha} \vec{V}_{ q\alpha} + R_{\alpha\beta} \vec{V}_{ q\beta} + c^2 Y_{\alpha\alpha}\vec{Q}_\alpha + c^2 Y_{\alpha\beta}\vec{Q}_\beta 
		\\
			& = n_{ q \alpha} \vec{u} + c^2 Y_{\alpha\alpha}\vec{Q}_\alpha + c^2 Y_{\alpha\beta}\vec{Q}_\beta + \Delta \vec{j}_\alpha.
			\label{eq:j_dif}
\end{align}
Here,  the expression in the first line 
is completely identical to that in Eq.\ \eqref{eq:j_nc}.	
In the second line the same result is presented in the form more suitable for establishing the 
connection with the phenomenological 
hydrodynamics.
Equation \eqref{eq:j_dif} introduces the diffusion currents 
according to definition
\begin{equation}
		\label{eq:i_def}
		\Delta \vec{j}_\alpha = R_{\alpha\alpha} \vec{V}_{i\alpha} + R_{\alpha\beta} \vec{V}_{i\beta}.
\end{equation}	
Inverting the relations \eqref{eq:i_def}, one obtains
\begin{equation}
		\label{eq:VnVp}
		\vec{ V}_{ in} = \frac{R_{ pp} \Delta \vec{j}_{ n} -  R_{ np} \Delta\vec{j}_{ p}}{\mathbb R} , \ \ \ \vec{ V}_{ ip} =  \frac{R_{ nn} \Delta\vec{j}_{ p} - R_{ pn}\Delta\vec{j}_{ n}}{\mathbb R} .
\end{equation}	

To complete the derivation one also needs to define the comoving reference frame in which 
$\vec{u}=0$.
}
This can be done in a number of ways.
We prefer to choose the { so-called} Landau-Lifshitz frame, in which the 
additional momentum { density} caused by the 
dissipative currents (diffusion currents in our case) equals zero \cite{LL6}. 
Using Eq.\ \eqref{eq:PJ}, this requirement translates into
\begin{equation}
 		\label{eq:LL_cond}
		 \sum_\alpha {\mu_\alpha \over c^2} \Delta \vec{j}_\alpha = 0.
\end{equation}
A similar condition is used in the study of nonsuperfluid mixtures in Refs.\ 
\cite{YakovlevShalybkov1991,DommesGusakovShternin2020}.

Let us now consider the right-hand side of Eqs.\ \eqref{eq:froce_bal_n} and \eqref{eq:froce_bal_p}. 
After substitution of the distribution functions \eqref{eq:mF_comp_tmp}  together with expressions 
\eqref{eq:f1} and \eqref{eq:phi_def} into the collision integral, it can be represented in the 
following form (see Appendix \ref{sec:Integral} for details):
\begin{equation}
		\label{eq:force}
		\sum_{{\vp} \sigma}  \vp  \, I_{\alpha} [\mF] = - J_{\alpha\beta} \left( {\vec V}_{ i\alpha} - {\vec V}_{ i\beta} \right),
\end{equation}
where $J_{\alpha\beta}$ are the momentum transfer rates 
($J_{\alpha\beta}=J_{\beta\alpha}$).%
%
\footnote{ Taking this property into account, one can easily see that 
Eqs.\ \eqref{eq:froce_bal_n} and \eqref{eq:froce_bal_p}, in fact, coincide.}
%
Use of Eqs.\ \eqref{eq:froce_bal_n}, \eqref{eq:froce_bal_p}, \eqref{eq:VnVp}, and \eqref{eq:force}
allows us to find:
\begin{equation}
		\label{eq:ip_res}
		\Delta {\vec j}_{ p} = - \frac{\mu_{ p} \mu_{ n}^2}{c^4{\rho}_{ q}^2}   \frac{\mathbb{R}^2}{ J_{ np}}\left( {\nabla {\mu}_{ p} \over \mu_{ p}} - {\nabla {\mu}_{ n} \over \mu_{ n}}\right),
		\ \ \ \ \ \
		\Delta {\vec j}_{ n} =  \frac{\mu_{ n} \mu_{ p}^2}{c^4{\rho}_{ q}^2}   \frac{\mathbb{R}^2}{ J_{ np}}\left( {\nabla {\mu}_{ p} \over \mu_{ p}} - {\nabla {\mu}_{ n} \over \mu_{ n}}\right).
\end{equation}
The determinant of the matrix $R_{\alpha\alpha'}$ can be represented, 
with the help of Eqs.\ \eqref{eq:gamma_aa}--\eqref{eq:S_det} and \eqref{eq:Naa_alt} 
as
\begin{equation}
		\label{eq:RR_expr}
		\mathbb{R} = n_{ n} n_{ p} \Phi_{ n} \Phi_{ p} {{\mathcal S}_{\rm nsf} \over \mathcal S},
\end{equation}
where we recall that ${\mathcal S}_{\rm nsf}$
is the value of the quantity ${\mathcal S}$ taken for a nonsuperfluid mixture [see Eq.\ 
(\ref{Snsf})].

The relation \eqref{eq:ip_res} allows us to express the diffusion coefficients arising in 
relativistic superfluid hydrodynamics \cite{DommesGusakovShternin2020,DommesGusakov2021}  (see 
Appendix \ref{ap:hydrodynamics}) in a practical case of low-temperature degenerate isothermal 
matter ($\nabla T = 0$). 
For uncharged binary mixture, the expression \eqref{eq:ap:dj_exp},  written down for both particle 
species ``$n$'' and ``$p$'', reads
\begin{align}
		\label{eq:in_thr_D}
		&\Delta \vec{j}_{ n} = - \mathcal{D}_{ nn}{ \nabla  \mu_{ n} \over T } - \mathcal{D}_{ np}{ \nabla  \mu_{ p} \over T }, \\
		\label{eq:ip_thr_D}
		&\Delta \vec{j}_{ p} = - \mathcal{D}_{ pn}{ \nabla  \mu_{ n} \over T } - \mathcal{D}_{ pp}{ \nabla  \mu_{ p} \over T }.
\end{align}
Plugging Eqs.\ \eqref{eq:LL_cond} and \eqref{eq:Onsager_rel} into the expression 
\eqref{eq:ip_thr_D} and comparing the result with Eq.\ \eqref{eq:in_thr_D}, we find
\begin{equation}
	\label{DnnDpp}
		\mathcal{D}_{ nn} = - {\mu_{ p} \over \mu_{ n}} \mathcal{D}_{ np}, \ \ \ \ 
		\mathcal{D}_{ pp} = - {\mu_{ n} \over \mu_{ p}} \mathcal{D}_{ np},
\end{equation}
and
\begin{equation}
	\label{last}
		\Delta \vec{j}_{ p} =  \frac{\mathcal{D}_{ np} \mu_{ n}}{T} \left( {\nabla {\mu}_{ p} \over \mu_{ p}} - {\nabla {\mu}_{ n} \over \mu_{ n}}\right).
\end{equation}
Finally, comparing Eq.\ \eqref{last} 
with \eqref{eq:ip_res}, we arrive at
\begin{equation}
		\label{eq:Dnp}
		\mathcal{D}_{ np}  = - \frac{\mu_{ p} \mu_{ n}T}{c^4{\rho}_{ q}^2}   \frac{\mathbb{R}^2}{ J_{ np}}.
\end{equation}	
{This is one of the central results of our work showing 
how the Fermi-liquid effects and superfluidity manifest themselves in the diffusion coefficients of 
the mixture.}
To 
expose
the Fermi-liquid effects in this formula, let us  consider a few special cases.
If one deals with a mixture of two independent Fermi liquids ($G_{\alpha\beta} = 0$), 
the diffusion coefficient takes the form
\begin{equation}
		\label{eq:Dnp_ind}
		\mathcal{D}_{ np}  = - \frac{\mu_{ p} \mu_{ n} n_{ qn}^2 n_{ qp}^2}{\left( 
		\mu_{ n} n_{ qn} +\mu_{ p} n_{ qp}\right)^2 }   \frac{ T }{ J_{ np}},
\end{equation}	
where (see Appendix \ref{ap:coeffs})
\begin{equation}
		n_{ q \alpha} = R_{\alpha\alpha}= \frac{n_\alpha m_{\alpha}^* c^2 
		\Phi_\alpha}{m_{\alpha}^* c^2 \Phi_\alpha + {\mu_\alpha}(1-\Phi_\alpha)}.
\end{equation}	
If all the Fermi-liquid effects can be neglected, Eq.\ \eqref{eq:Dnp_ind} remains valid, while the 
normal density reduces to 
\begin{equation}
		n_{ q \alpha} = {n_\alpha  \Phi_\alpha}.
\end{equation}	
Hence, the Fermi-liquid effects in the diffusion coefficient $\mathcal{D}_{np}$ arise by 
replacing $n_{qn}^2 n_{qp}^2$ with $\mathbb{R}^2$ in the numerator, 
while employing the general expression \eqref{eq:N_qa} for the normal densities in the 
denominator.
In the case of a nonsuperfluid mixture, the relation \eqref{eq:Dnp} turns into 
(cf.\ Eq.\ (113) in Ref.\ \cite{DommesGusakovShternin2020})%
%
\footnote{The expression in Ref.\  
\cite{DommesGusakovShternin2020} contains an additional constant $c$ in the denominator 
because the diffusion currents in that paper have the dimension of number density.}
%
%
\begin{equation}
		\label{eq:Dnp_nsf}
		\mathcal{D}_{ np}  = - \frac{\mu_{ p} \mu_{ n}n_{ n}^2 n_{p}^2 T}{(\mu_{ n} n_{ n} + \mu_{ p} n_{ p})^2J_{ np}}.
\end{equation}
%

\subsection{Entropy} 
\label{sec:entropy}

Let us now derive the equation for the entropy {density}.
Multiplying Eq.\ \eqref{eq:kin_eq} by $\ln \left[(1- \mF_{\vp + \vQa}^{(\alpha)})/\mF_{\vp + 
\vQa}^{(\alpha)}\right]$ and summing the result over the quantum states and particle species 
indices, one  obtains
\begin{equation}
		\label{eq:S_eq}
	 	\frac{\partial S}{\partial t} + \Div \left( \sum_{\vp \sigma \alpha}  \frac{\partial \mathfrak{E}_{\vp + \vQa}^{(\alpha)}}{\partial \vp} \sigma_{\vp + \vQa }^{(\alpha)}\right) = \Gamma_s,
\end{equation} 
where
\begin{equation}
		\label{eq:Gamma_s}
		\Gamma_{ s} = - \sum_{\vp \sigma \alpha} \ln \left( \frac{\mF_{\vp + \vQa}^{(\alpha)}}{1- \mF_{\vp + \vQa}^{(\alpha)}}\right) I_\alpha  [  \mF]
\end{equation}
is the entropy production rate.

Let us first consider the expression \eqref{eq:Gamma_s}.
Plugging the expansion \eqref{eq:mF_comp_tmp} together with Eq.\ \eqref{eq:f1} into it 
and linearizing the result with respect to $\phi_\alpha$, one gets
\begin{align} 
		\Gamma_{ s} &= - \sum_{\vp \sigma \alpha} \ln \left( \frac{\overline{\mF}_{\vp + \vQa,0}^{(\alpha)}  }{1- \overline{\mF}_{\vp + \vQa,0}^{(\alpha)}  }\right) I_\alpha  [  \mF] - \frac{1}{T} \sum_{\vp \sigma \alpha} \phi_\alpha I_\alpha  [  \mF]
		\nonumber
		\\
		&= \frac{1}{T}  \sum_{\vp \sigma \alpha} \left( \mathfrak{E}_{\vp + \vQa}^{(\alpha)} -\vp \vec{u} \right) I_\alpha   [  \mF] - \frac{1}{T} \sum_{\vp \sigma \alpha} \phi_\alpha I_\alpha [  \mF].
		\label{eq:Gamma_s_lin}
\end{align}
The first term  in this expression equals zero due to the energy and momentum conservation in 
collision events.
Substitution of Eq.\ \eqref{eq:kin_eq_mu} into the second term 
gives
\begin{align}
		\nonumber
		\Gamma_{ s} = &\frac{1}{T} 
	 	\sum_{\alpha\alpha'} \left[ \gamma_{\alpha\alpha'} - \left(\gamma_{\alpha\alpha}{\mu_\alpha \over c^2} + \gamma_{\alpha\beta}{\mu_\beta \over c^2}\right){n_{ q\alpha'} \over{\rho}_{ q}} \right]  \nabla\mu_{\alpha'} \sum_{\vp \sigma} \vp \phi_\alpha { \partial \fp \over \partial {E}_{\vp}^{(\alpha)}}  \\
		 = & \frac{1}{T} \frac{\mathbb R}{{\rho}_{ q}} {\mu_{ n} \mu_{ p} \over c^2} \left( {1 \over m_{ n}^* n_{ n} \Phi_{ n}} \sum_{\vp\sigma} \vp \phi_n   { \partial \mathfrak{f}_{\vp}^{(n)} \over \partial {E}_{\vp}^{ (n)}}  - {1 \over m_{ p}^* n_{ p} \Phi_{ p}} \sum_{\vp \sigma} \vp \phi_{ p}   { \partial \mathfrak{f}_{\vp}^{ (p)} \over \partial {E}_{\vp}^{(p)}}\right) \left( {\nabla {\mu}_{ n} \over \mu_{ n}} - {\nabla {\mu}_{ p} \over \mu_{ p}}\right),
		 \label{eq:Gamma_s_tmp}
\end{align}
where to obtain the second equality we  used Eq.\ \eqref{eq:Naa_alt}.
Now, plugging expression \eqref{eq:phi_def} together with Eq.\ \eqref{eq:VnVp} into 
\eqref{eq:Gamma_s_tmp}, and using Eqs.\  
{\eqref{eq:int_pp} and \eqref{eq:N_qa},}
one finds 
\begin{equation}
		\label{eq:Gamma_s_tmp2}
		\Gamma_{ s} = \frac{1}{T} \, {\mu_{ n} \mu_{ p} \over c^2} \, \frac{n_{ qn} \Delta \vec{j}_{ p} - n_{ qp} \Delta \vec{j}_{ n}}{\rho_q}  \left( {\nabla {\mu}_{ n} \over \mu_{ n}} - {\nabla {\mu}_{ p} \over \mu_{ p}}\right) .
\end{equation}
{In view of }
the definition \eqref{eq:rho_q} 
{and} the relation \eqref{eq:LL_cond}, one can 
transform Eq.\ \eqref{eq:Gamma_s_tmp2} into 
\begin{equation}
		\label{eq:Gamma_s_nc}
		\Gamma_{ s} =  - { \mu_{ p}  \over  T}  \left( {\nabla {\mu}_{ p} \over \mu_{ p}} - {\nabla {\mu}_{ n} \over \mu_{ n}}\right) \Delta \vec{j}_{ p}.
\end{equation}
Expressing
$\Delta \vec{j}_{ p}$ from
Eq.\ \eqref{eq:ip_res} 
one verifies
that $\Gamma_{ s} \geq 0$, as it should be.
Applying again the relation \eqref{eq:LL_cond} to the expression \eqref{eq:Gamma_s_nc}, one gets
\begin{equation}
		\Gamma_{ s} =  - {1 \over T} \sum_\alpha  \Delta \vec{j}_{\alpha}  \nabla  \mu_\alpha.
\end{equation}
In this form, the entropy production rate coincides with the phenomenological relation  
\eqref{eq:ap:Gamma_s_nr} for uncharged mixture at constant $T$.
In the limit of nonrelativistic equation of state, the entropy production rate reduces to the 
standard expression \cite{LL6}:
\begin{equation}
		\Gamma_{ s} = - { 1  \over  T}  \left( {\nabla {\mu}_{ n} \over m_{ n}} - {\nabla {\mu}_{ p} \over m_{ p}}\right) m_{ p} \Delta \vec{j}_{ p}.
\end{equation}

Let us now turn to the calculation of the entropy current density [see the term under divergence in 
Eq.\ \eqref{eq:S_eq}]. 
Linearizing it  using Eqs. \eqref{eq:entropy_dens}, \eqref{eq:E_exp}, \eqref{eq:F_exp_nc}, and 
\eqref{eq:dH_pn}, one obtains%
%
\footnote{We remind the reader that, as argued in Sec. \ref{sec:diffusion}, these equations can be 
used if one identifies $\vec{V}_{ q\alpha}$ with $\vec{u} + \vec{V}_{ i\alpha}$.}
%
\begin{align}
	&\sum_{\vp\sigma \alpha}  \frac{\partial \mathfrak{E}_{\vp + \vQa}^{(\alpha)}}{\partial \vp} \sigma_{\vp + \vQa}^{(\alpha)} 
	\approx
	\sum_{\vp \sigma \alpha}  \left[  \frac{\partial \Delta H_{\vp}^{(\alpha)}}{\partial \vp} \sigma_{\vp,0}^{(\alpha)}  +
	 {  {E}_{\vp}^{(\alpha)} \over T} \frac{\partial {E}_{\vp}^{(\alpha)}}{\partial \vp} \frac{\partial \fp}{ \partial {E}_{\vp}^{(\alpha)}}\left(  \Delta H_{\vp}^{(\alpha)} - \vp \vec{V}_{ q\alpha} \right)
	 \right] ,
	 \label{eq:S_flux_init}
\end{align}
where 
$\sigma_{\vp,0}^{(\alpha)}$ is the entropy density calculated for the distribution function 
\eqref{eq:fp_eq_cl}.
{
Substituting Eq.\ \eqref{eq:dH_pn} into \eqref{eq:S_flux_init} and 
}
accounting for Eqs.\ \eqref{eq:sigma_0_rel} 
one sees that
the first two terms in the expression \eqref{eq:S_flux_init} cancel out.
Thus, the entropy 
current
density 
becomes
\begin{equation}
		\sum_{{\vec p} \sigma \alpha}  \frac{\partial \mathfrak{E}_{\vp + \vQa }^{(\alpha)}}{\partial \vp} \sigma_{\vp + \vQa }^{(\alpha)} =
		\sum_{\alpha} S_{\alpha} \vec{V}_{ q\alpha},
\end{equation}
where $S_{\alpha}$
is the partial entropy {density} given by the expression \eqref{eq:part_entropy}.
Consequently, the entropy {generation} equation \eqref{eq:S_eq} takes the form 
\begin{equation}
		\frac{\partial S}{\partial t} + \Div \left( S_{ n} \vec{V}_{ qn} + S_{ p} \vec{V}_{ qp} 
		\right) = \Gamma_s.
\end{equation}
One sees that the vectors $\vec{V}_{ q\alpha}=\vec{u} + \vec{V}_{ i\alpha}$ have the 
meaning of the corresponding 
{\it partial entropy velocities}.
If the dissipative terms can be ignored, 
this equation reduces to the standard nondissipative form \cite{LL6}
\begin{equation}
		\label{eq:dS_ideal}
		\frac{\partial S}{\partial t} + \Div \left( S \vec{u} \right) = 0.
\end{equation}

{
For a deeper understanding of the nature of the velocities $\vec{V}_{ q\alpha}$
let us sum Eq. \eqref{eq:kin_eq} over the quantum states. 
As a result, we obtain the
``continuity equation'' for the Bogoliubov excitations that has the form 
\begin{equation}
	\label{cont2}
	\frac{\partial}{\partial t} \sum_{\vp \sigma} {\mF}_{\vp + \vQa}^{(\alpha)} 
	+ \Div  \left( \sum_{\vp \sigma}  \frac{\partial \mathfrak{E}_{\vp + \vQa}^{(\alpha)}}{\partial \vp} {\mF}_{\vp + \vQa}^{(\alpha)} \right)
	=  \sum_{\vp \sigma} I_\alpha[\mF].
\end{equation}
Note that the number of Bogoliubov excitations is not necessarily conserved 
(see Appendix \ref{sec:Integral} for details). 
Hence, the right-hand side of Eq.\ \eqref{cont2} is generally nonzero.
Linearizing the current density of Bogoliubov excitations, one gets 
\begin{equation}
	\sum_{\vp \sigma}  \frac{\partial \mathfrak{E}_{\vp + \vQa}^{(\alpha)}}{\partial \vp} {\mF}_{\vp + \vQa}^{(\alpha)} \approx
	\sum_{\vp \sigma} \left[ \frac{\partial \Delta H_{\vp}^{(\alpha)}}{\partial \vp} \fp  +
	  \frac{\partial {E}_{\vp}^{(\alpha)}}{\partial \vp} \frac{\partial \fp}{ \partial {E}_{\vp}^{(\alpha)}}\left(   \Delta H_{\vp}^{(\alpha)} - \vp \vec{V}_{ q\alpha} \right) \right].
\end{equation}
It is easy to see that the first two terms under the sum here can be combined into the total 
derivative over $\vp$. 
Hence, these terms  disappear  after the summation and one finally obtains 
\begin{equation}
	\sum_{\vp \sigma}  \frac{\partial \mathfrak{E}_{\vp + \vQa}^{(\alpha)}}{\partial \vp} {\mF}_{\vp + \vQa}^{(\alpha)} \approx
	\sum_{\vp \sigma} \fp \ \vec{V}_{ q\alpha}.
\end{equation}
Thus, one can say that the partial entropy moves together with the Bogoliubov excitations.
This result 
might be expected, since 
the distribution function ${\mF}_{\vp + \vQa}^{(\alpha)}$ of 
these 
excitations 
determines the partial entropy in the system 
[see Eqs.\ \eqref{eq:entropy} and \eqref{eq:part_entropy}].  
However, the nontrivial result is that in a strongly interacting mixture 
the partial entropy current is generally {\it not collinear} with the normal current of the 
corresponding particle species [see the first two terms in Eq.\ \eqref{eq:j_nc}].
This property is in sharp contrast with the case of a nonsuperfluid mixture, where these currents 
are 
always collinear, so that  
}	
\begin{equation}
	{\vec j}_{ \alpha} = n_\alpha  {\vec V}_{{ q\alpha}}.
\end{equation}
{To understand why the normal current densities in a superfluid mixture generally depend on both 
the 
velocities
${\vec V}_{{ qn}}$ and ${\vec V}_{{ qp}}$,
consider, for simplicity, 
a point in space where   
$\vec{Q}_{ n} = \vec{Q}_{ p} = 0$.}
Then, substituting Eqs.\ \eqref{eq:expand1} and \eqref{eq:N_exp_nc} into 
\eqref{eq:J} and linearizing the obtained expression with respect to $\vec{V}_{ q\alpha}$,
one gets
 \begin{align}
 	\nonumber
 	 \vec{j}_\alpha = \sum_{\vp \sigma} \left( {\frac{\partial \varepsilon_\vp^{(\alpha)}}{\partial 
 	 \vp}}  \frac{\partial \fp}{ \partial {E}_{\vp}}\Delta H_{\vp}^{(\alpha)} +  \frac{\partial 
 	 \Delta H_{\vp}^{(\alpha)}}{\partial \vp} \np  \right) -
	 \sum_{\vp \sigma} \frac{\partial \varepsilon_{\vp}^{(\alpha)}}{\partial \vp}  \frac{\partial \fp}{ \partial {E}_{\vp}} \vec{p}\vec{V}_{ q\alpha} = \\
	= \sum_{\vp \sigma} \left(  {\frac{\partial \varepsilon_\vp^{(\alpha)}}{\partial \vp}}  \frac{\partial \fp}{ \partial {E}_\vp}- \frac{\partial \np}{\partial \vp}\right) \Delta H_\vp^{(\alpha)} - 
	  \sum_{\vp \sigma} \frac{\partial \varepsilon_\vp^{(\alpha)}}{\partial \vp}  \frac{\partial 
	  \fp}{ \partial {E}_\vp} \vec{p}\vec{V}_{ q\alpha},
\end{align}
{where the second equality is obtained after integration by parts.}
In the nonsuperfluid mixture, the first sum in this equation vanishes 
and one arrives at the usual expression for the particle current density
(see, e.g., Ref.\ \cite{PinesNozieres}).
In fact, it is not necessary to know the energy correction $\Delta H_{\vp}^{(\alpha)}$ to calculate 
$\vec{j}_\alpha$
in 
this case.
{In contrast, in the superfluid mixture, this sum becomes comparable to the second one and
hence additional terms from $\Delta H_{{\vec p}}^{(\alpha)}$, containing both the vectors ${\vec 
V}_{{ qn}}$ and ${\vec V}_{{qp}}$,  come into play  [see Eq.\ \eqref{eq:dH_pn}].}

\section{Charged mixtures}
\label{sec:electric_field}

In this section we generalize the results of the previous section to charged mixtures 
(e.g., neutrons, protons, and electrons).
For simplicity, we assume that the magnetic field in the system is absent.
If the  particle species $\alpha$ possesses an electrical charge $e_\alpha$,
the ``superfluid'' equation  \eqref{eq:sf_eq}
should be replaced with \cite{Gusakov2010}
\begin{equation}
		\label{eq:sf_eq_charged}
		\frac{\partial {\vec{Q} _{\alpha}}}{\partial t} = - \nabla  \breve{\mu}_\alpha+   e_\alpha   \vec E, 
\end{equation}
where 
$\vec E$ is the electric field.
As before, the difference between $\breve{\mu}_\alpha$ and ${\mu}_\alpha$ will be ignored in the 
subsequent analysis.
For the sake of convenience, let us introduce the vectors $\vec{b}_\alpha =  \nabla  
{\mu}_\alpha-   e_\alpha   \vec E$, 
so that Eq.\ \eqref{eq:sf_eq_charged} 
becomes
\begin{equation}
		\label{eq:sf_eq_charged_2}
		\frac{\partial {\vec Q_{\alpha}}}{\partial t} = - \vec{b}_\alpha.
\end{equation}	

For our particular system $e_n = 0$, $e_p =q$, where $q$ is the elementary charge. 
To make the electrical quasineutrality possible, we should add a third constituent labeled ``$e$''
(e.g., electrons) with $e_{ e} = -q$. This constituent is assumed to be composed of 
noninteracting fermions (an ideal Fermi gas).
The kinetic equation for the ``$e$''  constituent takes the form \cite{Landau_Kin}
\begin{equation}
		\label{eq:kin_eq_e}
		\frac{\partial \mN_{\vp}^{ (e)}}{\partial t} + {\partial \varepsilon^{ (e)}_{\vp} \over \partial \vp}\frac{\partial \mN_{\vp}^{ (e)}}{\partial \vr} - e_{ e} \vec{E}  \frac{\partial \mN_\vp^{ (e)}}{\partial \vp} = I^{ (e)} [\mF,\mN],
\end{equation}	
where by $\mF$ and $\mN$ we denote, respectively, the set of distribution functions 
$\mF_{\vp+\vQa}^{(\alpha)}$ ($\alpha =  n, \ p$) and $\mN_{\vp}^{ (e)}$.
The partial entropy associated with the particle species ``$e$'' is given by 
\begin{equation}
		\label{eq:entropy_e}
		S_{ e} =   \sum_{{\vec p} \sigma} \sigma_{\vp}^{(e)} ,
\end{equation}	
where 
\begin{gather}
		\label{eq:entropy_dens_e}
		 \sigma_{\vp}^{(e)}  = 
  		  -\left( 1- \mN_{\vp}^{(e)} \right) \ln \left( 1- \mN_{\vp}^{(e)}\right)
  		  -  \mN_{{\vec p}}^{(e)}  \ln  \mN_{\vp}^{(e)}.
\end{gather}	
The approximate solution of Eq.\ (\ref{eq:kin_eq_e}) is sought in the form 
\begin{equation}
		\label{eq:mNe_comp}
		\mN_\vp^{ (e)} = \mN_{\vp, 0}^{ (e)}+ {\mathfrak{n}}_1^{ (e)} = \mN_{\vp, 0}^{(e)}- \frac{\partial \mN_{\vp, 0}^{ (e)}}{ \partial {\varepsilon}_\vp^{(e)}}\phi_{ e},
\end{equation}	
where the equilibrium distribution function is given by the usual Fermi-Dirac distribution,
\begin{equation}
		\mN_{\vec p, 0}^{(e)} = { 1\over 1+ e^{\left(\varepsilon^{(e)}_{\vec{p}} - \mu_{ e}- \vp\,\vec{u}\right)/T}},
\end{equation}	
Note that, in contrast to superfluids ``$n$'' and ``$p$'', 
the departure from the equilibrium does 
not modify the energy ${\varepsilon_\vp^{ (e)}}$,  
since it is independent of the distribution 
functions.

Repeating the derivation of Eq.\ \eqref{eq:mom_eq_a_lin} from Sec. \ref{sec:diffusion}, 
one obtains
for $\alpha=n, p$
\begin{equation}
		\label{eq:mom_eq_a_lin_charge}
		 \sum_{\alpha' =  n,p} R_{\alpha'\alpha} {\mu_{\alpha'} \over c^2}  {\partial \vec{u} 
		 \over \partial t}     +   \sum_{\alpha' =  n,p} R_{\alpha'\alpha} \vec{b}_{\alpha'} + 
		 S_\alpha  \nabla T = \sum_{\vec{p} \sigma} \vp  I_\alpha [  \mF, \mN].
\end{equation}	
In turn, similar derivation for particles ``$e$'' yields
\begin{equation}
		\label{eq:mom_eq_e_lin_charge}
		 n_{ e}{\mu_{ e} \over c^2}  {\partial \vec{u} \over \partial t}     + n_{ e}  
		 \vec{b}_{ e} + S_{ e}  \nabla T = \sum_{\vec{p} \sigma} \vp  I_{ e}  [  \mF, \mN].
\end{equation}		

Our purpose is to find the connection between the dissipative currents and the vectors 
$\vec{b}_{\alpha}$.
To this aim, one can again assume the ansatz \eqref{eq:phi_def} 
for the superfluid species ``$n$'' and ``$p$'', 
and the similar ansatz
	\begin{equation}
		\label{eq:phi_e}
		\phi_{ e} =  \vp \, \vec V_{ ie} 
	\end{equation}
for the species ``$e$''.	
Here, as before, the dependence of the vectors $\vec V_{ i\alpha}$ on 
the momentum variable $\vp$ is neglected.
Hence, following the derivation of Sec. \ref{currents}, 
one arrives at
the following expressions for 
the diffusion current densities [cf.\ Eq.\ \eqref{eq:i_def}]
\begin{align}
		\label{eq:I_n_charged}
		&\Delta \vec{j}_{ n} = R_{ nn} \vec{V}_{ in} +R_{ np} \vec{V}_{ ip}, \\
		\label{eq:I_p_charged}
		&\Delta \vec{j}_{ p} = R_{ pn} \vec{V}_{ in} +R_{ pp} \vec{V}_{ ip}, \\
		\label{eq:I_e_charged}
		&\Delta \vec{j}_{ e} = n_{ e} \vec{V}_{ ie}.
	\end{align}	
Plugging now expansions \eqref{eq:mF_comp_tmp} and \eqref{eq:mNe_comp} 
into the right-hand sides of Eqs.\ 
\eqref{eq:mom_eq_a_lin_charge} and \eqref{eq:mom_eq_e_lin_charge}, 
in which the temperature gradient is set to zero, one finds
\begin{align}
		\label{eq:Euler_n_charged}
			 &  \left(R_{ nn} {\mu_{ n} \over c^2} + R_{ pn} {\mu_{ p} \over c^2}  \right) {\partial \vec{u} \over \partial t}   +   R_{ nn} \vec{b}_{ n} + R_{ pn}\vec{b}_{ p}
		  	=  J_{ pn} \vec w_{ pn}  +  J_{en} \vec w_{ en},
		  	\\
		\label{eq:Euler_p_charged}
			 &  \left(R_{ pp} {\mu_{ p} \over c^2} + R_{ np} {\mu_{ n} \over c^2}  \right) {\partial \vec{u} \over \partial t}   +   R_{ pp}\vec{b}_{ p} + R_{ np}\vec{b}_{ n}
		  	=  -J_{ pn} \vec w_{ pn}  -  J_{ pe} \vec w_{ pe},	
		  	\\
		  	\label{eq:Euler_e_charged}
		  &	n_{ e} {\mu_{ e} \over c^2}  {\partial \vec{u} \over \partial t}   +  n_e 
		  	\vec{b}_{ e}
		  	=   - J_{ en} \vec w_{ en}   + J_{ pe} \vec w_{ pe},&
\end{align}
where we introduced a notation
\begin{equation}
		\label{eq:Vab}
		\vec w_{\alpha\beta} = \vec V_{ i\alpha} - \vec V_{ i\beta}
\end{equation}
and where $J_{\alpha\beta}$ are the momentum transfer rates 
(see Appendix \ref{sec:Integral}).	
To exclude the time derivative of the velocity $\vec{u}$ from these equations, one can sum them up, 
which results in
\begin{equation}
		\label{eq:mom_cons_charged}
		\rho_{ q} {\partial \vec u \over \partial t} =  -  n_{ qn}  \vec{b}_{ n}  -  n_{ qp}   \vec{b}_{ p}    -  n_{ e}\vec{b}_{ e},
\end{equation} 	
where the normal energy density $\rho_{ q}$ 
{now equals}
\begin{equation}
		{\rho}_{ q} = {\mu_{ n} \over c^2} n_{ qn} + {\mu_{ p} \over c^2} n_{ qp} + {\mu_{ e} \over c^2} n_{ e}.
\end{equation}
After  $\partial \vec{u}/\partial t$ is excluded, 
only two out of three of Eqs.\ \eqref{eq:Euler_n_charged}--\eqref{eq:Euler_e_charged} remain 
independent.
Taking, for instance, Eqs.\ \eqref{eq:Euler_p_charged} and \eqref{eq:Euler_e_charged}, 
using Eq.\ \eqref{eq:mom_cons_charged}, 
and
{
excluding $\vec{w}_{ pe} $ with the identity
}
 $\vec{w}_{ pe} \equiv \vec{w}_{ pn} - \vec{w}_{ en}$, one obtains
\begin{align} 
		  -  & \frac{\mu_{ p} \mu_{ n} \mathbb R + \mu_{ e} \mu_{ p} n_{ e} R_{ pp}}{ c^2{\rho}_{ q}} \left( {\vec{b}_{ n} \over \mu_{ n}} - {\vec{b}_{ p} \over \mu_{ p}}\right) 
		 + \frac{\mu_{ e} \mu_{ n} n_{ e} R_{ np} +  \mu_{ e} \mu_{ p} n_{ e} R_{ pp}}{c^2{\rho}_{ q}} \left( {\vec{b}_{ n} \over \mu_{ n}} - {\vec{b}_{ e} \over \mu_{ e}}\right)  
		 =  - (J_{ pn} + J_{ pe}) \vec w_{ pn}   + J_{ pe} \vec w_{ en}
		 \label{eq:froce_bal_p_charged_2}
		 \\
		& \frac{\mu_{ e} \mu_{ p} n_{ e} n_{ qp}}{c^2{\rho}_{ q}}   \left( {\vec{b}_{ n} \over \mu_{ n}} - {\vec{b}_{ p} \over \mu_{ p}}\right)
		 - \mu_{ e} n_{ e} \frac{\mu_{ n} n_{ qn} + \mu_{ p} n_{ qp}}{c^2 {\rho}_{ q}}   \left( {\vec{b}_{ n} \over \mu_{ n}} - {\vec{b}_{ e} \over \mu_{ e}}\right)
		 =   J_{ pe} \vec w_{ pn} - (J_{ en}+J_{ pe}) \vec w_{ en}. 		
	 	 \label{eq:froce_bal_e_charged_2} 
\end{align}	
Equations \eqref{eq:froce_bal_p_charged_2}--\eqref{eq:froce_bal_e_charged_2} allow us to express 
the vectors $\vec w_{\alpha\beta}$  through  the vectors $\vec{d}_\alpha$.
 In turn, Eqs.\ \eqref{eq:I_n_charged}--\eqref{eq:I_e_charged} together with the 
condition \eqref{eq:LL_cond} relate the vectors $\Delta \vec{j}_\alpha$ and $\vec{w}_{ \alpha 
\beta}$:
\begin{align}
		& \Delta \vec{j}_{ n} =- \frac{(\mu_{ p} \mathbb{R} - \mu_{ e} n_{ e} R_{ np} )\vec{w}_{ pn} + \mu_e n_{ e} n_{ qn} \vec{w}_{ en}}{c^2 {\rho}_{ q}}, \\
		\label{eq:i_p_charged}
		&\Delta \vec{j}_{ p} = \frac{(\mu_{ n} \mathbb{R} + \mu_{ e} n_{ e} R_{ pp} )\vec{w}_{ pn} - \mu_e n_{ e} n_{ qp} \vec{w}_{ en}}{c^2 {\rho}_{ q}},\\ 
		\label{eq:i_e_charged}
		&\Delta \vec{j}_{ e} = n_{ e} \frac{(c^2{\rho}_{ q} - \mu_{ e} n_{ e}) \vec{w}_{ en} - (\mu_p R_{ pp} + \mu_{ n} R_{ np})\vec{w}_{ pn}}{c^2 {\rho}_{ q}}.
\end{align}	

In the case of isothermal matter, the 	
phenomenological expression for the diffusion currents is [cf.\ Eq.\ \eqref{eq:ap:dj_exp}]
\begin{equation}
		\Delta \vec{j}_{\alpha} = - \sum_{\alpha'}\mathcal{D}_{\alpha\alpha'} {\vec{b}_{\alpha'} \over T }.
	\end{equation}
The relation \eqref{eq:LL_cond} allows us to exclude the diagonal diffusion coefficients 
$\mathcal{D}_{\alpha\alpha}$ and, after accounting for Eq.\ \eqref{eq:Onsager_rel}, one arrives at
\begin{equation}
		\label{eq:dj_exp_charged}
		\Delta \vec{j}_{\alpha} = \sum_{\beta\neq\alpha} \mathcal{D}_{\alpha\beta}\left({ 
		\vec{b}_\beta \over \mu_\beta}  -  { \vec{b}_\alpha \over \mu_\alpha}\right).
\end{equation}
Now, plugging the quantities $\vec{w}_{\alpha\alpha'}$ obtained as a solution to Eqs. 
\eqref{eq:froce_bal_p_charged_2}--\eqref{eq:froce_bal_e_charged_2} into Eqs.\
\eqref{eq:i_p_charged}--\eqref{eq:i_e_charged} 
and comparing the result with the phenomenological expression \eqref{eq:dj_exp_charged},  one can 
find the formulas for the  diffusion coefficients
$\mathcal{D}_{\alpha \alpha'}$.
We prefer not to present 
these
lengthy expressions 
here, but they 
can be easily found if necessary.
Clearly, the  procedure described above can be easily extended to a mixture of arbitrary number 
of species.
	
What remains to be done is to find an expression for the entropy production rate. 
Repeating the derivation from Sec. \ref{sec:entropy}, one gets
\begin{equation}
		\label{eq:gamma_s_lin_charged}
		\Gamma_s 
		=   -\frac{1}{T} \sum_{\vp \sigma \alpha} \vp \vec{V}_{ i \alpha} I_\alpha [  \mF, \mN] 
		=   -\frac{1}{T} \sum_{\vp \sigma}\left( \vec{w}_{ pn}\vp I_{ p} [  \mF, \mN]  +  \vec{w}_{ en} \vp I_{ e} [  \mF, \mN] \right) .
\end{equation}
Here,  the first equality
coincides with Eq.\ \eqref{eq:Gamma_s_lin} except that the summation is now performed over the 
three particle species, $\alpha = n,\,  p, \, e$.
In the second equality, we used the fact that 
$\sum_{\vp \sigma \alpha} \vp  I_\alpha [  \mF, \mN] = 0$.	
Plugging in the  left-hand side of Eqs.\ \eqref{eq:froce_bal_p_charged_2} and 
\eqref{eq:froce_bal_e_charged_2} instead of $\sum_{\vp \sigma } \vp  I_{\alpha} [  \mF, \mN] $ and 
using expressions \eqref{eq:i_p_charged} and \eqref{eq:i_e_charged}, we arrive at
\begin{equation}
 		\label{eq:Gamma_s_charged}
 		\Gamma_s = - {1\over T}\left[ \mu_{ p}  \Delta \vec{j}_{ p} \left( {\vec{b}_{ p} \over \mu_{ p}} - {\vec{b}_{ n} \over \mu_{ n}}\right) + 
	        \mu_{ e} \Delta \vec{j}_{ e}  \left( {\vec{b}_{ e} \over \mu_{ e}} - {\vec{b}_{ n} \over \mu_{ n}}\right) \right] .
\end{equation}	
To bring this expression into the form of the phenomenological Eq.\ \eqref{eq:ap:Gamma_s_nr}, one 
needs just to apply the relation \eqref{eq:LL_cond}.
To verify that the entropy production rate \eqref{eq:Gamma_s_charged}
is non-negative, 
one should substitute the right-hand sides of Eqs.\ 
\eqref{eq:froce_bal_p_charged_2} and \eqref{eq:froce_bal_e_charged_2} 
instead of $\sum_{\vp \sigma } \vp  I_{\alpha} [  \mF, \mN] $ into Eq.\ 
\eqref{eq:gamma_s_lin_charged}. {The resulting quadratic form}
\begin{equation}
		\Gamma_s =  {1\over T} \left( J_{ pn} \vec{w}_{ pn}^2 + J_{ en} \vec{w}_{ 
		en}^2  + J_{ pe} \vec{w}_{ pe}^2 \right)
\end{equation}
is obviously {positive-definite.}

\section{Summary}
\label{sec:conclusions}

In this paper we have developed a general formalism for studying particle diffusion in superfluid 
mixtures of strongly interacting Fermi liquids.	
Our results can be summarized as follows.

\begin{enumerate}
	
\item
The diffusion in superfluid mixtures is manifested through the modification of the 
expressions for the normal currents of all particle species $\alpha=n, p$ in the mixture.
These 
normal currents
can 
be introduced into the theory
by minimizing the thermodynamic potential \eqref{eq:F_potential_np}.
They can be expressed 
in terms of the velocities $\vec{V}_{ q\alpha}$, which are conjugate variables to 
the momenta of 
species $\alpha$. 
In the approximation of small velocities,
 the normal component of  each current density ${\vec j}_{\alpha}$
is a linear combination of the velocities $\vec{V}_{ qn}$ and $\vec{V}_{ qp}$.
The coefficients in this linear combinations constitute the normal entrainment matrix, $R_{\alpha 
\alpha'}$, introduced in this work for the first time.
		 
\item
The velocities $\vec{V}_{ q\alpha}$ can be interpreted as partial entropy velocities for 
particle species $\alpha$ {or, equivalently, as the velocities of the flow of Bogoliubov 
excitations $\alpha$}.
These velocities 
become independent and self-contained variables in superfluid mixtures, 
being generally not 
{ collinear}
to the corresponding normal current densities, 
as well as to the momentum densities, associated with the motion of thermal excitations of 
different particle species.
Here we find some resemblance to the Carter's formalism (see, e.g., Ref.\ 
\cite{AnderssonComer2021}),
in which the entropy current is treated on an equal footing with other conserved currents of a 
multifluid mixture.  
		
\item
To study diffusion effects in superfluid mixtures, 
we applied the Boltzmann-like 
kinetic
equation for Bogoliubov thermal excitations,
supplementing it with  
the continuity and superfluid equations for each particle species.
Using the Chapman-Enskog method and assuming the standard ansatz \eqref{eq:phi_def} for 
the nonequilibrium correction to the distribution function, we again arrived at the 
velocities $\vec{V}_{ q\alpha}$.
That is,
these velocities 
are natural 
variables for 
describing diffusion 
processes
in superfluid Fermi mixtures. 
In particular, the friction force between different particle species ``$n$'' and ``$p$'' 
appears to be directly proportional to $\vec{V}_{{ qn}}-\vec{V}_{{ qp}}$ 
[see, e.g., Eq.\ \eqref{eq:force}]. Somewhat loosely, one may say that the ``friction of entropy 
currents'' produces 
heat 
(the entropy velocity $\vec{V}_{{ q}\alpha}$ is generally not equal
to the normal fluid velocity of particle species $\alpha$, so that this result is  
nontrivial).

\item 
Using transport equations for Bogoliubov excitations, 
we obtained general expressions for the diffusion coefficients of a mixture of two
strongly-interacting Fermi superfluids [see Eqs.\ \eqref{DnnDpp} and \eqref{eq:Dnp}]. 
The diffusion coefficients depend on the matrix $R_{\alpha\alpha'}$ (which is responsible for the Fermi-liquid/entrainment effects in the mixture) and on the 
momentum transfer rates $J_{ \alpha\beta}$.
The general {expression} for $J_{ \alpha\beta}$ is presented in Appendix \ref{sec:Integral}
and formally has the same structure as 
$J_{ \alpha\beta}$ for
a mixture
of two weakly interacting superfluid Fermi gases.
		
\item
There is a quantitative difference between the procedure of calculation of diffusion coefficients  
for the nonsuperfluid and superfluid Fermi mixtures. 
In the first case, the inclusion of the Fermi-liquid effects is quite formal. 
Indeed, after 
expanding
the quasiparticle distribution function 
in powers of the Knudsen number [cf. Eq. \eqref{eq:mF_comp_tmp}],
\begin{equation}
			{\mN}_\vp^{(\alpha)} = \overline{\mN}_{\vp, 0}^{(\alpha)}+ \bar{\mathfrak{n}}_1^{ (\alpha)},
\end{equation}	
where [cf. Eq. \eqref{eq:ol_mF}]
\begin{equation}
		 	\overline{\mN}_{\vp, 0}^{(\alpha)} = 
			\frac{1}{ 1 + { e}^{(\varepsilon_{\vp}^{(\alpha)}  + \Delta H_{\vp}^{(\alpha)}-\vec{p}\, \vec{u})/T}},
\end{equation}
and substituting this representation into transport equations, 
the resulting (linearized) equations for the distribution function corrections  
$\bar{\mathfrak{n}}_1^{ (\alpha)}$ will be formally identical to the corresponding equations for a 
mixture of weakly interacting Fermi gases (see, e.g., Ref.\ \cite{SchmittShternin2018}). 
The expression for the current densities
is also identical to its counterpart in weakly interacting mixtures 
(see the discussion at the end of Sec.\ \ref{sec:entropy}).
As a consequence, there is no need for calculation of the energy corrections $ \Delta 
H_{\vp}^{(\alpha)}$, and 
the Landau parameters do not appear in the expression \eqref{eq:Dnp_nsf} for the diffusion 
coefficients.
In contrast, in the superfluid mixture one needs to determine the Bogoliubov excitation energy 
correction and the Landau parameters explicitly appear in the diffusion coefficients  [see  Eq.\ 
\eqref{eq:Dnp}]. 
The formal reason for this difference is discussed at the end of Sec.\ \eqref{sec:entropy}.
		
\item
The results discussed above were generalized to the case of charged mixtures in Sec.\
\ref{sec:electric_field}.
The extension of these results to arbitrary number of particle species in the mixture is
straightforward.
\end{enumerate}	

Summarizing, the framework for treating particle diffusion 
developed in the present work
opens the way 
for systematic calculations of diffusion coefficients in superfluid, strongly interacting Fermi 
mixtures, in particular, in superfluid neutron-star matter.

\section*{ACKNOWLEDGMENTS}
MEG is grateful to the Department of Particle Physics \& Astrophysics at the Weizmann Institute of Science for hospitality.
This research was supported by the Russian Science Foundation Grant \textnumero  22-12-00048.


\appendix

\section{Elements of matrices $\gamma_{\alpha\beta}$, $K_{\alpha\beta}$, $Y_{\alpha\beta}$, and
$R_{\alpha\beta}$ in various limiting cases}
\label{ap:coeffs}

\subsection{The case of one superfluid and one normal Fermi liquid ($\Phi_{ n}=1$)}

Assume that the species ``$p$'' is superfluid, while the species ``$n$'' is normal, 
i.e., the function $\Phi_{ n}=1$ [see Eq.\ \eqref{eq:Phi}].
In this limit, the elements of the matrix $\gamma_{\alpha\alpha'}$ 
reduce to
\begin{align}
		\gamma_{ nn} &=  {1 \over m_{ n}^*},
		\\
		\gamma_{ np} & = \frac{G_{ np}n_{ p}(1-\Phi_{ p})}{\mathcal S},
		\\
		\gamma_{ pp} &= {1 \over m_{ p}^*} \frac{\mathcal{S}_{\rm nsf}}{\mathcal S},
		\\
		\gamma_{ pn} &= 0.
\end{align}
In turn, the elements of the matrix $K_{\alpha\alpha'}$ become
\begin{align}
		K_{ nn} &= \frac{ G_{ nn} m^*_{ n} (n_{  p} + G_{ pp} m^*_{ p} \Phi_{ p}) - G_{ np}^2 m^*_{ n} m^*_{ p} \Phi_{ p} }{\mathcal S},
		\\
		K_{ np} &= \frac{G_{ np} m^*_{ p} n_{ p}\Phi_{p}}{\mathcal S},	
		\\
		K_{ pp} &= \frac{G_{ pp} m^*_{ p} \Phi_{ p} (n_{ n} + G_{ nn} m^*_{ n}) - G_{ np}^2 m^*_{ p} m^*_{ n} \Phi_{ p} }{\mathcal S},
		\\
		K_{ pn} &=  \frac{G_{ pn} m^*_{ n} n_{ n}}{\mathcal S}.
\end{align}

The superfluid entrainment matrix is given by
\begin{align}
		Y_{ nn} &= Y_{ np} = Y_{ pn} = 0,
		\\
		Y_{ pp} &=   {n_{ p} \over m_{ p}^* c^2} \frac{\mathcal{S}_{\rm nsf}}{\mathcal 
		S}(1-\Phi_{ p}),
\end{align}	
while the normal entrainment matrix equals
\begin{align}
	\label{R11}
		R_{ nn} &=  n_{ n},  \\
		R_{ np}  &= 0, \\
		R_{ pp} &= n_{ p} {\mathcal{S}_{\rm nsf} \over \mathcal {S}}\Phi_{ p}, \\  
		R_{ pn} &=  \frac{n_{ n} n_{ p} m_{ n}^* (1-\Phi_{ p}) G_{ np} }{\mathcal 
		S}.
		\label{R22}
\end{align}	 
In these expressions the function ${\mathcal S}$ 
is given by Eq.\ \eqref{eq:S_det}, which can be represented in the considered limiting case as
\begin{equation}
		\mathcal S = (n_{ n} + G_{ nn}  m^*_{ n}  )(n_{ p}  + G_{ pp} m^*_{ p} \Phi_{ p} ) - G_{ np} ^2 m^*_{ n}   m^*_{ p}  \Phi_{ p} .
\end{equation}
Similarly, the function $\mathcal{S}_{\rm nsf} $  coincides with ${\mathcal S}$ calculated for a 
completely nonsuperfluid mixture (i.e., assuming $\Phi_{ n}= \Phi_{ p} =1$)
\begin{align}
		\label{Snsf}
		\mathcal S_{\rm nsf} &= (n_{ n} + G_{ nn}  m^*_{ n}  )(n_{ p}  + G_{ pp} m^*_{ p} ) - G_{ np} ^2 m^*_{ n}   m^*_{ p} 
		   =  \frac{m_{ n}^*m_{ p}^* c^2}{\mu_{ n} \mu_{ p}} \left[  n_{ n} n_{ p} - G_{ np} \frac{\mu_{ n} n_{ n} + \mu_{ p} n_{ p}}{c^2} \right].
\end{align}	
Here, the second equality is obtained with the help of Eq.\ \eqref{eq:m_eff_eq}.
As follows from Eqs.\ \eqref{R11}--\eqref{R22}, in the case when both species 
are nonsuperfluid ($\Phi_{ n}=\Phi_{ p}=1$),
one has $R_{ nn}=n_{ n}$, $R_{ pp}=n_{ p}$, and $R_{ np}=R_{ pn}=0$.

\subsection{The case of strong superfluidity of one of the constituents 
($\Phi_{ p} \rightarrow 0$)}

Assume now that particle species ``$p$'' is strongly superfluid, $\Phi_{ p} \rightarrow 0$.
Then the matrix $\gamma_{\alpha\alpha'}$ simplifies to
\begin{align}
		\gamma_{ nn} &= \frac{n_{ n} + G_{ nn} m_{ n}^*}{m_{ n}^* (n_{ n} + G_{ nn} m_{ n}^*\Phi_{ n})}, 
		\\
		\gamma_{ np} &= \frac{G_{ np}}{n_{ n} + G_{ nn} m_{ n}^*\Phi_{ n}},
		\\
		\gamma_{ pp} &= \frac{(n_{ p} + G_{ pp} m_{ p}^*)(n_{ n} + G_{ nn} m_{ n}^*\Phi_{ n})- G_{ np}^2 m_{ n}^*m_{ p}^*\Phi_{ n}  }{m_{ p}^*n_{ p} (n_{ n} + G_{ nn} m_{ n}^*\Phi_{ n})},
		\\
		\gamma_{ pn} &= \frac{n_{ n} G_{ np} (1- \Phi_{ n} )}{n_{ p}( n_{ n} + 
		G_{ nn} m_{ n}^*\Phi_{ n})},
\end{align}	
while the matrix $K_{\alpha\alpha'}$ reduces to
\begin{align}
		K_{ nn} & = \frac{ G_{ nn} m^*_{ n} \Phi_{ n}}{( n_{ n} + G_{ nn} m_{ n}^*\Phi_{ n})},
		\\ 
		K_{ np} &= 0,
		\\
		K_{ pp} &= 0,
		\\ 
		K_{ pn} &= \frac{G_{ pn} m^*_{ n} n_{ n}\Phi_{ n}}{n_{ p}( n_{ n} + G_{ nn} m_{ n}^*\Phi_{ n})}.
\end{align}

The superfluid entrainment matrix becomes
\begin{align}
		Y_{ nn} &=  \frac{n_{ n} (n_{ n} + G_{ nn} m_{ n}^*)}{c^2 m_{ n}^* (n_{ n} + G_{ nn} m_{ n}^*\Phi_{ n})} (1-\Phi_{ n}),
		\\
		Y_{ pp} &=   \frac{(n_{ p} + G_{ pp} m_{ p}^*)(n_{ n} + G_{ nn} m_{ n}^*\Phi_{ n})- G_{ np}^2 m_{ n}^*m_{ p}^*\Phi_{ n}  }{c^2m_{ p}^* (n_{ n} + G_{ nn} m_{ n}^*\Phi_{ n})}, 
		\\
		Y_{ pn} & = Y_{ np} = \frac{n_{ n} G_{ np} (1- \Phi_{ n} )}{c^2( n_{ n} + G_{ nn} m_{ n}^*\Phi_{ n})},
\end{align}
while the normal entrainment matrix is
\begin{align}
		R_{ nn} &= \frac{n_{ n}(n_{ n} + G_{ nn} m_{ n}^*)\Phi_{ n}}{ (n_{ n} + G_{ nn} m_{ n}^*\Phi_{ n})},
		\\
		R_{ np} &= 0,
		\\
		R_{ pp} &= 0,
		\\
		\label{eq:Rnp_Tcp0}
		R_{ pn} &= 	\frac{n_{ n} m_{ n}^* G_{ np} \Phi_{ n}}{n_{ n} + G_{ nn} m_{ n}^*\Phi_{ n}}.
\end{align}
In the limit when both particle species are strongly superfluid, 
$\Phi_{ n}\rightarrow 0$ and $\Phi_{ p} \rightarrow 0$, one has from these expressions 
$R_{ nn}=R_{ pp}=R_{ np}=R_{ pn}=0$.

\subsection{The case of two independent Fermi liquids ($G_{ np} = G_{ pn} = 0$)}

Assume now that our mixture is composed of two superfluid Fermi liquids, which do not ``feel''
each other in a sense that the Landau parameters $f_1^{ np}=f_{1}^{ pn}=0$, 
i.e., $G_{ np} = G_{ pn} = 0$; see Eq.\ \eqref{eq:G_matrix}.

In this case,
\begin{align}
		\gamma_{\alpha\alpha} & = {1 \over m_{\alpha}^*} 
		\frac{n_\alpha + G_{\alpha\alpha} m_{\alpha}^*}{n_\alpha + G_{\alpha\alpha} m_{\alpha}^*\Phi_\alpha}
			= \frac{1}{m_{\alpha}^* \Phi_\alpha + {\mu_\alpha \over c^2}(1-\Phi_\alpha)},
			\\
		K_{\alpha\alpha} & =	
		\frac{ G_{\alpha\alpha} m_{\alpha}^* \Phi_\alpha}{n_\alpha + G_{\alpha\alpha} m_{\alpha}^*\Phi_\alpha}
			= \frac{(m_{\alpha}^* - {\mu_\alpha \over c^2})\Phi_\alpha }{m_{\alpha}^* \Phi_\alpha + {\mu_\alpha \over c^2}(1-\Phi_\alpha)},
			\\
		\gamma_{\alpha\beta}  &= 	
		  K_{\alpha\beta} 
		= 0,
\end{align}
where, we applied the relation \eqref{eq:m_eff_eq}.
In turn, the matrices $Y_{\alpha\alpha'}$ and $R_{\alpha\alpha'}$ reduce to
\begin{align}
		Y_{\alpha\alpha} &=  \frac{n_\alpha (1 - \Phi_\alpha)}{m_{\alpha}^* c^2\Phi_\alpha + \mu_\alpha (1-\Phi_\alpha)},
		\\
		R_{\alpha\alpha} &=  \frac{n_\alpha m_{\alpha}^* c^2 \Phi_\alpha}{m_{\alpha}^* c^2 \Phi_\alpha + \mu_\alpha(1-\Phi_\alpha)},
		\\
		Y_{\alpha\beta}  &= R_{\alpha\beta}  = 0.
\end{align}	
As it is expected, the non-diagonal elements of all matrices vanish in this case.

\section{Relativistic hydrodynamics of superfluid mixtures} 
\label{ap:hydrodynamics}

In this appendix, 
we briefly describe the phenomenological relativistic hydrodynamics of superfluid mixtures,
since we refer to some of its equations in the main text of the paper.
We use the version of hydrodynamics developed 
in Refs.\
\cite{Gusakov2007,Gusakov2016,GusakovDommes2016,DommesGusakovShternin2020,DommesGusakov2021}.
Despite this hydrodynamics  
taking into account 
an extremely rich set of various 
dissipative phenomena,  
here we restrict ourselves to considering its simplified version,
which only allows for diffusion as a dissipative mechanism.
The effects of diffusion have been incorporated into the relativistic hydrodynamics 
of normal and superfluid mixtures in Refs.\ \cite{DommesGusakovShternin2020,DommesGusakov2021}.
We will follow these works in what follows.
For simplicity, we assume that there are no vortices in the system and also ignore the effects 
related to polarization and magnetization of the medium.
In addition, we neglect possible chemical reactions converting different particle species into each 
other. 
With these reservations, the equations of superfluid relativistic hydrodynamics 
consist of

i) The energy-momentum conservation law
\begin{equation}
		\label{eq:ap:emcl}
		\frac{\partial T^{\mu\nu} } {\partial x^\nu} = 0,
\end{equation}
where 
\begin{equation}
		T^{\mu\nu} = \frac{E + P}{c^2} u^\mu u^\nu + P g^{\mu\nu} 
		 + \sum_{\alpha\alpha'} Y_{\alpha\alpha'} \left[ c^2 w_{(\alpha)}^\mu w_{(\alpha')}^\nu + \mu_\alpha w_{(\alpha')}^\mu u^\nu  +\mu_{\alpha'} w_{(\alpha)}^\nu u^\mu \right] 
		+  T_{\rm EM}^{\mu\nu}.
		\label{eq:ap:T_def}	 
\end{equation}
ii) The particle conservation law for each species $\alpha$
\begin{equation}
		\label{eq:ap:cont_eq}
		\frac{\partial j_{(\alpha)}^\mu } {\partial x^\mu} = 0,
\end{equation}
where
\begin{equation}
		\label{eq:ap:j_def}
		j_{(\alpha)}^\mu = n_{\alpha} u^\mu + \sum_{\alpha'} Y_{\alpha\alpha'} w_{(\alpha')}^\mu + \Delta j_{(\alpha)}^\mu.
\end{equation}
iii) The constraints on the four-velocities and four-currents:
\begin{equation}
		\label{eq:ap:uw_rel}
		u_\mu w_{(\alpha)}^\mu = 0,
\end{equation}
\begin{equation}
		\label{eq:ap:uj_rel}
		u_\mu \Delta j_{(\alpha)}^\mu = 0.
\end{equation}
iv) The second law of thermodynamics
\begin{align}
		dE = T dS + \sum_{\alpha} \mu_\alpha d n_\alpha + \sum_{\alpha\alpha'} {Y_{\alpha\alpha'} \over 2} d\left( w_{(\alpha)\mu} w_{(\alpha')}^\mu \right).
		\label{eq:ap:dE}	
\end{align}
v) The Maxwell equations
\begin{equation}
		\partial_\nu F^{\mu\nu} = {4 \pi \over c} J^{\mu}, \ \ \ \  \partial_\mu F^{\nu\xi} + \partial_\xi F^{\mu\nu} + \partial_\nu F^{\xi\mu} = 0.  
\end{equation}
In the formulas above, 
$T_{\rm EM}^{\mu\nu}$ is the standard electromagnetic energy-momentum tensor,
$F^{\mu\nu} $ is the electromagnetic field tensor,
$J^\nu = \sum_\alpha e_\alpha j_{(\alpha)}^\mu$ is the charge current density,
$e_\alpha$ is the charge of particle species $\alpha$,
$w_{(\alpha)}^\mu$ is the superfluid 
{four-vector given below,}
$\Delta j_{(\alpha)}^\mu$ is the dissipative correction to the particle current density,
$E$ is the energy density in the comoving frame of reference 
$[u^{\mu}=(c,0,0,0)]$, $P$ is the pressure
given by standard formula 
\begin{equation}
		\label{eq:ap:P_def}
		P  = -  E + T S + \sum_\alpha n_\alpha \mu_\alpha,
\end{equation}
and $g^{\mu\nu} = {\rm diag} (-1,1,1,1)$ is the Minkowski metrics. 
In the above expressions and in the text below, 
a summation is assumed over repeated space-time indices $\mu$, $\nu$, and $\xi$.
Note, however, that the sum over the 
{particle species}
indices $\alpha$ and $\alpha'$
will be written explicitly to facilitate comparison with the main text of the paper.
The superfluid four-velocity can be represented as 
\begin{equation}
		\label{eq:ap:w_def}
		w_{(\alpha)}^\mu = Q_{(\alpha)}^\mu  - \frac{\mu_\alpha }{c^2} u^\mu,	
\end{equation}
where we introduced the half Cooper-pair four-momentum 
\begin{equation}
		\label{eq:ap:Q_def}
		Q_{(\alpha)}^\mu = {1 \over 2} \partial^\mu \Phi_\alpha - {e_\alpha \over c} A^\mu,
\end{equation}
$\Phi_\alpha$ is the phase of the Cooper-pair condensate wave function, 
and $A^\mu$ is  the four-potential of the electromagnetic field.

Equations 
{\eqref{eq:ap:emcl}--\eqref{eq:ap:Q_def} }
allow one to obtain 
the entropy generation equation,
\begin{equation}
		\partial_\mu S^\mu = \Gamma_s,
\end{equation}
where
\begin{equation}
		S^\mu = S u^\mu - \sum_\alpha {\mu_\alpha \over T} \Delta j_{(\alpha)}^\mu
\end{equation}
is the entropy four-current and
\begin{equation}
		\label{eq:ap:Gamma_s}
		\Gamma_s = - \sum_\alpha \Delta j_{(\alpha)\mu} d_{(\alpha)}^\mu 
\end{equation}
is the entropy generation rate.
In this expression
\begin{equation}
		\label{eq:ap:d_a_def}
		d_{(\alpha)}^\mu =\left(  \partial^\mu  + u^\mu u^\nu \partial_\nu \right)\left( \mu_\alpha 
		\over T \right) - {e_\alpha E^\mu \over T},
\end{equation}
where 
\begin{equation}
		\label{eq:ap:Emu}
		E^\mu = {u_\nu \over c} F^{\mu\nu}.	
\end{equation}
If the gradients are small and there is no preferred direction, the current corrections 
$\Delta j_{(\alpha)}^\mu$ can be presented as 
\begin{equation}
		\label{eq:ap:j_Dd}
		\Delta j_{(\alpha)}^\mu = - \sum_{\alpha'} \mathcal{D}_{\alpha\alpha'} d_{(\alpha')}^\mu.
\end{equation}
It  follows from the Onsager principle that the diffusion coefficients 
$\mathcal{D}_{\alpha\alpha'}$ must be symmetric,
\begin{equation}
		\label{eq:Onsager_rel}
		\mathcal{D}_{\alpha\alpha'}  = \mathcal{D}_{\alpha'\alpha}.
\end{equation} 
%

\section{The nonrelativistic limit of superfluid hydrodynamics}
\label{ap:non-rel-hydr}

Let us consider a reference frame in which  
all the hydrodynamic velocities are nonrelativistic, 
i.e. $|u^i| , \, |c^2 Q_{\alpha}^i /\mu_\alpha | \ll c$, 
where the index $i$ runs over the spatial coordinates 
(as in Sec. \ref{sec:sf_currents}, we only consider fluid 
motions for which such reference frame does exist). 
Our aim is to find the form of hydrodynamic equations appropriate for this frame.
Using the relation \eqref{eq:ap:uw_rel}, 
one can rewrite the expression \eqref{eq:ap:w_def} as
\begin{align}
		\label{eq:ap:w0}
		&w_{(\alpha)}^0 = {\vec{u}  \vec{Q}_\alpha \over u^0}  - { \mu_\alpha \over c^2 }{ \vec{u}^2 \over u^0 }, \\
		\label{eq:ap:wspat}
		&\pmb{w}_\alpha = \vec{Q}_\alpha - \frac{\mu_\alpha}{c^2} \vec{u},
\end{align}
where $\vec{u}$ , $\pmb{w}_\alpha$, and $\vec{Q}_\alpha$ are the spatial parts of the four-vectors 
$u^\mu$, $w_{(\alpha)}^\mu$, and $Q_{(\alpha)}^\mu$. 
Substituting these expressions together with the expansion
\begin{equation}
		\label{eq:ap:u0}
		u^0 \approx c \left( 1 + {1\over 2}{\vec{u}^2 \over c^2}\right)
\end{equation}
into \eqref{eq:ap:T_def}
and neglecting the terms $\sim \vec{u} /c$ in that equation,
one gets
\begin{align}
		& T^{i0} \approx  c \rho_{ q}\vec{u} + c  \sum_{\alpha\alpha'} Y_{\alpha\alpha'}\mu_\alpha \vec{Q}_{\alpha'} + T_{\rm EM}^{i0},
		\label{eq:ap:T_i0_nr}
		 \\
		& T^{ij} \approx \rho_	{ q} u^i u^j  +  c^2 \sum_{\alpha\alpha'} Y_{\alpha\alpha'} {Q}_\alpha^i {Q}_{\alpha'}^j + P \delta^{ij}+ T_{\rm EM}^{ij},
		\label{eq:ap:T_ij_nr}
\end{align}
where $\delta^{ij}$ is the the Kronecker delta and the following notation has been introduced:
\begin{equation}
		\label{ap:eq:rho_q}
	 	\rho_{ q} = {TS \over c^2} + \sum_\alpha {\mu_\alpha \over c^2}\left(n_{\alpha}  -  \sum_{\alpha'} Y_{\alpha\alpha'} \mu_{\alpha'}  \right).
\end{equation}
This combination arises in expressions \eqref{eq:ap:T_i0_nr}--\eqref{eq:ap:T_ij_nr} after one 
accounts for the definition \eqref{eq:ap:P_def}.
Equation \eqref{eq:ap:emcl} can  be represented as
\begin{equation}
		\label{eq:ap:Euler} 
		\frac{1}{c} \frac{\partial T^{i0} } {\partial t}  + \frac{\partial T^{ij} } {\partial x^j} = 0,
\end{equation}
where one should employ the expressions \eqref{eq:ap:T_i0_nr} and \eqref{eq:ap:T_ij_nr}.
Using Eqs.\ \eqref{eq:ap:dE} and \eqref{eq:ap:P_def}, one gets the following 
{Gibbs-Duhem relation}
\begin{equation}
		\label{eq:ap:dP}
		d P \approx S dT + \sum_\alpha n_\alpha d \mu_\alpha + \sum_{\alpha \alpha'} 
		{Y_{\alpha\alpha'} \over 2} d( \pmb{w}_{\alpha} \pmb{w}_{\alpha'}),
\end{equation}
where we have neglected the terms containing  
${w}_{(\alpha)}^0 {w}_{(\alpha')}^0 
\sim (\vec{u}^2/c^2) \pmb{w}_{\alpha} \pmb{w}_{\alpha'} $ 
[see Eqs.\ \eqref{eq:ap:w0} and  \eqref{eq:ap:wspat}].

In the present paper, we work in the linear approximation in hydrodynamic velocities.
Thus, it is instructive to write down the phenomenological equations in the same approximation.	
Plugging Eqs.\ \eqref{eq:ap:T_i0_nr} and \eqref{eq:ap:T_ij_nr} 
into \eqref{eq:ap:Euler} and neglecting the terms quadratic in $\vec{u}$  and $\vQa$, we obtain
\begin{equation}
		\label{eq:ap:Euler_lin_tmp}
		 \rho_{ q} {\partial \vec{u} \over \partial t} + \sum_{\alpha\alpha'}{\mu_\alpha } Y_{\alpha\alpha'}{\partial \vec{Q}_{\alpha'}  \over \partial t}    +  {\nabla P } = - \frac{\partial T_{\rm EM}^{\mu\nu} } {\partial x^\nu}.
\end{equation}
Here we used the fact that the time derivatives of the quantities $\rho_{ q}$, 
$Y_{\alpha\alpha'}$, and $\mu_\alpha$ are of the linear order smallness in hydrodynamic velocities 
[see Eqs.\ \eqref{eq:dEdt_expand}--\eqref{eq:ap:dmu} and the discussion afterwards].
Expressing the pressure gradient using Eq.\ \eqref{eq:ap:dP}, 
and neglecting 
the small
terms $\propto \pmb{w}_{\alpha} \pmb{w}_{\alpha'}$, 
we present Eq.\ \eqref{eq:ap:Euler_lin_tmp} in the final form
\begin{equation}
		\label{eq:ap:Euler_lin}
		 \rho_{ q} {\partial \vec{u} \over \partial t} + \sum_{\alpha\alpha'}{\mu_\alpha } Y_{\alpha\alpha'}{\partial \vec{Q}_{\alpha'}  \over \partial t}   
		  + T {\nabla S } + \sum_{\alpha} n_{\alpha}\left( \nabla \mu_\alpha - e_\alpha 
		  \vec{E}\right)= 0.
\end{equation}
To derive this equation, we noted that the 
right-hand side of Eq.\ \eqref{eq:ap:Euler_lin_tmp} equals the 
Lorentz force (see, e.g., Ref.\ \cite{Landau_Class}).
Since 
the magnetic field in the present paper is neglected, we only keep
the electrical part of the Lorentz force in Eq.\ \eqref{eq:ap:Euler_lin}.

In the nonrelativistic limit the particle current density acquires the form
\begin{equation}
	\label{eqj}
		j_{(\alpha)}^0 = c n_\alpha, \ \ \ \ \vec{j}_\alpha =  Y_{\alpha\alpha} \vec{Q}_{\alpha}+ Y_{\alpha\beta} \vec{Q}_{\beta}+ n_{ q \alpha}  \vec{u}+ \Delta \vec{j}_{(\alpha)}.
\end{equation}	
To derive Eq.\ \eqref{eqj}, 
one needs to plug Eq.\ \eqref{eq:ap:w_def} into \eqref{eq:ap:j_def} 
and take into account Eqs.\ \eqref{eq:ap:uw_rel} and 
\eqref{eq:ap:uj_rel}. 
The continuity equation \eqref{eq:ap:cont_eq} transforms into 
\begin{equation} 
		\label{eq:ap:cont_eq_nr}
		{\partial n_\alpha \over \partial t} + \Div \vec{j}_\alpha = 0.
\end{equation}
The vectors $\Delta \vec{j}_{(\alpha)}$ are the linear functions of the spatial components of 
the four-vectors $d_{(\alpha)}^{\mu}$ [see Eq.\ \eqref{eq:ap:j_Dd}].
In the nonrelativistic limit, 
$E^\mu = ( 0, \vec{E})$, where $\vec{E}$ is the electric field. 
Hence, one can write  
\begin{equation}
		\label{eq:ap:d_nonrel}
		d_{(\alpha)}^0 = 0, \ \ \ \vec{d}_{(\alpha)} = \nabla \left( \mu_\alpha \over T \right) - {e_\alpha \vec{E} \over T} 
\end{equation}
and, in view of the relation  \eqref{eq:ap:j_Dd}, one has
\begin{equation}
		\label{eq:ap:dj_exp}
		 \Delta \vec{j}_{\alpha} = - \sum_{\alpha'} 	\mathcal{D}_{\alpha\alpha'} \left[ \nabla 
		 \left( \mu_\alpha \over T \right) - {e_\alpha \vec{E} \over T} \right].
\end{equation}

Since, according to the expressions \eqref{eq:ap:w_def} and \eqref{eq:ap:Q_def}, the superfluid 
four-velocity contains the gradient of the scalar function $\partial^\mu \Phi_\alpha$, it obviously 
satisfies the following constraint
\begin{equation}
		{\partial \over \partial x_\mu}  \left( w_{(\alpha)}^\nu + {e_\alpha \over c}A^\nu + \frac{\mu_\alpha }{c^2} u^\nu\right) - 
		{\partial \over \partial x_\nu}  \left( w_{(\alpha)}^\mu + {e_\alpha \over c}A^\mu + \frac{\mu_\alpha}{c^2} u^\mu  \right) = 0.
\end{equation}
Assuming that the index $\mu = 0$ and the index $\nu$ runs over the spatial coordinates, 
one gets from this equation,
after using Eq.\ \eqref{eq:ap:wspat},
\begin{equation}
	 \label{onemore}
		 {1 \over c} {\partial \over \partial t}  \left( \vec{Q}_\alpha + {e_\alpha \over c}\vec{A} \right) + \nabla \left( w_{(\alpha)}^0 + {e_\alpha \over c}A^0 + \frac{\mu_\alpha}{c^2} u^0  \right) = 0.
\end{equation}
Plugging Eqs.\ \eqref{eq:ap:w0} and \eqref{eq:ap:u0} into \eqref{onemore}
one finally arrives at
\begin{equation}
		\label{eq:ap:superfluid_eq}
	  	{\partial \vec{Q}_\alpha  \over \partial t}  + \nabla \left[ \mu_\alpha \left(1-{1 \over 2}{\vec{u}^2 \over c^2} \right) + \vec{u}\vec{Q}_\alpha \right] -e_\alpha \vec{E} = 0,
\end{equation}
where the electric field is expressed through the components of the four-potential:
\begin{equation}
		\vec{E} = - \nabla A^0 - {1 \over c} {\partial \vec{A} \over \partial t}. 
\end{equation}
In the linear approximation in hydrodynamic velocities
this equation reduces to%
%
\footnote{
The superfluid equation \eqref{eq:ap:superfluid_eq_lin}	should be compared with
the more general, but similar Eq.\ \eqref{eq:sf_eq_charged}.
Note that Eq.\ \eqref{eq:sf_eq_charged} was obtained 
without linearization with respect to hydrodynamic velocities.
The discrepancy between Eqs.\
\eqref{eq:sf_eq_charged} and \eqref{eq:ap:superfluid_eq_lin} arises for two reasons. 
First,  the 
chemical potentials in two equations are defined in different ways 
(measured in different reference frames). 
However, as  already discussed, this difference is of the second-order smallness in 
the hydrodynamic velocities and can be ignored in the linear approximation. 
Second, the nonequilibrium chemical potential $\breve{\mu}_\alpha$ also contains a dissipative 
correction.
For small deviations from the thermodynamic equilibrium this correction 
is exclusively due to the
bulk viscosity.
However, all the viscous terms were 
omitted
in the phenomenological 
hydrodynamics of Appendix \ref{ap:hydrodynamics},
on which the derivation of Eq.\ \eqref{eq:ap:superfluid_eq_lin} is based.
The form of the superfluid equation with the dissipative correction 
{is given}
in 
Ref.\ \cite{Gusakov2007}.}
%
\begin{equation}
		\label{eq:ap:superfluid_eq_lin}
	  	{\partial \vec{Q}_\alpha  \over \partial t}  + \nabla  \mu_\alpha -e_\alpha \vec{E} = 0.
\end{equation}	

Substituting the vector \eqref{eq:ap:d_nonrel} into Eq.\ \eqref{eq:ap:Gamma_s}, one obtains the 
nonrelativistic expression for the entropy production rate
\begin{equation}
		\label{eq:ap:Gamma_s_nr}
		\Gamma_s = - \sum_\alpha \Delta \vec{j}_{\alpha} \left[ \nabla \left( \mu_\alpha \over T 
		\right) - {e_\alpha \vec{E} \over T} \right].
\end{equation}
%

\section{The effective interaction Hamiltonian}
\label{sec:col_mat}

Let  us consider a mixture of two superfluid Fermi liquids.
It is assumed that the scattering of the Landau-liquid quasiparticles can be described by an 
effective Hamiltonian of the following form
\begin{equation}
	\label{eq:Hv_1}
	\hat{H}_v = {1\over 2} \sum _{1,2,3,4} \left( 1,2| {V} |3,4\right)  \hat{a}_1^\dag 
	\hat{a}_2^\dag \hat{a}_3 \hat{a}_4.
\end{equation}	
Here  
$\hat{a}_k$ and $\hat{a}_k^\dag$ are, respectively,  
the destruction and creation operators for a quasiparticle in a 
quantum state $k$.
Each quantum state $k$ is characterized by the set of quantum numbers
$(\vec{Q}_{\mathfrak{a}_k}+ \vec{k}_k, s_k, 
\mathfrak{a}_k)$, where
$\vec{Q}_{\mathfrak{a}_k}+\vec{k}_k$ is the quasiparticle momentum, 
$s_k = \pm 1$ is the spin index,  
and $\mathfrak{a}_k = \pm 1$ is the isospin index 
($\mathfrak{a}_k =  1$ can be associated  with the particle species ``$p$'',
while $\mathfrak{a}_k = - 1$  with the species ``$n$'').
In what follows we also make use of the notation 
$- k = (\vec{Q}_{\mathfrak{a}_k} - \vec{k}_k, -s_k,  \mathfrak{a}_k)$.
We do not specify the dependence of the matrix elements $\left( 1,2| {V} |3,4\right)$ on the 
quantum states variables, but assume that they contain the following Kronecker deltas:
\begin{equation}
	(1,2| V|3,4) \rightarrow  ( 1,2| V|3,4)  \delta_{\mathfrak{a}_1 + \mathfrak{a}_2, 
	\mathfrak{a}_3 + \mathfrak{a}_4} \delta_{\vec{k}_1 + \vec{k}_2, \vec{k}_3 + \vec{k}_4 },
\end{equation}
ensuring conservation of the total isospin and momentum in particle collisions.
Note that, strictly speaking, in our notation the momentum of a quasiparticle equals 
$\vec{Q}_{\mathfrak{a}_k} + \vec{k}_k$ and, consequently, the momentum conservation law reads 
$\vec{Q}_{\mathfrak{a}_1} + \vec{k}_1 + \vec{Q}_{\mathfrak{a}_2} + \vec{k}_2 = 
\vec{Q}_{\mathfrak{a}_3} + \vec{k}_3 + \vec{Q}_{\mathfrak{a}_4} + \vec{k}_4$.
However, it is easy to see that, in view of the isospin conservation, 
the vectors $\vec{Q}_{\mathfrak{a}_i}$ cancel out.

Taking into account the commutation rules for operators $\hat{a}_k$ and $\hat{a}_k^\dag$, one can 
antisymmetrize the expression \eqref{eq:Hv_1}  and present it as 
\begin{equation}
	\label{eq:Hv}
	\hat{H}_v = {1\over 4} \sum _{1,2,3,4} \langle 1,2| {V} |3,4\rangle_a  \hat{a}_1^\dag 
	\hat{a}_2^\dag \hat{a}_3 \hat{a}_4, 
\end{equation}
where the matrix elements $\langle 1,2| {V} |3,4\rangle_a$ are given by%
%
\footnote{Note that in Eq.\ \eqref{angle} the matrix elements 
$\left( 2,1| {V} |4,3\right)$ and $\left( 2,1| {V} |3,4\right)$ 
are equal, respectively, to
$\left( 2,1| {V} |4,3\right)=\left( 1,2| {V} |3,4\right)$ and 
$\left( 2,1| {V} |3,4\right)=\left( 1,2| {V} |4,3\right)$, 
so that \eqref{angle} can actually be represented as
$\langle 1,2| {V} |3,4\rangle_a=\left( 1,2| {V} |3,4\right) 
-  \left( 1,2| {V} |4,3\right)$ \cite{BlaizotRipkaBook}.
 }
%
%
\begin{equation}
	\label{angle}
	\langle 1,2| {V} |3,4\rangle_a  = \frac{1}{2} \Big[ \left( 1,2| {V} |3,4\right) -  \left( 1,2| 
	{V} |4,3\right) -  \left( 2,1| {V} |3,4\right) +  \left( 2,1| {V} |4,3\right) \Big]
\end{equation}
and have the following property:
\begin{align}
	\label{eq:V_rel_1}
	& \langle 1,2| V|3,4\rangle_a = -\langle 1,2| V|4,3\rangle_a = - \langle 2,1| V|3,4\rangle_a = 
	\langle 2,1| V|4,3\rangle_a,
\end{align}
Note that these elements contain the same Kronecker deltas 
as $\left( 1,2| {V} |3,4\right)$, i.e. ,
\begin{equation}
	\label{eq:V_delta}
	\langle 1,2| V|3,4\rangle_a \rightarrow  \langle 1,2| V|3,4\rangle_a  \delta_{\mathfrak{a}_1 + 
	\mathfrak{a}_2, \mathfrak{a}_3 + \mathfrak{a}_4} \delta_{\vec{k}_1 + \vec{k}_2, \vec{k}_3 + 
	\vec{k}_4 }.
\end{equation}	
We added a subscript ``$a$'' to the matrix element $\langle 1,2| {V} |3,4\rangle_a$ in order to 
distinguish it from the elements \eqref{eq:ap:matr_scat}--\eqref{eq:ap:matr_destr} introduced 
below.

Let us now make the
Bogoliubov transformation for the Landau quasiparticle operators,
\begin{equation}
	\label{eq:Bog_tr}
	\hat{a}_k = u_k \hat{b}_k + s_k v_{-k} \hat{b}_{-k}^\dag,
\end{equation} 
where the coherence factors are given by Eqs.\ \eqref{eq:upvp1} and \eqref{upQ}, and the 
Bogolubov thermal excitation operators $\hat{b}_k$ and $\hat{b}_{k}^\dag$ obey the 
canonical Fermi commutation relations.
Substituting \eqref{eq:Bog_tr} into \eqref{eq:Hv}, one can represent the 
Hamiltonian in the form
\begin{align}
	\hat{H}_v &= \sum_{1,2,3,4} \left\{
	{1\over4} \langle 1, 2| V| 3,4\rangle_b \hat{b}_1^\dag \hat{b}_2^\dag \hat{b}_3 \hat{b}_4
	+ {1 \over 6}  \langle 1, 2,3| V| 4\rangle_b \hat{b}_1^\dag \hat{b}_2^\dag \hat{b}_3^\dag 
	\hat{b}_4
	+ {1 \over 6}  \langle 1| V|2, 3, 4\rangle_b \hat{b}_1^\dag \hat{b}_2 \hat{b}_3 \hat{b}_4 
	\right.
	\nonumber
	\\
	& \left. + {1 \over 24}  \langle 1, 2, 3, 4| V|{ 0} \rangle_b \hat{b}_1^\dag \hat{b}_2^\dag  
	\hat{b}_3^\dag  \hat{b}_4^\dag 
	+ {1 \over 24}  \langle { 0 }  | V|1, 2, 3, 4 \rangle_b \hat{b}_1 \hat{b}_2  \hat{b}_3  
	\hat{b}_4,
	\right\},
	\label{eq:ap:Hv_decomp}		
\end{align}
where the following coefficients have been introduced 
\begin{align}
	\langle 1, 2| V| 3,4\rangle_b & = 
	u_1 u_2 u_3 u_4 \langle 1,2| V|3,4\rangle_a
	+ s_1 s_2 s_3 s_4 v_1 v_2 v_3 v_4   \langle -4,-3| V|-2,-1\rangle_a
	\nonumber
	\\
	&-s_2 s_4 u_1 v_{2} u_3 v_{4}   \langle 1,-4| V|3,-2\rangle_a
	-s_2 s_3 u_1 v_{2} v_3 u_{4}   \langle 1,-3| V|-2,4\rangle_a
	\nonumber
	\\
	&-s_1 s_4 v_1 u_{2} u_3 v_{4}   \langle -4,2| V|3,-1\rangle_a
	-s_1 s_3 v_1 u_{2} v_3 u_{4}    \langle -3,2| V|-1,4\rangle_a,
	\label{eq:ap:matr_scat}
	\\
	\nonumber
	\langle 1, 2,3| V| 4\rangle_b & =
	\\
	\nonumber
	&        s_2 u_1 v_2 u_3 u_4 \langle 1,3| V|-2,4\rangle_a  + s_1 s_3 s_4 v_1 u_2 v_3 v_4   
	\langle -4,2| V|-3,-1\rangle_a 			
	\\
	\nonumber
	&+  s_1 v_1 u_2 u_3 u_4 \langle 3,2| V|-1,4\rangle_a  + s_2 s_3 s_4 u_1 v_2 v_3 v_4   \langle 
	-4,1| V|-2,-3\rangle_a 	
	\\
	&-  s_3 u_1 u_2 v_3 u_4 \langle 1,2| V|-3,4\rangle_a  - s_1 s_2 s_4 v_1 v_2 u_3 v_4   \langle 
	-4,3| V|-2,-1\rangle_a 	,
	\label{eq:ap:matr_dec}
	\\
	\nonumber
	\nonumber
	\langle 1| V|2, 3, 4\rangle_b & =
	\\
	\nonumber
	&         s_4 u_1 u_2 u_3 v_4 \langle 1,-4| V|3,2\rangle_a  + s_1 s_2 s_3 v_1 v_2 v_3 u_4   
	\langle -2,-3| V|4,-1\rangle_a		
	\\
	\nonumber
	&+    s_3 u_1 u_2 v_3 u_4 \langle 1,-3| V|2,4\rangle_a  + s_1 s_2 s_4 v_1 v_2 u_3 v_4   \langle 
	-4,-2| V|3,-1\rangle_a	
	\\
	&-  s_2 u_1 v_2 u_3 u_4 \langle 1,-2| V|3,4\rangle_a  - s_1 s_3 s_4 v_1 u_2 v_3 v_4   \langle 
	-4,-3| V|2,-1\rangle_a ,	
	\label{eq:ap:matr_coal}
	\\
	\langle 1, 2, 3, 4| V|{ 0} \rangle_b  & = 6 s_3 s_4 u_1 u_2 v_3 v_4 \langle 1,2| 
	V|-3,-4\rangle_a,
	\\
	\langle { 0}  | V|1, 2, 3, 4 \rangle_b  & = 6  s_1 s_2 v_1 v_2 u_3 u_4 \langle -1,-2| 
	V|3,4\rangle_a. 	
	\label{eq:ap:matr_destr}				  
\end{align}
To obtain these expressions, one should use the relation \eqref{eq:V_rel_1} 
together with the commutation relations for the operators $\hat{b}_i$ and $\hat{b}_i^\dag$.
Collecting, for example, the terms with three destruction and one creation operators, 
one gets
\begin{align}
	& \frac{1}{4}\sum_{1,2,3,4} 
	\left\{ s_2 u_{1} v_{-2} u_{3} u_4 \hat{b}_{1}^\dag \hat{b}_{-2} \hat{b}_3 \hat{b}_4  
	+ s_1 v_{-1} u_{2} u_{3} u_4 \hat{b}_{-1} \hat{b}_{2}^\dag \hat{b}_3 \hat{b}_4  
	\right.
	\nonumber
	\\
	&
	\left.
	\ \ \ \ \ \ \ \ 	+ s_1 s_2 s_3 v_{-1} v_{-2} v_{-3} u_{4}   \hat{b}_{-1} \hat{b}_{-2} 
	\hat{b}_{-3}^\dag \hat{b}_4  
	+ s_1 s_2 s_4 v_{-1} v_{-2} u_{3} u_{-4}   \hat{b}_{-1} \hat{b}_{-2} 
	\hat{b}_{3}\hat{b}_{-4}^\dag   
	\right\}  \langle1,2|V|3,4\rangle_a
	\nonumber
	\\
	&= \frac{1}{2} \sum_{1,2,3,4}  \Big\{- s_2 u_{1} v_{2} u_{3} u_4 \langle1,- 2|V|3,4\rangle_a  
	+ s_1 s_2 s_3 v_1 v_2 v_3 u_4  \langle -3,-2|V|-1,4\rangle_a  		
	\Big\} \hat{b}_1^\dag \hat{b}_2 \hat{b}_3 \hat{b}_4
	\nonumber
	\\
	\nonumber
	&=  \frac{1}{6} \sum_{1,2,3,4}  \Big\{
	s_4 u_1 u_2 u_3 v_4 \langle 1,-4| V|3,2\rangle_a    + s_1 s_2 s_3 v_1 v_2 v_3 u_4   \langle 
	-3,-2| V|-1,4\rangle_a		
	\\
	\nonumber
	&\ \ \ \ \ \ \ \ \ \ \ \ \ \ \ \ +    s_3 u_1 u_2 v_3 u_4 \langle 1,-3| V|2,4\rangle_a  +      
	s_1 s_2 s_4 v_1 v_2 u_3 v_4   \langle -4,-2| V|3,-1\rangle_a
	\\
	&\ \ \ \ \ \ \ \ \ \ \ \ \ \ \ \ -  s_2 u_1 v_2 u_3 u_4 \langle 1,-2| V|3,4\rangle_a  - s_1 s_3 
	s_4 v_1 u_2 v_3 v_4   \langle -4,-3| V|2,-1\rangle_a 
	\Big\} \hat{b}_1^\dag \hat{b}_2 \hat{b}_3 \hat{b}_4 
	\nonumber\\
	&\equiv  \frac{1}{6} \sum_{1,2,3,4} \langle 1| V|2, 3, 4\rangle_b \, \hat{b}_1^\dag \hat{b}_2 
	\hat{b}_3 \hat{b}_4.&
\end{align}
In the second equality,
accounting for the fact that there is a summation over all quantum numbers, 
we antisymmetrize the sum in order to make the resulting matrix element
$\langle 1| V|2, 3, 4\rangle_b$
antisymmetric 
with respect to permutations of second, third, and fourth quantum states.
The same procedure allows us to obtain the (antisymmetric) matrix element 
$\langle 1, 2,3| V| 4\rangle_b $.
The coefficients in expression \eqref{eq:ap:Hv_decomp} can be considered as matrix elements for
different processes involving  Bogoliubov excitations: 
the element $\langle 1, 2| V| 3,4\rangle_b$ describes scattering $3,4 \rightarrow 1,2$, 
the elements $\langle 1, 2,3| V| 4\rangle_b$ and   $\langle 1| V|2, 3, 4\rangle_b $
describe decay $ 4 \rightarrow 1,2,3$ and coalescence $ 2,3,4, \rightarrow 1$, while the matrix 
elements  	
$ \langle 1, 2, 3, 4| V|0 \rangle_b$  and  $\langle 0  | V|1, 2, 3, 4 \rangle_b$ describe creation 
and destruction of four Bogoliubov excitations.
However, since these excitations have positive energy,
the last two processes are 
forbidden by the energy conservation.
Hence, these terms can be ignored, in particular, 
in calculations of Appendix \ref{sec:Integral}.

Note once again that the matrix elements \eqref{eq:ap:matr_scat}--\eqref{eq:ap:matr_coal} are 
constructed in such a way to make them 
antisymmetric with respect to permutations of the excitations in the initial state, 
as well as in the final state.
It can also be verified that the element $\langle 1, 2,3| V| 4\rangle_b$ can be obtained from 
$\langle 1| V|2, 3, 4\rangle_b $ by complex conjugation and interchanging 
the states
$1\leftrightarrow 4$ and $2\leftrightarrow 3$.

Substituting Eq.\ \eqref{eq:V_delta} into 
\eqref{eq:ap:matr_scat}--\eqref{eq:ap:matr_coal}, 
one can verify that these 
matrix elements contain the
following momentum Kronecker deltas
\begin{align}
	\langle 1, 2| V| 3,4\rangle_b \rightarrow \langle 1, 2| V| 3,4\rangle_b  \delta_{\vec{k}_1 + 
	\vec{k}_2, \vec{k}_3 + \vec{k}_4 },
\label{111}
	\\
	\langle 1| V| 2, 3,4\rangle_b \rightarrow \langle 1| V| 2, 3,4\rangle_b  \delta_{\vec{k}_1 -  
	\vec{k}_2 ,  \vec{k}_3 + \vec{k}_4 },
	\\
	\langle 1, 2, 3| V| 4\rangle_b \rightarrow \langle 1, 2,3| V| 4\rangle_b  \delta_{\vec{k}_1 + 
	\vec{k}_2 ,  \vec{k}_4 -  \vec{k}_3}.
\label{333}
\end{align}
At the same time, one cannot 
factor out the isospin Kronecker deltas from these matrix elements
because the Bogoliubov transformation involves two operators with 
different values of momentum, but with the same isospin index $\mathfrak{a}_k$.
The important isospin-related property of the matrix elements (\ref{111})--(\ref{333}),
which holds true in superfluid mixtures,
is that they remain nonzero only 
for transitions with
even number of particles of each species.
This obvious property is used in Appendix \ref{sec:Integral} 
to obtain the collision integral \eqref{eq:col_int}.

\section{Momentum transfer rates}
\label{sec:Integral}

{
The total collision integral for particle species $\alpha$ can be represented as a sum
\begin{equation}
		I_{\alpha} = \sum_{\alpha'} I_{\alpha\alpha'},
\end{equation}
where $I_{\alpha\alpha'}$ is the part of the integral  
describing collisions 
with particle species $\alpha'$.
In contrast to 
Landau
quasiparticles,
the number of Bogoliubov excitations 
is not necessarily conserved in the collisions. 
That is why, in addition to scattering, the collision integral 
contains also the terms
describing decay and coalescence events. 
In terms of the Bogoliubov thermal excitations, 
the collision integral can be written as
\begin{align}
		\small
		\nonumber
		I_{\alpha\alpha'} &=  \sum_{\vp_2,\vp_3,\vp_4} \Big\{ 
		\frac{1}{1+\delta_{\alpha,\alpha'}}W_{\rm scat.1}(4_\alpha 3_{\alpha'}|2_{\alpha'} 
		1_\alpha) 
		\delta( \mathfrak{E}_{\vp_{4}+ {\vec Q}_{\alpha}}^{(\alpha)} + \mathfrak{E}_{\vp_{3}+ {\vec 
		Q}_{\alpha'}}^{(\alpha')} -  \mathfrak{E}_{\vp_{2}+ {\vec Q}_{\alpha'}}^{(\alpha')} - 
		\mathfrak{E}_{\vp_{1}+ {\vec Q}_{\alpha}}^{(\alpha)})
		\\
		\nonumber
		&\times \left[\mF_{\vp_{4}+ {\vec Q}_{\alpha}}^{(\alpha)} \mF_{\vp_{3}+ {\vec 
		Q}_{\alpha'}}^{(\alpha')} (1-\mF_{\vp_{2}+ {\vec Q}_{\alpha'}}^{(\alpha')}) 
		(1-\mF_{\vp_{1}+ {\vec Q}_{\alpha}}^{(\alpha)}) - \mF_{\vp_{1}+ {\vec 
		Q}_{\alpha}}^{(\alpha)} \mF_{\vp_{2}+ {\vec Q}_{\alpha'}}^{(\alpha')} (1-\mF_{\vp_{3}+ 
		{\vec Q}_{\alpha'}}^{(\alpha')}) (1-\mF_{\vp_{4}+ {\vec Q}_{\alpha}}^{(\alpha)})\right]  
		\\
		\nonumber
		\\
		\nonumber
		& +\frac{1-\delta_{\alpha,\alpha'}}{2}W_{\rm scat.2}(4_{\alpha'} 3_{\alpha'}|2_{\alpha} 
		1_\alpha) 
		\delta( \mathfrak{E}_{\vp_{4}+ {\vec Q}_{\alpha'}}^{(\alpha')} + \mathfrak{E}_{\vp_{3}+ 
		{\vec Q}_{\alpha'}}^{(\alpha')} -  \mathfrak{E}_{\vp_{2}+ {\vec Q}_{\alpha}}^{(\alpha)} - 
		\mathfrak{E}_{\vp_{1}+ {\vec Q}_{\alpha}}^{(\alpha)})
		\\
		\nonumber
		&\times \left[\mF_{\vp_{4}+ {\vec Q}_{\alpha'}}^{(\alpha')} \mF_{\vp_{3}+ {\vec 
		Q}_{\alpha'}}^{(\alpha')} (1-\mF_{\vp_{2}+ {\vec Q}_{\alpha}}^{(\alpha)}) (1-\mF_{\vp_{1}+ 
		{\vec Q}_{\alpha}}^{(\alpha)}) - \mF_{\vp_{1}+ {\vec Q}_{\alpha}}^{(\alpha)} \mF_{\vp_{2}+ 
		{\vec Q}_{\alpha}}^{(\alpha)} (1-\mF_{\vp_{3}+ {\vec Q}_{\alpha'}}^{(\alpha')}) 
		(1-\mF_{\vp_{4}+ {\vec Q}_{\alpha'}}^{(\alpha')})\right] 
		\\
		\nonumber
		\\
		\nonumber
		&+  \frac{1-\delta_{\alpha\alpha'}}{2}W_{\rm dec. 1}( 4_\alpha | 3_{\alpha'} 2_{\alpha'} 
		1_{\alpha} )
		\delta( \mathfrak{E}_{\vp_{1}+ {\vec Q}_{\alpha}}^{(\alpha)} + \mathfrak{E}_{\vp_{2}+ {\vec 
		Q}_{\alpha'}}^{(\alpha')} +  \mathfrak{E}_{\vp_{3}+ {\vec Q}_{\alpha'}}^{(\alpha')} - 
		\mathfrak{E}_{\vp_{4}+ {\vec Q}_{\alpha}}^{(\alpha)})
		 \\
		\nonumber
		&\times \left[  \mF_{\vp_{4}+ {\vec Q}_{\alpha}}^{(\alpha)} (1-\mF_{\vp_{1}+ {\vec 
		Q}_{\alpha}}^{(\alpha)}) (1-\mF_{\vp_{2}+ {\vec Q}_{\alpha'}}^{(\alpha')}) (1-\mF_{\vp_{3}+ 
		{\vec Q}_{\alpha'}}^{(\alpha')}) - \mF_{\vp_{1}+ {\vec Q}_{\alpha}}^{(\alpha)} 
		\mF_{\vp_{2}+ {\vec Q}_{\alpha'}}^{(\alpha')} \mF_{\vp_{3}+ {\vec Q}_{\alpha'}}^{(\alpha')} 
		(1-\mF_{\vp_{4}+ {\vec Q}_{\alpha}}^{(\alpha)})\right]	
		\\
		\nonumber
		\\
		\nonumber	
		&+ \frac{1}{1+\delta_{\alpha\alpha'}}  W_{\rm dec. 2}(4_{\alpha'} |  3_{\alpha'}  
		2_{\alpha} 1_\alpha)  
		\delta( \mathfrak{E}_{\vp_{1}+ {\vec Q}_{\alpha}}^{(\alpha)} + \mathfrak{E}_{\vp_{2}+ {\vec 
		Q}_{\alpha}}^{(\alpha)} +  \mathfrak{E}_{\vp_{3}+ {\vec Q}_{\alpha'}}^{(\alpha')} - 
		\mathfrak{E}_{\vp_{4}+ {\vec Q}_{\alpha'}}^{(\alpha')})
		\\
		\nonumber
		&\times \left[  \mF_{\vp_{4}+ {\vec Q}_{\alpha'}}^{(\alpha')} (1-\mF_{\vp_{1}+ {\vec 
		Q}_{\alpha}}^{(\alpha)}) (1-\mF_{\vp_{2}+ {\vec Q}_{\alpha}}^{(\alpha)}) (1-\mF_{\vp_{3}+ 
		{\vec Q}_{\alpha'}}^{(\alpha')}) - \mF_{\vp_{1}+ {\vec Q}_{\alpha}}^{(\alpha)} 
		\mF_{\vp_{2}+ {\vec Q}_{\alpha}}^{(\alpha)} \mF_{\vp_{3}+ {\vec Q}_{\alpha'}}^{(\alpha')} 
		(1-\mF_{\vp_{4}+ {\vec Q}_{\alpha'}}^{(\alpha')})\right]
		\\
		\nonumber
		\\
		\nonumber	
		&+ \frac{1}{2 + 4\delta_{\alpha,\alpha'}} W_{\rm coal.} (4_\alpha3_{\alpha'} 2_{\alpha'} | 
		1_\alpha)
		\delta( \mathfrak{E}_{\vp_{4}+ {\vec Q}_{\alpha}}^{(\alpha)} + \mathfrak{E}_{\vp_{3}+ {\vec 
		Q}_{\alpha'}}^{(\alpha')} +  \mathfrak{E}_{\vp_{2}+ {\vec Q}_{\alpha'}}^{(\alpha')} - 
		\mathfrak{E}_{\vp_{1}+ {\vec Q}_{\alpha}}^{(\alpha)})
		 \\
		&\times \left[\mF_{\vp_{4}+ {\vec Q}_{\alpha}}^{(\alpha)} \mF_{\vp_{3}+ {\vec 
		Q}_{\alpha'}}^{(\alpha')} \mF_{\vp_{2}+ {\vec Q}_{\alpha'}}^{(\alpha')} (1-\mF_{\vp_{1}+ 
		{\vec Q}_{\alpha}}^{(\alpha)}) - \mF_{\vp_{1}+ {\vec Q}_{\alpha}}^{(\alpha)} 
		(1-\mF_{\vp_{2}+ {\vec Q}_{\alpha'}}^{(\alpha')}) (1-\mF_{\vp_{3}+ {\vec 
		Q}_{\alpha'}}^{(\alpha')}) (1-\mF_{\vp_{4}+ {\vec Q}_{\alpha}}^{(\alpha)})\right]	
		\Big\},
		\label{eq:col_int}
\end{align}
Here $W_{ q}(i|f)$ is the differential transition probability for the collision event $i 
\rightarrow f$.
In Eq.\ \eqref{eq:col_int} we use a slightly different notation for the Bogoliubov excitation 
quantum state in comparison to that used in Appendix \ref{sec:col_mat}. 
Namely,
by writing $i_\alpha$, we explicitly indicate that a given excitation is in the isospin state 
$\alpha$, while $i$ stands for the momentum state $\vQa + \vec{p}_i$ 
[the spin quantum number is not included in $i_\alpha$ since we only consider spin-averaged 
quantities in Eq.\ \eqref{eq:col_int}].
The differential transition probabilities are already summed over the spin states of the 
Bogoliubov excitations  $2_{\alpha}$, $3_{\alpha}$, $4_\alpha$ and averaged over the spin 
states of the excitation $1_\alpha$.
The factors $(1+\delta_{\alpha\alpha'})^{-1}$, $(1-\delta_{\alpha\alpha'})/2$, and 
$(2+4\delta_{\alpha\alpha'})^{-1}$ are added to prevent double 
or sixfold counting of the same collision event.
In the case of 
$\alpha'  \neq \alpha$, 
the sets of the scattering and decay events
split into two subsets,
corresponding to
the transition probabilities $W_{\rm scat. 1}$, $W_{\rm scat. 2}$,  and $W_{\rm dec. 1}$, 
$W_{\rm dec. 2}$, respectively.
If $\alpha' = \alpha$, both  $W_{\rm scat. 1}$ and  $W_{\rm scat. 2}$
as well as both $W_{\rm dec. 1}$ and $W_{\rm dec. 2}$ 
describe the same sets of scattering and decay events.  
In view of that, the multiplier  $(1-\delta_{\alpha\alpha'})/2$ is added to $W_{\rm scat. 2}$ and 
$W_{\rm dec. 1}$. The delta functions in the expression \eqref{eq:col_int} ensure the conservation 
of energy. 	
The differential transition probabilities can be represented in the following form:
\begin{align}
		\label{eq:ap:W_scat}
		&W_{\rm scat.1}(4_\alpha 3_{\alpha'}|2_{\alpha'} 1_\alpha) 
		\nonumber
		\\
		& = {1 \over 2}\sum_{\sigma_4, \sigma_3, \sigma_2, \sigma_1}  
		2\pi | \langle \vec{Q}_{\alpha} + \vec{p}_1, \sigma_1, \alpha; \   \vec{Q}_{\alpha'} + 
		\vec{p}_2, \sigma_2, \alpha' | V| \vec{Q}_{\alpha'} + \vec{p}_3, \sigma_3, \alpha' ; \   
		\vec{Q}_{\alpha} + \vec{p}_4, \sigma_4, \alpha \rangle_b |^2 \delta_{\vp_1+\vp_2, 
		\vp_3+\vp_4},
\end{align}	
\begin{align}
		\label{eq:ap:W_scat2}
		&W_{\rm scat.2}(4_{\alpha'} 3_{\alpha'}|2_{\alpha} 1_\alpha)   
		\nonumber
		\\
		& = {1 \over 2}\sum_{\sigma_4, \sigma_3, \sigma_2, \sigma_1}  
		2\pi | \langle \vec{Q}_{\alpha} + \vec{p}_1, \sigma_1, \alpha; \   \vec{Q}_{\alpha} + 
		\vec{p}_2, \sigma_2, \alpha | V| \vec{Q}_{\alpha'} + \vec{p}_3, \sigma_3, \alpha'; \   
		\vec{Q}_{\alpha'} + \vec{p}_4, \sigma_4, {\alpha'} \rangle_b |^2 \delta_{\vp_1+\vp_2, 
		\vp_3+\vp_4},
\end{align}		
\begin{align}
		\label{eq:ap:W_dec1}
		&W_{\rm dec. 1}(4_\alpha | 3_{\alpha'} 2_{\alpha'}  1_\alpha)   
		\nonumber
		\\
		& = {1 \over 2}\sum_{\sigma_4, \sigma_3, \sigma_2, \sigma_1}  
		2\pi | \langle \vec{Q}_{\alpha} + \vec{p}_1, \sigma_1, \alpha; \ \vec{Q}_{\alpha'} + 
		\vec{p}_2, \sigma_2, \alpha'; \  \vec{Q}_{\alpha'} + \vec{p}_3, \sigma_3, \alpha'  | V|   
		\vec{Q}_{\alpha} + \vec{p}_4, \sigma_4, \alpha \rangle_b |^2  \delta_{\vp_1+ \vp_2, \vp_4 - 
		\vp_3},
\end{align}	
\begin{align}
		\label{eq:ap:W_dec2}
		&W_{\rm dec. 2}(4_{\alpha'} | 3_{\alpha'} 2_{\alpha}  1_\alpha) 
		\nonumber
		\\
		& = {1 \over 2}\sum_{\sigma_4, \sigma_3, \sigma_2, \sigma_1}  
		2\pi | \langle \vec{Q}_{\alpha} + \vec{p}_1, \sigma_1, \alpha; \ \vec{Q}_{\alpha} + 
		\vec{p}_2, \sigma_2, \alpha; \  \vec{Q}_{\alpha'} + \vec{p}_3, \sigma_3, \alpha'  | V|   
		\vec{Q}_{\alpha'} + \vec{p}_4, \sigma_4, \alpha' \rangle_b |^2  \delta_{\vp_1+ \vp_2, \vp_4 
		- \vp_3},
\end{align}	
\begin{align}
		\label{eq:ap:W_coal}
		&W_{\rm coal.}(4_\alpha 3_{\alpha'} 2_{\alpha'} | 1_\alpha)   
		\nonumber
		\\
		& = {1 \over 2}\sum_{\sigma_4, \sigma_3, \sigma_2, \sigma_1}  
		2\pi | \langle \vec{Q}_{\alpha} + \vec{p}_1, \sigma_1, \alpha  | V| \vec{Q}_{\alpha'} + 
		\vec{p}_2, \sigma_2, \alpha'; \  \vec{Q}_{\alpha'} + \vec{p}_3, \sigma_3, \alpha'; \   
		\vec{Q}_{\alpha} + \vec{p}_4, \sigma_4, \alpha \rangle_b |^2  \delta_{\vp_1-\vp_2, 
		\vp_3+\vp_4},
\end{align}	
where  $\langle f | V| i \rangle_b$ are the matrix elements of the effective interaction 
Hamiltonian in the basis of Bogoliubov excitations given in Appendix \ref{sec:col_mat},
and the factor $1/2$ arises due to averaging over the spin index $\sigma_1$.
The energy delta functions are already taken into account in the expression \eqref{eq:col_int}.
}	
The functions $W_{ q}(i|f)$ are symmetric with respect to permutations of quantum numbers of the 
Bogoliubov excitations in the initial state, 
as well as in the final states [e.g., $W_{\rm scat.}(4_\alpha 
3_{\alpha}|2_{\alpha} 1_\alpha) = W_{\rm scat.}(3_\alpha 4_{\alpha}|2_{\alpha} 1_\alpha) = W_{\rm 
scat.}(4_\alpha 3_{\alpha}|1_{\alpha} 2_\alpha)$ and so on]	
since the matrix elements $\langle f | V| i \rangle_b$ are antisymmetric with respect to such 
transformations  (see Appendix \ref{sec:col_mat}).
The function $W_{ q}(i|f)$ is also symmetric with respect to interchange of initial 
and 
final states [e.g., $W_{\rm scat.}(4_\alpha 3_{\alpha}|2_{\alpha} 1_\alpha) = W_{\rm 
scat.}(2_{\alpha} 
1_\alpha|4_\alpha 3_{\alpha})$]. This symmetry follows from the Hermiticity of the matrix 
$\langle f | V| i \rangle_b$.

{
Considering integrals $I_{\alpha\alpha'}$, let us
analyze, for example,  
the expression in the first square brackets in \eqref{eq:col_int}.
It can be represented as
\begin{equation}
		\label{eq:ap:sq_br_ex}
		\mF_{\vp_{1}+ {\vec Q}_{\alpha}}^{(\alpha)} \mF_{\vp_{2}+ {\vec Q}_{\alpha'}}^{(\alpha')}  
		\mF_{\vp_{3}+ {\vec Q}_{\alpha'}}^{(\alpha')} \mF_{\vp_{4}+ {\vec Q}_{\alpha}}^{(\alpha)} 
		\left( \frac{1-\mF_{\vp_{2}+ {\vec Q}_{\alpha'}}^{(\alpha')}}{ \mF_{\vp_{2}+ {\vec 
		Q}_{\alpha'}}^{(\alpha')}}    \frac{1-\mF_{\vp_{1}+ {\vec Q}_{\alpha}}^{(\alpha)}}{ 
		\mF_{\vp_{1}+ {\vec Q}_{\alpha}}^{(\alpha)}} -
		\frac{1-\mF_{\vp_{3}+ {\vec Q}_{\alpha'}}^{(\alpha')}}{ \mF_{\vp_{3}+ {\vec 
		Q}_{\alpha'}}^{(\alpha')}}    \frac{1-\mF_{\vp_{4}+ {\vec Q}_{\alpha}}^{(\alpha)}}{ 
		\mF_{\vp_{4}+ {\vec Q}_{\alpha}}^{(\alpha)}} 
		\right).
\end{equation}
Using Eqs.\ \eqref{eq:mF_comp_tmp}, \eqref{eq:ol_mF}, and \eqref{eq:f1},
one can write 
\begin{equation}
	\label{ff}
		\frac{1-\mF_{\vp+ {\vec Q}_{\alpha}}^{(\alpha)}}{ \mF_{\vp+ {\vec Q}_{\alpha}}^{(\alpha)}} 
		\approx \frac{1-\overline{\mF}_{\vp+ {\vec Q}_{\alpha},0}^{(\alpha)}}{ \overline{\mF}_{\vp+ 
		{\vec Q}_{\alpha},0}^{(\alpha)}} \left(1 - \frac{\phi_\alpha}{T} \right) = 
		\exp\left(\mathfrak{E}_{\vp + \vQa}^{(\alpha)}-\vec{p}\, \vec{u} \over T\right)  \left(1 - 
		\frac{ \phi_\alpha}{T} \right),
\end{equation}
where we linearized this expression with respect to explicitly written function 
$\phi_\alpha$ 
but keep it 
untouched
inside the distribution functions $\overline{\mF}_{\vp+ {\vec Q}_{\alpha}}^{(\alpha)}$.%
\footnote{Recall that the distribution function $\overline{\mF}_{\vp+ 
{\vec Q}_{\alpha}}^{(\alpha)}$ depends on the local Bogoliubov excitation energy 
$\mathfrak{E}_{\vp + \vQa}^{(\alpha)}$, which, in turn, depends on the distribution functions  
${\mF}_{\vp+ {\vec Q}_{\alpha}}^{(\alpha)}$.}
%
Plugging Eq.\ \eqref{ff} into \eqref{eq:ap:sq_br_ex} 
and taking into account the energy 
delta function from the integral \eqref{eq:col_int} and the Kronecker delta 
from the expression \eqref{eq:ap:W_scat}, one obtains%
%
\footnote{It should be  emphasized that the true energy $\mathfrak{E}_{\vp 
+ \vQa}^{(\alpha)}$ is conserved during the collision event, not the equilibrium energy 
$\mathfrak{E}_{\vp + \vQa,0}^{(\alpha)}$.} 
%
\begin{equation}
		\overline{\mF}_{\vp_{1}+ {\vec Q}_{\alpha},0}^{(\alpha)} \overline{\mF}_{\vp_{2}+ {\vec 
		Q}_{\alpha'},0}^{(\alpha')}  (1-\overline{\mF}_{\vp_{3}+ {\vec Q}_{\alpha'},0}^{(\alpha')}) 
		(1-\overline{\mF}_{\vp_{4}+ {\vec Q}_{\alpha},0}^{(\alpha)})
		\frac{\phi_{4\alpha} + \phi_{3\alpha'} - \phi_{2\alpha'} - \phi_{1\alpha} }{T}.
\end{equation}
Now one can complete the linearization with respect to the Knudsen number $\mathcal{K}$ 
by replacing the distribution functions $\overline{\mF}_{\vp+ {\vec Q}_{\alpha},0}^{(\alpha)}$ with
the equilibrium distributions $\mF_{\vp+ {\vec Q}_{\alpha},0}^{(\alpha)}$.
Repeating similar procedure for all the expressions in the square brackets in Eq.\ 
\eqref{eq:col_int}, one gets	
%
\begin{align}
		\small
		\nonumber
		I_{\alpha\alpha'} &=  \frac{1}{T} \sum_{\vp_2,\vp_3,\vp_4} \Big\{ 
		\frac{1}{1+\delta_{\alpha,\alpha'}}W_{\rm scat.1}(4_\alpha 3_{\alpha'}|2_{\alpha'} 
		1_\alpha) 
		\delta( \mathfrak{E}_{\vp_{4}+ {\vec Q}_{\alpha},0}^{(\alpha)} + \mathfrak{E}_{\vp_{3}+ 
		{\vec Q}_{\alpha'},0}^{(\alpha')} -  \mathfrak{E}_{\vp_{2}+ {\vec 
		Q}_{\alpha'},0}^{(\alpha')} - \mathfrak{E}_{\vp_{1}+ {\vec Q}_{\alpha},0}^{(\alpha)})\times 
		\\
		\nonumber
		&\times \delta_{\vp_1+\vp_2,  \vp_3 + \vp_4}  \mF_{\vp_{1}+ {\vec Q}_{\alpha},0}^{(\alpha)} 
		\mF_{\vp_{2}+ {\vec Q}_{\alpha'},0}^{(\alpha')} (1-\mF_{\vp_{3}+ {\vec 
		Q}_{\alpha'},0}^{(\alpha')}) (1-\mF_{\vp_{4}+ {\vec Q}_{\alpha},0}^{(\alpha)}) 
		\left[ \phi_{4\alpha} + \phi_{3\alpha'} - \phi_{2\alpha'} - \phi_{1\alpha} \right]+ 
		\\
		\nonumber
		\\
		\nonumber
		+ &\frac{1- \delta_{\alpha,\alpha'}}{2}W_{\rm scat.2}(4_{\alpha'} 3_{\alpha'}|2_{\alpha} 
		1_\alpha) 
		\delta( \mathfrak{E}_{\vp_{4}+ {\vec Q}_{\alpha'},0}^{(\alpha')} + \mathfrak{E}_{\vp_{3}+ 
		{\vec Q}_{\alpha'},0}^{(\alpha')} -  \mathfrak{E}_{\vp_{2}+ {\vec Q}_{\alpha},0}^{(\alpha)} 
		- \mathfrak{E}_{\vp_{1}+ {\vec Q}_{\alpha},0}^{(\alpha)})\times 
		\\
		\nonumber
		&\times \delta_{\vp_1+\vp_2,  \vp_3 + \vp_4}  \mF_{\vp_{1}+ {\vec Q}_{\alpha},0}^{(\alpha)} 
		\mF_{\vp_{2}+ {\vec Q}_{\alpha},0}^{(\alpha)} (1-\mF_{\vp_{3}+ {\vec 
		Q}_{\alpha'},0}^{(\alpha')}) (1-\mF_{\vp_{4}+ {\vec Q}_{\alpha'},0}^{(\alpha')}) 
		\left[ \phi_{4\alpha'} + \phi_{3\alpha'} - \phi_{2\alpha} - \phi_{1\alpha} \right]+ 
		\\
		\nonumber
		\\
		\nonumber
		&+   \frac{1- \delta_{\alpha,\alpha'}}{2} W_{\rm dec. 1}( 4_\alpha | 3_{\alpha'} 
		2_{\alpha'} 1_{\alpha} )
		\delta( \mathfrak{E}_{\vp_{4}+ {\vec Q}_{\alpha},0}^{(\alpha)} - \mathfrak{E}_{\vp_{3}+ 
		{\vec Q}_{\alpha'},0}^{(\alpha')} -  \mathfrak{E}_{\vp_{2}+ {\vec 
		Q}_{\alpha'},0}^{(\alpha')} - \mathfrak{E}_{\vp_{1}+ {\vec Q}_{\alpha},0}^{(\alpha)})\times 
		\\
		\nonumber
		&\times  \delta_{\vp_1+\vp_2,  \vp_4 - \vp_3}  \mF_{\vp_{1}+ {\vec 
		Q}_{\alpha},0}^{(\alpha)} \mF_{\vp_{2}+ {\vec Q}_{\alpha'},0}^{(\alpha')} \mF_{\vp_{3}+ 
		{\vec Q}_{\alpha'},0}^{(\alpha')} (1-\mF_{\vp_{4}+ {\vec Q}_{\alpha},0}^{(\alpha)})\left[ 
		\phi_{4\alpha} - \phi_{3\alpha'} - \phi_{2\alpha'} - \phi_{1\alpha} \right]
		\\
		\nonumber
		\\
		\nonumber
		&+ \frac{1}{1+\delta_{\alpha,\alpha'}} W_{\rm dec. 2}(4_{\alpha'} |  3_{\alpha'}  
		2_{\alpha} 1_\alpha)  
		\delta( \mathfrak{E}_{\vp_{4}+ {\vec Q}_{\alpha'},0}^{(\alpha')} - \mathfrak{E}_{\vp_{3}+ 
		{\vec Q}_{\alpha'},0}^{(\alpha')} -  \mathfrak{E}_{\vp_{2}+ {\vec Q}_{\alpha},0}^{(\alpha)} 
		- \mathfrak{E}_{\vp_{1}+ {\vec Q}_{\alpha},0}^{(\alpha)})\times \\
		\nonumber
		&\times  \delta_{\vp_1+\vp_2,  \vp_4 - \vp_3}  \mF_{\vp_{1}+ {\vec 
		Q}_{\alpha},0}^{(\alpha)} \mF_{\vp_{2}+ {\vec Q}_{\alpha},0}^{(\alpha)} \mF_{\vp_{3}+ {\vec 
		Q}_{\alpha'},0}^{(\alpha')} (1-\mF_{\vp_{4}+ {\vec Q}_{\alpha'},0}^{(\alpha')})\left[ 
		\phi_{4\alpha'} - \phi_{3\alpha'} - \phi_{2\alpha} - \phi_{1\alpha} \right]	
		\\
		\nonumber
		\\
		\nonumber	
		&+ \frac{1}{2 + 4\delta_{\alpha,\alpha'}} W_{\rm coal.} (4_\alpha3_{\alpha'} 2_{\alpha'} | 
		1_\alpha)
		\delta( \mathfrak{E}_{\vp_{4}+ {\vec Q}_{\alpha},0}^{(\alpha)} + \mathfrak{E}_{\vp_{3}+ 
		{\vec Q}_{\alpha'},0}^{(\alpha')} +  \mathfrak{E}_{\vp_{2}+ {\vec 
		Q}_{\alpha'},0}^{(\alpha')} - \mathfrak{E}_{\vp_{1}+ {\vec Q}_{\alpha},0}^{(\alpha)})\times 
		\\
		&\times \delta_{\vp_1-\vp_2,  \vp_3 + \vp_4}    
		(1 - \mF_{\vp_{1}+ {\vec Q}_{\alpha},0}^{(\alpha)} )\mF_{\vp_{2}+ {\vec 
		Q}_{\alpha'},0}^{(\alpha')} \mF_{\vp_{3}+ {\vec Q}_{\alpha'},0}^{(\alpha')} \mF_{\vp_{4}+ 
		{\vec Q}_{\alpha},0}^{(\alpha)}\left[ \phi_{4\alpha} + \phi_{3\alpha'} + \phi_{2\alpha'} - 
		\phi_{1\alpha} \right]
		\Big\}.
		\label{eq:col_int_2}
\end{align}	
Here we  replaced  
$\mathfrak{E}_{\vp+{\vec Q}_{\alpha}}^{(\alpha)}$ with  
$\mathfrak{E}_{\vp+ {\vec Q}_{\alpha},0}^{(\alpha)}$ in the 
energy delta functions, which is justifiable in the linear approximation, 
and extracted the momentum Kronecker deltas from the transition 
probabilities \eqref{eq:ap:W_scat}--\eqref{eq:ap:W_coal}. 
One can see that, if $\phi_\alpha = 0$ ($\mF_{\vp+ {\vec Q}_{\alpha}}^{(\alpha)} = \overline 
\mF_{\vp+ {\vec Q}_{\alpha},0}^{(\alpha)} $), the collision integral vanishes. 
Thus, the distribution functions $\overline \mF_{\vp+ {\vec Q}_{\alpha},0}^{(\alpha)}$ can be 
considered as solutions to Eq.\ \eqref{eq:col_int_eq}.
	
Our next aim will be to find the expression for the integral $\sum_{\vp_1, \sigma_1} \vp_1 
I_{\alpha\alpha'}$.
To that end,  we multiply Eq.\ \eqref{eq:col_int_2} by $\vp_1$, 
sum the result over the quantum states $1_\alpha$,
and substitute the expression \eqref{eq:phi_def}.
The result is
%
\begin{align}
		\small
		\nonumber
		&\sum_{\vp_1, \sigma_1} \vp_1 I_{\alpha\alpha'} 
		\nonumber
		\\
		&  = - \frac{2}{T} \sum_{\vp_1,\vp_2,\vp_3,\vp_4} \Big\{
		\frac{1}{1+\delta_{\alpha,\alpha'}}W_{\rm scat.1}(4_\alpha 3_{\alpha'}|2_{\alpha'} 
		1_\alpha) 
		\delta( \mathfrak{E}_{\vp_{4}+ {\vec Q}_{\alpha},0}^{(\alpha)} + \mathfrak{E}_{\vp_{3}+ 
		{\vec Q}_{\alpha'},0}^{(\alpha')} -  \mathfrak{E}_{\vp_{2}+ {\vec 
		Q}_{\alpha'},0}^{(\alpha')} - \mathfrak{E}_{\vp_{1}+ {\vec Q}_{\alpha},0}^{(\alpha)})
		\nonumber
		\\
		&\times \delta_{\vp_1+\vp_2,  \vp_3 + \vp_4}  \mF_{\vp_{1}+ {\vec Q}_{\alpha},0}^{(\alpha)} 
		\mF_{\vp_{2}+ {\vec Q}_{\alpha'},0}^{(\alpha')} (1-\mF_{\vp_{3}+ {\vec 
		Q}_{\alpha'},0}^{(\alpha')}) (1-\mF_{\vp_{4}+ {\vec Q}_{\alpha},0}^{(\alpha)}) 
		\vp_1 \left[ (\vp_1 - \vp_4) (\vec{V}_{ i \alpha} - \vec{V}_{ i \alpha'} )\right] 
		\nonumber
		\\
		\nonumber\\
		& + \frac{1-\delta_{\alpha,\alpha'}}{2}W_{\rm scat.2}(4_{\alpha'} 3_{\alpha'}|2_{\alpha} 
		1_\alpha) 
		\delta( \mathfrak{E}_{\vp_{4}+ {\vec Q}_{\alpha'},0}^{(\alpha')} + \mathfrak{E}_{\vp_{3}+ 
		{\vec Q}_{\alpha'},0}^{(\alpha')} -  \mathfrak{E}_{\vp_{2}+ {\vec Q}_{\alpha},0}^{(\alpha)} 
		- \mathfrak{E}_{\vp_{1}+ {\vec Q}_{\alpha},0}^{(\alpha)})
		\nonumber\\
		&\times \delta_{\vp_1+\vp_2,  \vp_3 + \vp_4}  \mF_{\vp_{1}+ {\vec Q}_{\alpha},0}^{(\alpha)} 
		\mF_{\vp_{2}+ {\vec Q}_{\alpha},0}^{(\alpha)} (1-\mF_{\vp_{3}+ {\vec 
		Q}_{\alpha'},0}^{(\alpha')}) (1-\mF_{\vp_{4}+ {\vec Q}_{\alpha'},0}^{(\alpha')}) 
		\vp_1 \left[ (\vp_1 + \vp_2) (\vec{V}_{ i \alpha} - \vec{V}_{ i \alpha'} )\right] 
		\nonumber\\
		\nonumber\\
		&+\frac{1-\delta_{\alpha\alpha'}}{2} W_{\rm dec. 1}( 4_\alpha | 3_{\alpha'} 2_{\alpha'} 
		1_{\alpha} )
		\delta( \mathfrak{E}_{\vp_{4}+ {\vec Q}_{\alpha},0}^{(\alpha)} - \mathfrak{E}_{\vp_{3}+ 
		{\vec Q}_{\alpha'},0}^{(\alpha')} -  \mathfrak{E}_{\vp_{2}+ {\vec 
		Q}_{\alpha'},0}^{(\alpha')} - \mathfrak{E}_{\vp_{1}+ {\vec Q}_{\alpha},0}^{(\alpha)}) 
		\nonumber\\
		&\times  \delta_{\vp_1+\vp_2,  \vp_4 - \vp_3}  \mF_{\vp_{1}+ {\vec 
		Q}_{\alpha},0}^{(\alpha)} \mF_{\vp_{2}+ {\vec Q}_{\alpha'},0}^{(\alpha')} \mF_{\vp_{3}+ 
		{\vec Q}_{\alpha'},0}^{(\alpha')} (1-\mF_{\vp_{4}+ {\vec Q}_{\alpha},0}^{(\alpha)})
		\vp_1 \left[ (\vp_1 - \vp_4) (\vec{V}_{ i \alpha} - \vec{V}_{ i \alpha'} )\right]
		\nonumber
	\end{align}
	\begin{align}
		&+  \frac{1}{1+\delta_{\alpha,\alpha'}} W_{\rm dec. 2}(4_{\alpha'} |  3_{\alpha'}  
		2_{\alpha} 1_\alpha)  
		\delta( \mathfrak{E}_{\vp_{4}+ {\vec Q}_{\alpha'},0}^{(\alpha')} - \mathfrak{E}_{\vp_{3}+ 
		{\vec Q}_{\alpha'},0}^{(\alpha')} -  \mathfrak{E}_{\vp_{2}+ {\vec Q}_{\alpha},0}^{(\alpha)} 
		- \mathfrak{E}_{\vp_{1}+ {\vec Q}_{\alpha},0}^{(\alpha)}) 
		\nonumber\\
		&\times  \delta_{\vp_1+\vp_2,  \vp_4 - \vp_3}  \mF_{\vp_{1}+ {\vec 
		Q}_{\alpha},0}^{(\alpha)} \mF_{\vp_{2}+ {\vec Q}_{\alpha},0}^{(\alpha)} \mF_{\vp_{3}+ {\vec 
		Q}_{\alpha'},0}^{(\alpha')} (1-\mF_{\vp_{4}+ {\vec Q}_{\alpha'},0}^{(\alpha')})
		\vp_1 \left[ (\vp_1 + \vp_2) (\vec{V}_{ i \alpha} - \vec{V}_{ i \alpha'} )\right]
		\nonumber\\
		\nonumber\\
		&+ \frac{1}{2 + 4\delta_{\alpha,\alpha'}} W_{\rm coal.} (4_\alpha3_{\alpha'} 2_{\alpha'} | 
		1_\alpha)
		\delta( \mathfrak{E}_{\vp_{4}+ {\vec Q}_{\alpha},0}^{(\alpha)} + \mathfrak{E}_{\vp_{3}+ 
		{\vec Q}_{\alpha'},0}^{(\alpha')} +  \mathfrak{E}_{\vp_{2}+ {\vec 
		Q}_{\alpha'},0}^{(\alpha')} - \mathfrak{E}_{\vp_{1}+ {\vec Q}_{\alpha},0}^{(\alpha)})
			\nonumber\\
		& \times   \delta_{\vp_1-\vp_2,  \vp_3 + \vp_4}  
		(1 - \mF_{\vp_{1}+ {\vec Q}_{\alpha},0}^{(\alpha)}) \mF_{\vp_{2}+ {\vec 
		Q}_{\alpha'},0}^{(\alpha')} \mF_{\vp_{3}+ {\vec Q}_{\alpha'},0}^{(\alpha')} \mF_{\vp_{4}+ 
		{\vec Q}_{\alpha},0}^{(\alpha)}
		\vp_1 \left[ (\vp_1 - \vp_4) (\vec{V}_{ i \alpha} - \vec{V}_{ i \alpha'} )\right]	
		\Big\},
		\label{eq:ap:pIab_tmp}
\end{align}	
where we make use of the momentum conservation in particle collisions. 
The factor 2 in Eq.\ \eqref{eq:ap:pIab_tmp} arises from the summation over the spin index 
$\sigma_1$. 
Note that the sum \eqref{eq:ap:pIab_tmp} vanishes identically if $\alpha'  = \alpha$ 
(i.e., when,$\vec{V}_{ i \alpha'} = \vec{V}_{ i \alpha}$). 
This is a consequence of the fact that a given particle species cannot lose or gain momentum
through interaction with itself.
In what follows,	
we only consider a nontrivial case when $\alpha' = \beta \neq \alpha$.
Note that an arbitrary vector $\vec{a}$ can be added to the velocities $\vec{V}_{ i \alpha}$ 
without 
affecting the collision integrals.
This is a direct consequence of the ambiguity related to the definition of the velocity 
$\vec{u}$, see Sec. \ref{sec:diffusion}.}

Before proceeding further 
let us notice that, 
in view of the
properties of the matrix 
$\langle f | V| i \rangle_b$ 
discussed at the end of Appendix \ref{sec:col_mat},
the transition probabilities 
$W_{\rm coal.} (4_\alpha3_{\beta} 2_{\beta} | 1_\alpha)$ 
and $W_{\rm dec. 1}( 1_\alpha | 2_{\beta} 3_{\beta} 
4_{\alpha} )$
coincide,
$W_{\rm coal.} (4_\alpha3_{\beta} 2_{\beta} | 1_\alpha) = W_{\rm dec. 1}( 1_\alpha | 2_{\beta} 
3_{\beta} 4_{\alpha} )$.	
Accounting for this fact and using other symmetries of the 
transition probabilities $W_{ q}(i|f)$ [see a passage after Eq.\ \eqref{eq:ap:W_coal}]
one gets, after some redefinitions of running  variables, 
\begin{align}
		\nonumber
		&\sum_{\vp_1 \sigma_1} \vp_1  I_{\alpha\beta}
		\\
		\nonumber
		& =  - {1 \over T} \sum_{\vp_1,\vp_2,\vp_3,\vp_4}  \Big\{ 
		W_{\rm scat.1}(4_\alpha 3_{\beta}|2_{\beta} 1_\alpha) 
		\delta( \mathfrak{E}_{\vp_{4}+ \vQa,0}^{(\alpha)} + \mathfrak{E}_{\vp_{3}+ 
		\vQb,0}^{(\beta)} -  \mathfrak{E}_{\vp_{2}+ \vQb,0}^{(\beta)} - \mathfrak{E}_{\vp_{1}+ 
		\vQa,0}^{(\alpha)})
		\delta_{\vp_1 + \vp_2,  \vp_3 + \vp_4} 
		\\
		\nonumber	
		&\times {\mF}_{\vp_{1}+\vQa,0}^{(\alpha)} {\mF}_{\vp_{2}+ \vQb,0}^{(\beta)} 
		(1-{\mF}_{\vp_{3}+\vQb,0}^{(\beta)}) (1-{\mF}_{\vp_{4}+ \vQa,0}^{(\alpha)})
		(\vp_1 - \vp_4)\left[ (\vp_1 - \vp_4) (\vec{V}_{ i \alpha} - \vec{V}_{ i \beta} 
		)\right]
		\\
		\nonumber
		\\
		\nonumber
		& +\frac{1}{2} W_{\rm scat.2}(4_\beta 3_{\beta}|2_\alpha 1_\alpha) 
		\delta( \mathfrak{E}_{\vp_{4}+ \vQb,0}^{(\beta)} + \mathfrak{E}_{\vp_{3}+ 
		\vQb,0}^{(\beta)} -  \mathfrak{E}_{\vp_{2}+ \vQa,0}^{(\alpha)} - \mathfrak{E}_{\vp_{1}+ 
		\vQa,0}^{(\alpha)})
		\delta_{\vp_1 + \vp_2,  \vp_3 + \vp_4}  
		\\
		\nonumber
		&\times {\mF}_{\vp_{1}+\vQa,0}^{(\alpha)} {\mF}_{\vp_{2}+ \vQa,0}^{(\alpha)} 
		(1-{\mF}_{\vp_{3}+\vQb,0}^{(\beta)}) (1-{\mF}_{\vp_{4}+ \vQb,0}^{(\beta)})
		(\vp_1 + \vp_2)\left[ (\vp_1 + \vp_2) (\vec{V}_{ i \alpha} - \vec{V}_{ i \beta} 
		)\right]
		\\
		\nonumber
		\\
		\nonumber	
		& +  W_{\rm dec.1}(4_\alpha |  3_{\beta} 2_{\beta} 1_\alpha )  
		\delta( \mathfrak{E}_{\vp_{4}+ \vQa,0}^{(\alpha)} - \mathfrak{E}_{\vp_{3}+ 
		\vQb,0}^{(\beta)} -  \mathfrak{E}_{\vp_{2}+ \vQb,0}^{(\beta)} - \mathfrak{E}_{\vp_{1}+ 
		\vQa,0}^{(\alpha)})
		\delta_{\vp_1+\vp_2,  \vp_4 - \vp_3} 
		\\
		\nonumber
		&\times {\mF}_{\vp_{1}+\vQa,0}^{(\alpha)} {\mF}_{\vp_{2}+ \vQb,0}^{(\beta)} 
		{\mF}_{\vp_{3}+\vQb,0}^{(\beta)} (1-{\mF}_{\vp_{4}+ \vQa,0}^{(\alpha)})
		(\vp_1- \vp_4) \left[ (\vp_1 - \vp_4) (\vec{V}_{ i \alpha} - \vec{V}_{ i \beta} 
		)\right]
		\\
		\nonumber
		\\
		\nonumber
		& + W_{\rm dec.2} (4_\beta | 3_{\beta} 2_\alpha 1_{\alpha} )  
		\delta( \mathfrak{E}_{\vp_{4}+ \vQb,0}^{(\beta)} - \mathfrak{E}_{\vp_{3}+ \vQb,0}^{(\beta)} 
		-  \mathfrak{E}_{\vp_{2}+ \vQa,0}^{(\alpha)} - \mathfrak{E}_{\vp_{1}+ \vQa,0}^{(\alpha)})
		\delta_{\vp_1+\vp_2,  \vp_4 - \vp_3}
		\\
		&\times {\mF}_{\vp_{1}+\vQa,0}^{(\alpha)} {\mF}_{\vp_{2}+ \vQa,0}^{(\alpha)} 
		{\mF}_{\vp_{3}+\vQb,0}^{(\beta)} (1-{\mF}_{\vp_{4}+ \vQb,0}^{(\beta)})
		(\vp_1 + \vp_2)\left[ (\vp_1 + \vp_2) (\vec{V}_{ i \alpha} - \vec{V}_{ i \beta} 
		)\right] \Big\} .
		\label{eq:ap:pIab_tmp2}
\end{align}
This formula can be further simplified provided that the hydrodynamic velocities are small.
Since Eq.\ \eqref{eq:ap:pIab_tmp2} already contains a small difference $(\vec{V}_{ i \alpha} - 
\vec{V}_{ i \beta} )$, 
one just needs to set $\vec{u} = \vQa = \vQb = 0$ in all other functions in this expression.
The final result is 
\begin{equation}
		\sum_{\vp_1 \sigma_1} \vp_1  I_{\alpha\beta} = -J_{\alpha\beta}(\vec{V}_{ i \alpha} - 
		\vec{V}_{ i \beta} ),
\end{equation}
where 
\begin{align}
		\small
		&J_{\alpha\beta} =    {1 \over 3 T} \sum_{\vp_1,\vp_2,  \vp_3,\vp_4} 
		\nonumber
		\\
		\nonumber
		&  \Big[
		W_{\rm scat.1}(4_\alpha 3_{\beta}|2_{\beta} 1_\alpha)  
		\delta({E}_{\vp_{4}}^{(\alpha)} + {E}_{\vp_{3}}^{(\beta)} -  {E}_{\vp_{2}}^{(\beta)} - 
		{E}_{\vp_{1}}^{(\alpha)})
		\delta_{\vp_1+\vp_2,  \vp_3 + \vp_4}
		\mathfrak{f}_{\vp_{1}}^{(\alpha)} \mathfrak{f}_{\vp_{2}}^{(\beta)} 
		(1-\mathfrak{f}_{\vp_{3}}^{(\beta)}) (1-\mathfrak{f}_{\vp_{4}}^{(\alpha)})
		(\vp_1 - \vp_4)^2 
		\\
		\nonumber
		\\
		\nonumber
		& + 
		\frac{1}{2} W_{\rm scat.2}(4_\beta 3_{\beta}|2_{\alpha} 1_\alpha)  
		\delta({E}_{\vp_{4}}^{(\beta)} + {E}_{\vp_{3}}^{(\beta)} -  {E}_{\vp_{2}}^{(\alpha)} - 
		{E}_{\vp_{1}}^{(\alpha)})
		\delta_{\vp_1+\vp_2,  \vp_3 + \vp_4}
		\mathfrak{f}_{\vp_{1}}^{(\alpha)} \mathfrak{f}_{\vp_{2}}^{(\alpha)} 
		(1-\mathfrak{f}_{\vp_{3}}^{(\beta)}) (1-\mathfrak{f}_{\vp_{4}}^{(\beta)})
		(\vp_1 + \vp_2)^2 
		\\
		\nonumber
		\\
		\nonumber
		& +  W_{\rm dec.1}(4_\alpha |  3_{\beta} 2_{\beta} 1_\alpha ) 
		\delta( {E}_{\vp_{4}}^{(\alpha)} - {E}_{\vp_{3}}^{(\beta)} -  {E}_{\vp_{2}}^{(\beta)} - 
		{E}_{\vp_{1}}^{(\alpha)})
		\delta_{\vp_1+\vp_2,  \vp_4 - \vp_3}
		\mathfrak{f}_{\vp_{1}}^{(\alpha)} \mathfrak{f}_{\vp_{2}}^{(\beta)} 
		\mathfrak{f}_{\vp_{3}}^{(\beta)} (1-\mathfrak{f}_{\vp_{4}}^{(\alpha)})
		(\vp_1- \vp_4)^2 
		\\
		\nonumber
		\\
		& + W_{\rm dec.2} (4_\beta | 3_{\beta} 2_\alpha 1_{\alpha} ) 
		\delta( {E}_{\vp_{4}}^{(\beta)} - {E}_{\vp_{3}}^{(\beta)} -  {E}_{\vp_{2}}^{(\alpha)} - 
		{E}_{\vp_{1}}^{(\alpha)})
		\delta_{\vp_1+\vp_2,  \vp_4 - \vp_3}
		\mathfrak{f}_{\vp_{1}}^{(\alpha)}  \mathfrak{f}_{\vp_{2}}^{(\alpha)} 
		\mathfrak{f}_{\vp_{3}}^{(\beta)} (1- \mathfrak{f}_{\vp_{4}}^{(\beta)})
		(\vp_1 + \vp_2)^2 \Big]
		\label{eq:ap:mtr}
\end{align}
is the momentum transfer rate.
	
Since the total momentum of the mixture is conserved in collisions,  
the momentum transfer rate 
should be symmetric 
with respect to interchanging of particle species indices:
$J_{\alpha\beta} = J_{\beta\alpha}$.
It is easy to verify that Eq.\ \eqref{eq:ap:mtr} satisfies this property. 
Indeed, the scattering terms in Eq.\ \eqref{eq:ap:mtr} are obviously symmetric,
while the ``dec.1'' term turns into the  ``dec.2'' term (and vice versa) 
after the replacement $\alpha \leftrightarrow \beta$.
To see this, one should compare the expression \eqref{eq:ap:W_dec1} with \eqref{eq:ap:W_dec2}
after interchanging the indices ($\alpha \leftrightarrow \beta$) and
running variables ($\vp_1 \leftrightarrow \vp_3$).

The expression for the momentum transfer rate was obtained under assumption that both particle 
species are superfluid. If one of them (say, the species $\alpha$) is normal, 
the result should be modified in two ways. 
First, the corresponding distribution function $\fp$ and energy $E_{\vp}^{(\alpha)}$ 
of Bogoliubov excitations
should be replaced with the (quasi)particle  distribution function $\np$ and the energy 
$\varepsilon_{\vp}^{(\alpha)}$, respectively. 
Second, the collisions that do not conserve the number of 
(quasi)particles of species $\alpha$ should be disregarded,
i.e., 
one should set: 
$W_{\rm scat.2}(4_\beta 3_{\beta}|2_{\alpha} 1_\alpha)   =  W_{\rm 
dec.2} (4_\beta | 3_{\beta} 2_\alpha 1_{\alpha} ) = 0$.


\bibliography{mn-jour,paper}   

\end{document}